\documentclass{article}

\usepackage{arxiv}

\title{Semi-static Conditions in Low-latency C++ for High Frequency Trading: \\ Better than Branch Prediction Hints}

\date{August 27, 2023}

\newif\ifuniqueAffiliation
\uniqueAffiliationtrue

\author{ \href{https://orcid.org/0000-0001-6846-6649}{\includegraphics[scale=0.06]{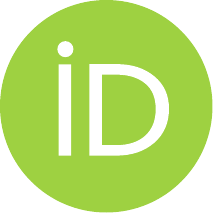}\hspace{1mm}Paul Alexander Bilokon} \\
	Department of Computing\\
        Department of Mathematics\\
	Imperial College London\\
	South Kensington Campus\\
        Exhibition Road\\
        London SW7 2AZ\\
	\texttt{paul.bilokon@imperial.ac.uk} \\
	\And
	Maximilian Lucuta \\
	Department of Computing\\
	Imperial College London\\
	South Kensington Campus\\
        Exhibition Road\\
        London SW7 2AZ\\
	\texttt{maxlucuta.17@gmail.com} \\
	\And
	Erez Shermer \\
	qSpark LLC\\
	3422 Old Capitol Trl\\
	Wilmington, DE, 19808-6124 \\
        United States\\
	\texttt{eshermer@qspark.co}\\
}

\renewcommand{\shorttitle}{Semi-static Conditions}

\PassOptionsToPackage{table}{xcolor}
\usepackage{textpos}
\usepackage[numbers]{natbib} 
\usepackage{tabularx,longtable,multirow,subfigure,caption}
\usepackage{fncylab} 
\usepackage{fancyhdr} 
\usepackage{url} 
\usepackage[english]{babel}
\usepackage{amsmath}
\usepackage{graphicx}
\usepackage{dsfont}
\usepackage{epstopdf} 
\usepackage{backref} 
\usepackage{array}
\usepackage{latexsym}
\usepackage[pdftex,pagebackref,hypertexnames=false,colorlinks]{hyperref} 
\usepackage{listings}
\usepackage{caption,subcaption}
\usepackage{circuitikz}
\usepackage{pgfplots}
\usepackage{booktabs}
\usepackage[table]{xcolor}
\usepackage{graphicx}
\usepackage{amssymb}
\usepackage{pifont}

\hypersetup{pdftitle={},
  pdfsubject={}, 
  pdfauthor={},
  pdfkeywords={}, 
  pdfstartview=FitH,
  pdfpagemode={UseOutlines},
  bookmarksnumbered=true, bookmarksopen=true, colorlinks,
    citecolor=black,%
    filecolor=black,%
    linkcolor=black,%
    urlcolor=black}

\usepackage[all]{hypcap}

\renewcommand*{\ttdefault}{cmtt}
\definecolor{lightgray}{RGB}{242,242,242}

\def\@makechapterhead#1{%
  \vspace*{10\p@}%
  {\parindent \z@ \raggedright \sffamily
    \interlinepenalty\@M
    \Huge\bfseries \thechapter \space\space #1\par\nobreak
    \vskip 30\p@
  }}

\def\@makeschapterhead#1{%
  \vspace*{10\p@}%
  {\parindent \z@ \raggedright
    \sffamily
    \interlinepenalty\@M
    \Huge \bfseries  #1\par\nobreak
    \vskip 30\p@
  }}

\allowdisplaybreaks

\hypersetup{
pdftitle={Semi-static Conditions in Low-latency C++ for High Frequency Trading: Better than Branch Prediction Hints},
pdfsubject={q-bio.NC, q-bio.QM},
pdfauthor={Paul Alexander Bilokon, Maximilian Lucuta, Erez Shermer},
pdfkeywords={semi-static conditions, high-performance computing, low latency, high-frequency trading, C++},
}

\begin{document}
\maketitle

\begin{abstract}
Conditional branches pose a challenge for code optimisation, particularly in low latency settings. For better performance, processors leverage dedicated hardware to predict the outcome of a branch and execute the following instructions speculatively, a powerful optimisation. Modern branch predictors employ sophisticated algorithms and heuristics that utilise historical data and patterns to make predictions, and often, are extremely effective at doing so. Consequently, programmers may inadvertently underestimate the cost of misprediction when benchmarking code with synthetic data that is either too short or too predictable. While eliminating branches may not always be feasible, C++20 introduced the [[likely]] and [[unlikely]] attributes that enable the compiler to perform spot optimisations on assembly code associated with likely execution paths. Can we do better than this?

This work presents the development of a novel language construct, referred to as a semi-static condition, which enables programmers to dynamically modify the direction of a branch at run-time by modifying the assembly code within the underlying executable. Subsequently, we explore scenarios where the use of semi-static conditions outperforms traditional conditional branching, highlighting their potential applications in real-time machine learning and high-frequency trading. Throughout the development process, key considerations of performance, portability, syntax, and security were taken into account. The resulting construct is open source and can be accessed at \url{https://github.com/maxlucuta/semi-static-conditions}.
\end{abstract}

\section*{Acknowledgments}
Thank you to Jonathan Keinan, Lior Keren, Nataly Rasovsky, Nimrod Sapir, Michael Stevenson, and other colleagues at qSpark for many constructive suggestions.

\section{Introduction}

\subsection{Aims and Approach}

The implementation of semi-static conditions, a language construct that can programmatically alter the direction of a branch at execution time, and the identification of applications where it outperforms conditional branching constitute the primary objectives of this project. To achieve this goal, minimal run-time overhead associated with branch-taking must be ensured. A key challenge in the development process is to create a language construct that mimics the behavior of direct method invocations while simultaneously providing the ability for dynamic switching, in order to offer competitive performance with existing software solutions.

The research contribution of this paper is structured into two stages. In the first stage, the focus is on the development of the construct, with an emphasis on the strategies employed to facilitate low-latency branch-taking through binary editing and addressing the associated considerations of syntax, performance and portability. The second stage shifts the focus to exploring instances where this construct surpasses conditional statements, and understanding the behaviour and implications of semi-static conditions at the hardware level. Additionally, a comprehensive software archive showcasing examples of usage will be created and made readily accessible.

\subsection{Research Context}

Current research in branch prediction optimization has predominantly concentrated on hardware-based solutions that aim to enhance speculative efficiencies and overall performance of modern CPU's. However, despite these efforts, the problem of branch prediction remains unresolved \cite{lin2019branch}, with limited attention given to software-based optimizations. For typical commercial applications, the pursuit of such micro-optimizations is often unnecessary and may introduce additional complexity. Nonetheless, in industries like High Frequency Trading, even slight improvements in execution latencies on the clock cycle level are highly valued. Consequently, these optimizations are highly sought-after and can provide significant competitive advantages. Due to their crucial role in determining a firm's profitability and success, cutting-edge research on software-based optimizations in such industries is typically shrouded in secrecy.

Several books have attempted to bridge the divide between computational and financial research, aiming to combine mathematical modeling and the development of algorithmic trading strategies (e.g., \cite{aldridge2013high, cartea2015algorithmic}). Additionally, numerous public conferences are available, focusing on the development of low-latency execution systems (e.g., \cite{nimrod2019cpp, carl2017cpp, chandler2014cpp, david2022cpp}), emphasizing topics such as data structures, atomics, and low-latency design patterns. Notably, there has been some emphasis on branchless design \cite{fedor2021cpp}, which showcases some common alternatives to conditional statements with high misprediction rates. However, the strategies employed in this context lack flexibility and rely on the assumption that branches can be pre-computed without incurring significant costs.

In contrast, there is a wealth of literature and extensively documented resources available for C++ \cite{stroustrup1986overview, stroustrup1994design}, which is widely used as the primary language in the development of low-latency trading systems. However, when it comes to the development of semi-static conditions and strategies involving the modification of running executables using C++, the scope becomes more specialized. Nevertheless, these techniques are well-documented and find applications in various areas such as debuggers, profilers, hot patching software, and security tools (e.g., \cite{hunt1999detours, security2005, gnu2023}).

There is a significant scarcity of ultra-low latency C++ tools, particularly those specifically designed to address control flow problems, offering both rigorous application verification and superior performance. While it is possible to come across online posts outlining small-scale experiments that focus on minor branch optimizations in specific scenarios (as demonstrated in \cite{fedor2021cpp}), such micro-optimizations are often overlooked and left to the compiler and hardware to handle. Interestingly, extensive research has been conducted on the true cost of branch misprediction \cite{branchMis2006, infoq2023}, highlighting its significant contribution to performance bottlenecks in low-latency systems. While these articles provide in-depth analysis of benchmark data and performance losses, the strategies proposed for preventing branch mispredictions remain lackluster or non-existent.

In light of the existing literature and its insights into software-based branch optimization problems, it becomes apparent that a research gap exists, which this study aims to fill. The motivation behind this research lies in providing solutions to the aforementioned void and contributing to the understanding and resolution of software-based branch optimization.

\subsection{Outline}

In the research review portion of this work, the focus is on encapsulating and critically analyzing research pertaining to the problem at hand. The section begins with an outline of modern CPU pipelined architectures and the implications and cost of conditional branching. Next, attention is shifted to advancements in hardware-based solutions, outlining the strengths and weaknesses of various schemes when encountering branches of different predictability. The focus is then redirected to software, providing an outline of C++ and its importance in the development of low-latency trading systems. Language features that exist exclusively to optimize branch prediction are also discussed. Finally, discussions are concluded with HFT, examining economic effects, known technical advancements and the problems that remain to be solved in the industry.

The next two sections dedicated to the research contribution, which can be conceptualized as consisting of two stages: the development of semi-static conditions and data-backed applications. In the development stage, a sequential approach is adopted to identify the requirements and challenges associated with designing the language construct. The solution to the problem is outlined and demonstrated, providing an overview of key theory with subsequent design decisions and optimizations. The proceeding section demonstrates the instances where semi-static conditions offer superior performance compared to conditional branching, accompanied by detailed analyses and supporting benchmark data to substantiate the findings, with novel investigations into effects of binary editing on modern hardware. Furthermore, discussions are conducted on how the outlined scenarios can be incorporated into a commercial trading system, considering their suitability and practical implications.

Next, the software contribution alongside the various experimental methods employed to benchmark semi-static checks are critically evaluated. Detailed examples of usage are provided, along with recommendations for maximizing the security and reliability of the language construct. We then provide some concluding remarks and future directions, discussing potential areas for further research and development.

\subsection{Summary of Achievements}

The project has achieved significant milestones across multiple dimensions. The research contribution offers valuable insights and approaches for developing software-based branch optimizations, with comprehensive and rigorous investigations into the resulting hardware level behaviours, filling an important research gap in the academe. By employing unconventional yet effective binary editing strategies, the project enables ultra-low latency branch execution through the decoupling of condition evaluation logic and branch taking, controlled directly by the programmer. Through extensive benchmarking and detailed performance analysis in pseudo-realistic scenarios, the proof-of-concept semi-static checks demonstrate their real-world applicability, particularly within real-time systems and high-frequency trading.

The research contribution is encapsulated within a open-source library that allows programmers to utilize the language construct for both commercial and experimental purposes. With a focus on syntax, security, efficiency, and portability, the semi-static conditions can be seamlessly integrated into high-performance real-time systems, as exemplified in domains such as high-frequency trading and real-time machine learning. The availability of this library further enhances the practical usability and potential impact of the project's achievements.

\section{Background}
\subsection{Pipelining and Conditional Branches}

The microprocessor is an integrated circuit responsible for executing arithmetic, logic, control, and input/output operations in a digital system \cite{hennessy2017computer}. In the early 1970s, the Intel 4004 emerged as the first commercially available microprocessor, initially designed as a 4-bit central processing unit (CPU) with a clock speed of 740 kHz for early printing calculators \cite{intel4004, s2015modern}. Over the past three decades, advancements in integrated circuit technology have enabled microprocessor manufacturers to develop increasingly sophisticated CPUs, driven by the exponential growth in transistor density dictated by Moore's law \cite{moore1965cramming}. Alongside these developments, the introduction of modern instruction sets and standardized operating systems has propelled the computational capabilities of contemporary computers to unprecedented heights.

Modern CPU's utilize pipelining as an implementation technique to exploit the inherent parallelism in instruction execution to improve overall throughput \cite{hennessy2017computer}. Similar to cars on an assembly line, pipelining allows for the overlapping execution of multiple instructions, with each step in the assembly line constituting a pipe stage that represents a phase in the fetch-decode-execute cycle. If all stages take the same amount of time, a pipeline with \emph{n} stages will achieve a throughput \emph{n} times faster than an un-pipelined counterpart, with the bottleneck stage bounding the number of processor cycles required for a single execution \cite{hennessy2017computer}. Whilst we provide a simplified schematic of instruction pipelining, it is important to note that modern processors vastly more complex pipeline architectures, often super-scalar with complex instruction sets and addressing modes to prevent memory-access conflicts in instruction/data memory, optimize register handling, and maximize instruction level-parallelism \cite{stallings2017computer}. 

\begin{figure}[!ht]
\centering
\resizebox{\textwidth}{!}{%
\begin{circuitikz}
\tikzstyle{every node}=[font=\huge]

\draw [ color={rgb,255:red,242; green,242; blue,242} , fill={rgb,255:red,242; green,242; blue,242}, line width=0.2pt ] (-11.5,20) rectangle (30,18);
\draw [ color={rgb,255:red,242; green,242; blue,242} , fill={rgb,255:red,242; green,242; blue,242}, line width=0.2pt ] (-11.5,16) rectangle (30,14.5);
\draw [ color={rgb,255:red,242; green,242; blue,242} , fill={rgb,255:red,242; green,242; blue,242}, line width=0.2pt ] (-11.5,13) rectangle (30,11.5);
\draw [ color={rgb,255:red,242; green,242; blue,242} , fill={rgb,255:red,242; green,242; blue,242}, line width=0.2pt ] (-11.5,10) rectangle (30,8);

\draw [, line width=5pt](-11.5,20) to[short] (30,20);
\draw [, line width=2pt](-2.5,18) to[short] (30,18);
\draw [, line width=2pt](-11.5,16) to[short] (30,16);
\draw [, line width=2pt](-11.5,14.5) to[short] (30,14.5);
\draw [, line width=2pt](-11.5,13) to[short] (30,13);
\draw [, line width=2pt](-11.5,11.5) to[short] (30,11.5);
\draw [, line width=2pt](-11.5,10) to[short] (30,10);
\draw [, line width=5pt](-11.5,8) to[short] (30,8);

\node [font=\Huge] at (14.25,19) {\textbf{Clock number}};
\node [font=\Huge] at (-7.5,17) {\textbf{Instruction number}};
\node [font=\Huge] at (-0.75,17) {\textbf{1}};
\node [font=\Huge] at (2.9,17) {\textbf{2}};
\node [font=\Huge] at (6.45,17) {\textbf{3}};
\node [font=\Huge] at (9.95,17) {\textbf{4}};
\node [font=\Huge] at (13.45,17) {\textbf{5}};
\node [font=\Huge] at (16.95,17) {\textbf{6}};
\node [font=\Huge] at (20.45,17) {\textbf{7}};
\node [font=\Huge] at (23.95,17) {\textbf{8}};
\node [font=\Huge] at (27.45,17) {\textbf{9}};

\node [font=\huge] at (-8.25,15.25) {Instruction \emph{i} + 1};
\node [font=\huge] at (-8.25,13.75) {Instruction \emph{i} + 2};
\node [font=\huge] at (-8.25,12.25) {Instruction \emph{i} + 3};
\node [font=\huge] at (-8.25,10.75) {Instruction \emph{i} + 4};
\node [font=\huge] at (-8.25,9) {Instruction \emph{i} + 5};

\node [font=\huge] at (-0.75,15.25) {IF};
\node [font=\huge] at (2.9,15.25) {ID};
\node [font=\huge] at (6.45,15.25) {EX};
\node [font=\huge] at (9.95,15.25) {MEM};
\node [font=\huge] at (13.45,15.25) {WB};

\node [font=\huge] at (2.9,13.75) {IF};
\node [font=\huge] at (6.45,13.75) {ID};
\node [font=\huge] at (9.95,13.75) {EX};
\node [font=\huge] at (13.45,13.75) {MEM};
\node [font=\huge] at (16.95,13.75) {WB};

\node [font=\huge] at (6.45,12.25) {IF};
\node [font=\huge] at (9.95,12.25) {ID};
\node [font=\huge] at (13.45,12.25) {EX};
\node [font=\huge] at (16.95,12.25) {MEM};
\node [font=\huge] at (20.45,12.25) {WB};

\node [font=\huge] at (9.95,10.75) {IF};
\node [font=\huge] at (13.45,10.75) {ID};
\node [font=\huge] at (16.95,10.75) {EX};
\node [font=\huge] at (20.45,10.75) {MEM};
\node [font=\huge] at (23.95,10.75) {WB};

\node [font=\huge] at (13.45,9) {IF};
\node [font=\huge] at (16.95,9) {ID};
\node [font=\huge] at (20.45,9) {EX};
\node [font=\huge] at (23.95,9) {MEM};
\node [font=\huge] at (27.45,9) {WB};
\end{circuitikz}
}

\label{Figure 2.1}
\vspace*{5mm}
\caption{Simplified representation of 5 stage pipeline using RISC instruction set. On each clock cycle, a new instruction is fetched and begins the 5 stage fetch-decode-execute cycle, with older instructions in the 5\textsuperscript{th} stage being retired, maintaining a throughput five times greater than non-pipeline processor. IF = instruction fetch, ID = instruction decode, EX = execution, MEM = memory access, and WB = write back. Table has been adapted from \cite{hennessy2017computer}.}
\end{figure}

Pipelining in CPUs, while a powerful optimization technique, introduces hazards that can impact the overall performance. These hazards can be broadly categorized into three types: structural hazards, data hazards, and control hazards. (1) Structural hazards occur when the hardware resources are incompatible with the sequence of instructions, leading to conflicts in resource allocation and pipeline stalls. However, advancements in super-scalar architectures and out-of-order instruction execution have made structural hazards less prevalent to virtually non-existent \cite{hennessy2017computer}. (2) Data hazards arise when an instruction depends on the completion of a previous instruction that has not finished executing. These dependencies can cause conflicts and hinder parallel execution of instructions, but are broadly mitigated through the use of virtual registers (register renaming) \cite{stallings2017computer}. (3) Control hazards are caused by conditional branch instructions that change the program counter. These instructions introduce uncertainty into the execution flow since the branch target needs to be determined prior to the next instruction fetch \cite{hennessy2017computer}. Branches constitute around 12-30\% of all instructions executed on modern instruction sets and are widely regarded the most significant barrier to achieving single cycle executions \cite{mcfarling1986reducing}, and as a consequence, has become a large area of focus for optimisations by hardware and software engineers.

Microprocessors employ various strategies to mitigate control hazards in the CPU instruction pipeline. The simplest and potentially the most costly approach is a full pipeline stall/freeze, where instructions following a branch instruction are ignored until the target of the branch is known, resulting in a fixed cycle penalty \cite{hennessy2017computer, smith1995microarchitecture}. Improving upon this, processors can make static predictions about the branch target instead of discarding subsequent instructions, maintaining sequential execution of instructions pertaining to either the \emph{taken} or \emph{not-taken} branches. By leveraging compiler static analysis and optimizing likely paths of execution, static prediction becomes a powerful optimization technique in pipelined processors, providing non-zero probabilities of correctly predicted branch targets and thus minimises throughput loss from flushes \cite{mcfarling1986reducing, deitrich1998improving}. Though an improvement, this approach is rather inflexible particularly for branch targets that change. Static prediction schemes lack the ability to adapt at runtime to changing patterns in branch target execution which is a common theme for the majority of conditional branches. With modern CPU's employing increasingly more speculative architectures and deeper pipelines to maximise instruction throughput \cite{hennessy2017computer}, branch penalties become more significant, scaling monotonically with pipeline depth \cite{emma1987characterization}. With this in account, it is clear that processors require even more aggressive optimisation techniques beyond simple static analysis to minimise idle cycles from branch mispredictions.

\begin{figure}[t]
\centering
\resizebox{\textwidth}{!}{%
\begin{circuitikz}
\tikzstyle{every node}=[font=\huge]

\draw [ color={rgb,255:red,242; green,242; blue,242} , fill={rgb,255:red,242; green,242; blue,242}, line width=0.2pt ] (-11.5,13) rectangle (30,11.5);
\draw [ color={rgb,255:red,242; green,242; blue,242} , fill={rgb,255:red,242; green,242; blue,242}, line width=0.2pt ] (-11.5,10) rectangle (30,8.5);
\draw [ color={rgb,255:red,242; green,242; blue,242} , fill={rgb,255:red,242; green,242; blue,242}, line width=0.2pt ] (-11.5,7) rectangle (30,5.5);
\draw [ color={rgb,255:red,242; green,242; blue,242} , fill={rgb,255:red,242; green,242; blue,242}, line width=0.2pt ] (-11.5,4) rectangle (30,2.5);

\draw [, line width=5pt](-11.5,14.5) to[short] (30,14.5);
\draw [, line width=2pt](-11.5,13) to[short] (30,13);
\draw [, line width=2pt](-11.5,11.5) to[short] (30,11.5);
\draw [, line width=2pt](-11.5,10) to[short] (30,10);
\draw [, line width=2pt](-11.5,8.5) to[short] (30,8.5);
\draw [, line width=2pt](-11.5,7) to[short] (30,7);
\draw [, line width=2pt](-11.5,5.5) to[short] (30,5.5);
\draw [, line width=2pt](-11.5,4) to[short] (30,4);
\draw [, line width=2pt](-11.5,2.5) to[short] (30,2.5);
\draw [, line width=5pt](-11.5,1) to[short] (30,1);

\node [font=\huge] at (-6.25,13.75) {Untaken branch instruction};
\node [font=\huge] at (-8.25,12.25) {Instruction \emph{i} + 1};
\node [font=\huge] at (-8.25,10.75) {Instruction \emph{i} + 2};
\node [font=\huge] at (-8.25,9.25) {Instruction \emph{i} + 3};

\node [font=\huge] at (-6.65,6.25) {Taken branch instruction};
\node [font=\huge] at (-8.25,4.75) {Instruction \emph{i} + 1};
\node [font=\huge] at (-8.85,3.25) {Branch target};
\node [font=\huge] at (-8,1.75) {Branch target + 1};

\node [font=\huge] at (2.9,13.75) {IF};
\node [font=\huge] at (6.45,13.75) {ID};
\node [font=\huge] at (9.95,13.75) {EX};
\node [font=\huge] at (13.45,13.75) {MEM};
\node [font=\huge] at (16.95,13.75) {WB};

\node [font=\huge] at (2.9,6.25) {IF};
\node [font=\huge] at (6.45,6.25) {ID};
\node [font=\huge] at (9.95,6.25) {EX};
\node [font=\huge] at (13.45,6.25) {MEM};
\node [font=\huge] at (16.95,6.25) {WB};

\node [font=\huge] at (6.45,12.25) {IF};
\node [font=\huge] at (9.95,12.25) {ID};
\node [font=\huge] at (13.45,12.25) {EX};
\node [font=\huge] at (16.95,12.25) {MEM};
\node [font=\huge] at (20.45,12.25) {WB};

\node [font=\huge] at (6.45,4.75) {IF};
\node [font=\huge] at (9.95,4.75) {\textbf{idle}};
\node [font=\huge] at (13.45,4.75) {\textbf{idle}};
\node [font=\huge] at (16.95,4.75) {\textbf{idle}};
\node [font=\huge] at (20.45,4.75) {\textbf{idle}};

\node [font=\huge] at (9.95,10.75) {IF};
\node [font=\huge] at (13.45,10.75) {ID};
\node [font=\huge] at (16.95,10.75) {EX};
\node [font=\huge] at (20.45,10.75) {MEM};
\node [font=\huge] at (23.95,10.75) {WB};

\node [font=\huge] at (9.95,3.25) {IF};
\node [font=\huge] at (13.45,3.25) {ID};
\node [font=\huge] at (16.95,3.25) {EX};
\node [font=\huge] at (20.45,3.25) {MEM};
\node [font=\huge] at (23.95,3.25) {WB};

\node [font=\huge] at (13.45,9.25) {IF};
\node [font=\huge] at (16.95,9.25) {ID};
\node [font=\huge] at (20.45,9.25) {EX};
\node [font=\huge] at (23.95,9.25) {MEM};
\node [font=\huge] at (27.45,9.25) {WB};

\node [font=\huge] at (13.45,1.75) {IF};
\node [font=\huge] at (16.95,1.75) {ID};
\node [font=\huge] at (20.45,1.75) {EX};
\node [font=\huge] at (23.95,1.75) {MEM};
\node [font=\huge] at (27.45,1.75) {WB};
\end{circuitikz}
}

\label{fig:my_label}
\vspace*{5mm}
\caption{Example of a \emph{predicted-not-taken} scheme and the instances when the branch is taken (top) and the branch is not taken (bottom). Correct prediction (top) results in subsequent instructions to fall through, wheres misprediction (bottom) results instrcutions pertaining to the branch to be flushed, incurring a single cycle penalty. Table has been adapted from \cite{hennessy2017computer}.}
\end{figure}

The final mitigation technique utilised commonly on classical 5-stage pipeline MIPS architectures, but are typically non-existent on modern processors, are \emph{branch delay slots} which interleave instructions independent to the branch prior to branch target deduction at decode or execute time \cite{flynncomputer}. The concept is rooted in the observation that not all instructions in a program depend on the outcome of a branch instruction, thereby in theory, allowing the compiler to schedule instructions before the branch is taken and hide some of the branches latency. However, on modern processors, branch delay slots are generally avoided. Utilizing branch delay slots effectively requires the compiler to identify and schedule instructions that can fill the slots, adding complexity to the compiler design and optimization process. Compiler writers need to analyze dependencies, identify independent instructions, and rearrange the code to take advantage of the delay slots. This additional burden makes it more challenging to generate efficient code and can increase compilation time, and are generally poor at doing so \cite{mcfarling1986reducing, flynncomputer}. With deeply complex instruction pipelines on modern processors, the scheduling task becomes exponentially more complex and interferes with modern hardware solutions, deeming the once performance enhancing technique as a complication in modern processors.

Whilst this section offers a fundamental overview of conditional branches and their implications at the processor level, it is crucial to recognize the significance of primitive branch penalty mitigation techniques. These early forms of mitigation laid the groundwork for the modern solutions discussed in the subsequent sections, and how the evolution of modern speculative processors played a pivotal role in driving the development of powerful hardware-based solutions, such as the branch predictor.

\subsection{Dynamic Branch Prediction}

The advancement of super-scalar processors has introduced intricate speculative architectures and deep instruction pipelines, which effectively maximize throughput to meet the performance requirements of contemporary computers. Consequently, the issue of branch misprediction has emerged as a significant impediment to sequential instruction execution \cite{hennessy2017computer, emma1987characterization}. As briefly mentioned earlier, static branch prediction techniques rely on predetermined rules or assumptions rather than leveraging runtime information. To enhance static prediction in modern architectures, compilers play a vital role in making profile-guided decisions based on historical execution patterns. Notably, the binomial distribution of simple branches renders static prediction an effective strategy \cite{hennessy2017computer, pan1992improving}. Supporting this notion, Fisher and Freudenberger's work \cite{fisher1992predicting} demonstrates that applications with statically predictable branches exhibit commendable performance under the current paradigm. The overarching limitation of the former schemes however is the inability to adapt to runtime changes or varying input conditions, and the lack of ability to capture complex patterns in branch behavior. Whilst previous execution patterns can be used as a proxy to determine the likelihood of a branch, in a real time system with non-deterministic data patterns, solely relying on such a scheme would likely lead to higher mispredictions rates and performance degradation. In light of these issues, a plethora of dynamic branch predictors where developed with the ability to adapt, leverage runtime information, and capture complex branch patterns to improve prediction accuracy and runtime performance.

Dynamic branch predictors (BP) can be broadly categorized into one-level \emph{local} BP's and two-level \emph{global} BP's, with more modern BP schemes incorporating the strengths of both. One-level BP's typically utilize a one-dimensional \emph{branch prediction buffer} or \emph{branch history table}, acting as a cache indexed by the lower bits of the program counter related to a branch instruction in memory \cite{hennessy2017computer, mittal2019survey}. 1-bit prediction schemes have a single bit entry in the branch history table, indicating whether the branch has recently been taken or not. In the event of a misprediction, the prediction bit is inverted. While this scheme offers simplicity and low hardware overhead, the limited historical information stored in a single bit often leads to frequent mispredictions \cite{hennessy2017computer}. To address this limitation, \emph{N-bit} saturated counter schemes are statically assigned to branches with distinct addresses. When a branch is about to be taken, the counter value associated with the branch and the direction of the branch is determined based on a predetermined threshold \cite{pan1992improving}. The count is incremented if the branch is taken, and vice versa. It may seem intuitive that increasing the number of bits would improve prediction accuracy by storing more information about the branch history. However, Smith \cite{smith1998study} demonstrated that strategies employing larger counter sizes than classical 2-bit schemes do not consistently result in higher prediction accuracies. For loop insensitive programs with biased branches, 2-bit counter schemes have been found to be effective with branch misprediction rates averaging at ~11\%, but performance was found to deteriorate with integer based programs with more complex branch dependencies \cite{hennessy2017computer}. The limitations of one-level BP schemes are multifaceted. Relying on a single branch history table restricts the ability to capture intricate patterns and dependencies between branches, hampering the prediction accuracy of one-level BPs \cite{smith1998study}. Additionally, interference can occur with branch buffer accesses, as finite space necessitates the use of hashing schemes to access the bit-counters for predictions. This can lead to collisions, with negative aliasing occurring more prominently than positive aliasing \cite{mittal2019survey}.

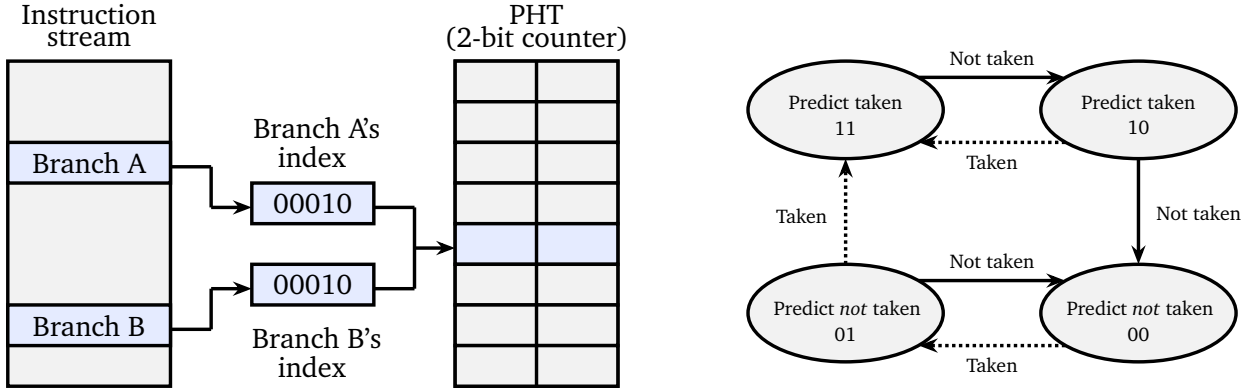
\begin{figure}[t]
\centering
\resizebox{1\textwidth}{!}{%
\begin{circuitikz}
\tikzstyle{every node}=[font=\Large]
\draw [ fill={rgb,255:red,242; green,242; blue,242} , line width=3pt ] (-10,20) rectangle (-5,10);
\draw [, line width=3pt ] (-7.25,16.75) rectangle (-7.25,16.75);
\draw [ fill={rgb,255:red,229; green,235; blue,255} , line width=3pt ] (-10,17.5) rectangle  node {\Huge Branch A} (-5,16.25);
\draw [ fill={rgb,255:red,229; green,235; blue,255} , line width=3pt ] (-10,12.5) rectangle  node {\Huge Branch B} (-5,11.25);
\draw [ fill={rgb,255:red,229; green,235; blue,255} , line width=3pt ] (-2.5,16.25) rectangle  node {\Huge 00010} (1.25,15);
\draw [ fill={rgb,255:red,229; green,235; blue,255} , line width=3pt ] (-2.5,13.75) rectangle  node {\Huge 00010} (1.25,12.5);
\draw [, line width=3pt ] (3.75,20) rectangle (8.75,10);
\draw [ fill={rgb,255:red,242; green,242; blue,242} , line width=3pt ] (3.75,20) rectangle (8.75,18.75);
\draw [ fill={rgb,255:red,242; green,242; blue,242} , line width=3pt ] (3.75,18.75) rectangle (8.75,17.5);
\draw [ fill={rgb,255:red,242; green,242; blue,242} , line width=3pt ] (3.75,17.5) rectangle (8.75,16.25);
\draw [ fill={rgb,255:red,242; green,242; blue,242} , line width=3pt ] (3.75,16.25) rectangle (8.75,15);
\draw [ fill={rgb,255:red,229; green,235; blue,255} , line width=3pt ] (3.75,15) rectangle (8.75,13.75);
\draw [ fill={rgb,255:red,242; green,242; blue,242} , line width=3pt ] (3.75,13.75) rectangle (8.75,12.5);
\draw [ fill={rgb,255:red,242; green,242; blue,242} , line width=3pt ] (3.75,12.5) rectangle (8.75,11.25);
\draw [ fill={rgb,255:red,242; green,242; blue,242} , line width=3pt ] (3.75,11.25) rectangle (8.75,10);
\draw [, line width=3pt](6.25,20) to[short] (6.25,10);
\draw [, line width=3pt](-5,11.75) to[short] (-3.75,11.75);
\draw [, line width=3pt](-5,16.75) to[short] (-3.75,16.75);
\draw [, line width=3pt](-3.75,16.75) to[short] (-3.75,15.5);
\draw [, line width=3pt](-3.75,11.75) to[short] (-3.75,13);
\draw [ line width=3pt, -Stealth] (-3.75,15.5) -- (-2.5,15.5);
\draw [ line width=3pt, -Stealth] (-3.75,13) -- (-2.5,13);
\draw [, line width=3pt](1.25,15.5) to[short] (2.5,15.5);
\draw [, line width=3pt](1.25,13) to[short] (2.5,13);
\draw [, line width=3pt](2.5,15.5) to[short] (2.5,13);
\draw [ line width=3pt, -Stealth] (2.5,14.25) -- (3.75,14.25);
\node [font=\Huge] at (6.25,21.45) {PHT};
\node [font=\Huge] at (6.25,20.7) {(2-bit counter)};
\node [font=\Huge] at (-7.5,21.45) {Instruction};
\node [font=\Huge] at (-7.5,20.7) {stream};
\node [font=\Huge] at (-0.5,17.95) {Branch A's};
\node [font=\Huge] at (-0.5,11.5) {Branch B's};
\node [font=\Huge] at (-0.65,10.6) {index};
\node [font=\Huge] at (-0.65,17.05) {index};
\draw [ fill={rgb,255:red,242; green,242; blue,242} , line width=3pt ] (15.75,18.5) ellipse (3cm and 1.5cm);
\draw [ fill={rgb,255:red,242; green,242; blue,242} , line width=3pt ] (15.75,12.25) ellipse (3cm and 1.5cm);
\draw [ fill={rgb,255:red,242; green,242; blue,242} , line width=3pt ] (24.75,12.25) ellipse (3cm and 1.5cm);
\draw [ fill={rgb,255:red,242; green,242; blue,242} , line width=3pt ] (24.75,18.5) ellipse (3cm and 1.5cm);
\node [font=\LARGE] at (15.75,17.95) {11};
\node [font=\LARGE] at (24.75,17.95) {10};
\node [font=\LARGE] at (15.75,11.6) {01};
\node [font=\LARGE] at (24.75,11.6) {00};
\node [font=\LARGE] at (15.75,18.75) {Predict taken};
\node [font=\LARGE] at (24.75,18.75) {Predict taken};
\node [font=\LARGE] at (15.8,12.4) {Predict \emph{not} taken};
\node [font=\LARGE] at (24.8,12.4) {Predict \emph{not} taken};

\draw [ line width=3pt, -Stealth] (18,19.5) -- (22.5,19.5);%
\draw [ line width=3pt, -Stealth] (24.75,17) -- (24.75,13.75);
\draw [ line width=3pt, -Stealth] (18,13.25) -- (22.5,13.25);%

\draw [line width=3pt, -Stealth, dashed] (22.5,11.25) -- (18,11.25);%
\draw [line width=3pt, -Stealth, dashed] (22.5,17.5) -- (18,17.5);%
\draw [line width=3pt, -Stealth, dashed] (15.75,13.75) -- (15.75,17);

\node [font=\LARGE] at (14.4,15.25) {Taken};
\node [font=\LARGE] at (20.25,10.65) {Taken};%
\node [font=\LARGE] at (20.25,16.9) {Taken};%

\node [font=\LARGE] at (20.25,20.1) {Not taken};%
\node [font=\LARGE] at (26.6,15.35) {Not taken};
\node [font=\LARGE] at (20.25,13.85) {Not taken};
\end{circuitikz}
}%

\label{fig:my_label}
\vspace*{5mm}
\caption{Diagrammatic representation of a one-level branch history table with aliasing interference between two branches, adapted from \cite{mittal2019survey} (left). State machine for 2-bit branch prediction, adapted from \cite{hennessy2017computer} (right).}
\end{figure}

In light of these limitations, Yeh and Patt \cite{yeh1991two} proposed the first two-level adaptive branch prediction scheme that utilises a \emph{global branch history table} which maintains a shared history of branch outcomes, allowing it to capture patterns and dependencies between branches. Implementation wise, two level BP's comprise of a branch history register (BHR) which tracks recent outcomes of branches and a global pattern history table (PHT) which stores patterns and outcomes associated with specific branch instructions. In this scheme both the BHR and PHT work in collaborative fashion; most recent branch results are shifted into the BHR with branch instruction addresses used to index a BHR table, the content of the BHR is used to index the global PHT for making predictions, with mispredictions updating both the PHT and BHR \cite{yeh1991two, yeh1992two, mittal2019survey}. Whilst the two-level adaptive predictor improves prediction accuracy and captures branch dependencies, it still faces challenges related to aliasing conflicts similar to saturated counter-based branch predictors. Yeh and Patt explored several alternative branch prediction schemes based on the original two-level approach \cite{yeh1992two}, and whilst branch correlation algorithms could be improved it was found that inherit trade-offs existed between efficient storage capacity, memory access overhead and reduced interference, often bounded by physical constraints. As a consequence, current state-of-the-art BP research is focused primarily on the development and optimisation of the prediction algorithms that build off the foundational work originally done by \cite{yeh1991two, yeh1992two}, and have been immensely successful at doing so with the emergence of incredibly powerful BP's such as the competition winning TAGE-L with 3-4 mispredictions per 1000 instructions \cite{seznec2016tage, zangeneh2020branchnet}. 

\begin{figure}[!ht]
\centering
\resizebox{1\textwidth}{!}{%
\begin{circuitikz}
\tikzstyle{every node}=[font=\Huge]
\draw [, line width=4pt ] (-30,19.75) rectangle (-27.5,10);
\draw [, line width=4pt ] (-22.5,20) rectangle (-17.5,10);
\draw [, line width=4pt ] (-10,20) rectangle (-5,10);
\draw [, line width=4pt ] (5,20) rectangle (10,10);
\draw [ fill={rgb,255:red,229; green,235; blue,255} , line width=4pt ] (-22.5,17.5) rectangle (-17.5,16.25);
\draw [ fill={rgb,255:red,229; green,235; blue,255} , line width=4pt ] (-10,13.75) rectangle (-5,12.5);
\draw [ fill={rgb,255:red,229; green,235; blue,255} , line width=4pt ] (5,16.25) rectangle (10,15);
\draw [ fill={rgb,255:red,229; green,235; blue,255} , line width=4pt ] (-30,13.75) rectangle (-27.5,12.5);
\draw [, line width=4pt](-30,18.75) to[short] (-27.5,18.75);
\draw [, line width=4pt](-30,17.5) to[short] (-27.5,17.5);
\draw [, line width=4pt](-30,16.25) to[short] (-27.5,16.25);
\draw [, line width=4pt](-30,15) to[short] (-27.5,15);
\draw [, line width=4pt](-30,11.25) to[short] (-27.5,11.25);
\draw [, line width=4pt](-22.5,18.75) to[short] (-17.5,18.75);
\draw [, line width=4pt](-22.5,15) to[short] (-17.5,15);
\draw [, line width=4pt](-22.5,13.75) to[short] (-17.5,13.75);
\draw [, line width=4pt](-22.5,12.5) to[short] (-17.5,12.5);
\draw [, line width=4pt](-22.5,11.25) to[short] (-17.5,11.25);
\draw [, line width=4pt](-10,18.75) to[short] (-5,18.75);
\draw [, line width=4pt](-10,17.5) to[short] (-5,17.5);
\draw [, line width=4pt](-10,16.25) to[short] (-5,16.25);
\draw [, line width=4pt](-10,15) to[short] (-5,15);
\draw [, line width=4pt](-10,11.25) to[short] (-5,11.25);
\draw [, line width=4pt](5,18.75) to[short] (10,18.75);
\draw [, line width=4pt](5,17.5) to[short] (10,17.5);
\draw [, line width=4pt](5,13.75) to[short] (10,13.75);
\draw [, line width=4pt](5,12.5) to[short] (10,12.5);
\draw [, line width=4pt](5,11.25) to[short] (10,11.25);
\draw [ fill={rgb,255:red,242; green,242; blue,242} , line width=4pt , rounded corners, ] (-23.75,23.75) rectangle  node {\Huge Index} (-20,22.5);
\draw [ fill={rgb,255:red,242; green,242; blue,242} , line width=4pt , rounded corners, ] (-18.75,23.75) rectangle  node {\Huge Tag} (-15,22.5);
\draw [ fill={rgb,255:red,242; green,242; blue,242} , line width=4pt , rounded corners, ] (-11.25,23.75) rectangle  node {\Huge Index} (-7.5,22.5);
\draw [ fill={rgb,255:red,242; green,242; blue,242} , line width=4pt , rounded corners, ] (-6.25,23.75) rectangle  node {\Huge Tag} (-2.5,22.5);
\draw [ fill={rgb,255:red,242; green,242; blue,242} , line width=4pt , rounded corners, ] (3.75,23.75) rectangle  node {\Huge Index} (7.5,22.5);
\draw [ fill={rgb,255:red,242; green,242; blue,242} , line width=4pt , rounded corners, ] (8.75,23.75) rectangle  node {\Huge Tag} (12.5,22.5);
\draw [, line width=4pt](-33.75,26.25) to[short] (10,26.25);
\draw [, line width=4pt](-32.5,26.25) to[short] (-32.5,13);
\draw [ line width=4pt, -Stealth] (-32.5,13) -- (-30,13);
\draw [ fill={rgb,255:red,0; green,0; blue,0} , line width=0.2pt ] (-32.5,26.25) circle (0.25cm);
\draw [, line width=4pt](-23.75,28.75) to[short] (-16.25,28.75);
\draw [, line width=4pt](-11.25,28.75) to[short] (-3.75,28.75);
\draw [, line width=4pt](3.75,28.75) to[short] (11.25,28.75);
\draw [ line width=4pt, -Stealth] (-21.25,28.75) -- (-21.25,23.75);
\draw [ line width=4pt, -Stealth] (-16.25,28.75) -- (-16.25,23.75);
\draw [ line width=4pt, -Stealth] (-8.75,28.75) -- (-8.75,23.75);
\draw [ line width=4pt, -Stealth] (-3.75,28.75) -- (-3.75,23.75);
\draw [ line width=4pt, -Stealth] (6.25,28.75) -- (6.25,23.75);
\draw [ line width=4pt, -Stealth] (11.25,28.75) -- (11.25,23.75);
\draw [ line width=4pt, -Stealth] (10,26.25) -- (10,23.75);
\draw [ line width=4pt, -Stealth] (5,26.25) -- (5,23.75);
\draw [ line width=4pt, -Stealth] (-5,26.25) -- (-5,23.75);
\draw [ line width=4pt, -Stealth] (-10,26.25) -- (-10,23.75);
\draw [ line width=4pt, -Stealth] (-17.5,26.25) -- (-17.5,23.75);
\draw [ line width=4pt, -Stealth] (-22.5,26.25) -- (-22.5,23.75);
\draw [ fill={rgb,255:red,0; green,0; blue,0} , line width=0.2pt ] (-22.5,26.25) circle (0.25cm);
\draw [ fill={rgb,255:red,0; green,0; blue,0} , line width=0.2pt ] (-17.5,26.25) circle (0.25cm);
\draw [ fill={rgb,255:red,0; green,0; blue,0} , line width=0.2pt ] (-10,26.25) circle (0.25cm);
\draw [ fill={rgb,255:red,0; green,0; blue,0} , line width=0.2pt ] (-5,26.25) circle (0.25cm);
\draw [ fill={rgb,255:red,0; green,0; blue,0} , line width=0.2pt ] (5,26.25) circle (0.25cm);
\draw [, line width=4pt](-25,23) to[short] (-25,16.75);
\draw[, line width=4pt] (-23.75,23) to[short] (-25,23);
\draw [ line width=4pt, -Stealth] (-25,16.75) -- (-22.5,16.75);
\draw [ fill={rgb,255:red,242; green,242; blue,242} , line width=4pt , rounded corners, ] (-20,7.5) rectangle  node {\Huge Tag match?} (-15,6.25);
\draw [ fill={rgb,255:red,242; green,242; blue,242} , line width=4pt , rounded corners, ] (-7.5,7.5) rectangle  node {\Huge Tag match?} (-2.5,6.25);
\draw [ fill={rgb,255:red,242; green,242; blue,242} , line width=4pt , rounded corners, ] (7.5,7.5) rectangle  node {\Huge Tag match?} (12.5,6.25);
\draw [ line width=4pt, -Stealth] (-18.75,10) -- (-18.75,7.5);
\draw [ line width=4pt, -Stealth] (-16.25,22.5) -- (-16.25,7.5);
\draw [ line width=4pt, -Stealth] (-6.25,10) -- (-6.25,7.5);
\draw [ line width=4pt, -Stealth] (-3.75,22.5) -- (-3.75,7.5);
\draw [ line width=4pt, -Stealth] (8.75,10) -- (8.75,7.5);
\draw [ line width=4pt, -Stealth] (11.25,22.5) -- (11.25,7.5);
\draw[, line width=4pt] (-11.25,23) to[short] (-12.5,23);
\draw [, line width=4pt](-12.5,23) to[short] (-12.5,13);
\draw [ line width=4pt, -Stealth] (-12.5,13) -- (-10,13);
\draw[, line width=4pt] (3.75,23) to[short] (2.5,23);
\draw [, line width=4pt](2.5,23) to[short] (2.5,15.5);
\draw [ line width=4pt, -Stealth] (2.5,15.5) -- (5,15.5);
\draw [, line width=4pt ] (-1,12.75) circle (0.25cm);
\draw [, line width=4pt ] (-0.25,12.75) circle (0.25cm);
\draw [, line width=4pt ] (0.5,12.75) circle (0.25cm);

\draw [, line width=4pt](-27.5,5) to[short] (-22.5,5);
\draw [, line width=4pt](-26.25,3.5) to[short] (-23.75,3.5);
\draw [, line width=4pt](-27.45,5) to[short] (-26.2,3.5);
\draw [, line width=4pt](-22.55,5) to[short] (-23.8,3.5);

\draw [, line width=4pt](-28.75,10) to[short] (-28.75,6.75);
\draw [, line width=4pt](-28.75,6.75) to[short] (-26.25,6.75);
\draw [, line width=4pt](-21.25,10) to[short] (-21.25,6.75);
\draw[, line width=4pt] (-21.25,6.75) to[short] (-23.75,6.75);
\draw [, line width=4pt](-17.5,6.25) to[short] (-17.5,4.25);
\draw [ line width=4pt, -Stealth] (-26.25,6.75) -- (-26.25,5);
\draw [ line width=4pt, -Stealth] (-23.75,6.75) -- (-23.75,5);
\draw [ line width=4pt, -Stealth] (-17.5,4.25) -- (-23,4.25);
\draw [, line width=4pt](-25,3.5) to[short] (-25,1.75);
\draw [, line width=4pt](-25,1.75) to[short] (-13.75,1.75);

\draw [, line width=4pt](-15,0) to[short] (-10,0);
\draw [, line width=4pt](-13.75,-1.5) to[short] (-11.25,-1.5);
\draw [, line width=4pt](-14.95,0) to[short] (-13.7,-1.5);
\draw [, line width=4pt](-10.05,0) to[short] (-11.3,-1.5);

\draw [, line width=4pt](-8.75,10) to[short] (-8.75,8.5);
\draw[, line width=4pt] (-8.75,8.5) to[short] (-11.25,8.5);
\draw [ line width=4pt, -Stealth] (-11.25,8.5) -- (-11.25,0);
\draw [ line width=4pt, -Stealth] (-13.75,1.75) -- (-13.75,0);
\draw [, line width=4pt](-5,6.25) to[short] (-5,-0.75);
\draw [ line width=4pt, -Stealth] (-5,-0.75) -- (-10.5,-0.75);
\draw [, line width=4pt ] (-5.25,-3.25) circle (0.25cm);
\draw [, line width=4pt ] (-4.5,-3.25) circle (0.25cm);
\draw [, line width=4pt ] (-3.75,-3.25) circle (0.25cm);

\draw[, line width=4pt] (5,-5) to[short] (0,-5);
\draw [, line width=4pt](1.25,-6.5) to[short] (3.75,-6.5);

\draw[, line width=4pt] (4.95,-5) to[short] (3.7,-6.5);
\draw [, line width=4pt](0.05,-5) to[short] (1.3,-6.5);

\draw [, line width=4pt](-12.5,-3.25) to[short] (-9.25,-3.25);
\draw [line width=4pt, dashed] (-9.25,-3.25) -- (-6.25,-3.25);
\draw [, line width=4pt](-12.5,-1.5) to[short] (-12.5,-3.25);
\draw [, line width=4pt](-1.25,-3.25) to[short] (1.25,-3.25);
\draw [, line width=4pt](6.25,10) to[short] (6.25,8.5);
\draw[, line width=4pt] (6.25,8.5) to[short] (3.75,8.5);
\draw [, line width=4pt](10,6.25) to[short] (10,-6);
\draw [ line width=4pt, -Stealth] (1.25,-3.25) -- (1.25,-5);
\draw [ line width=4pt, -Stealth] (3.75,8.5) -- (3.75,-5);
\draw [ line width=4pt, -Stealth] (10,-6) -- (4.25,-6);
\draw [ line width=4pt, -Stealth] (2.5,-6.5) -- (2.5,-8.25);
\node [font=\Huge] at (-21.9,29.35) {hist[L\textsubscript{1}:0]};
\node [font=\Huge] at (-9.4,29.35) {hist[L\textsubscript{2}:0]};
\node [font=\Huge] at (5.65,29.35) {hist[L\textsubscript{N}:0]};
\node [font=\Huge] at (-32.5,27) {PC};
\node [font=\Huge] at (5.25,-7.75) {Prediction};
\end{circuitikz}
}%

\label{fig:my_label}
\vspace*{5mm}
\caption{Organisation of the TAGE (Tagged Geometric) BP with N-tagged tables. TAGE features a base binomial predictor which provides a basic prediction and a number of partially tagged tables that store branch history information and associated predictions. At prediction time, each of the tagged tables are indexed using different history lengths that form a geometric series, with the longest history generally being chosen \cite{sadooghi2012toward}. The TAGE-L predictor builds on this and uses a dynamic table organization with varying number of tables and table lengths which better captures branch outcomes and history lengths \cite{seznectageL}. TAGE-SC-L further builds on this and incorporates a statistical correlator to further refine predictions \cite{seznec2016tage}. Figure adapted from \cite{sadooghi2012toward}.}
\end{figure}
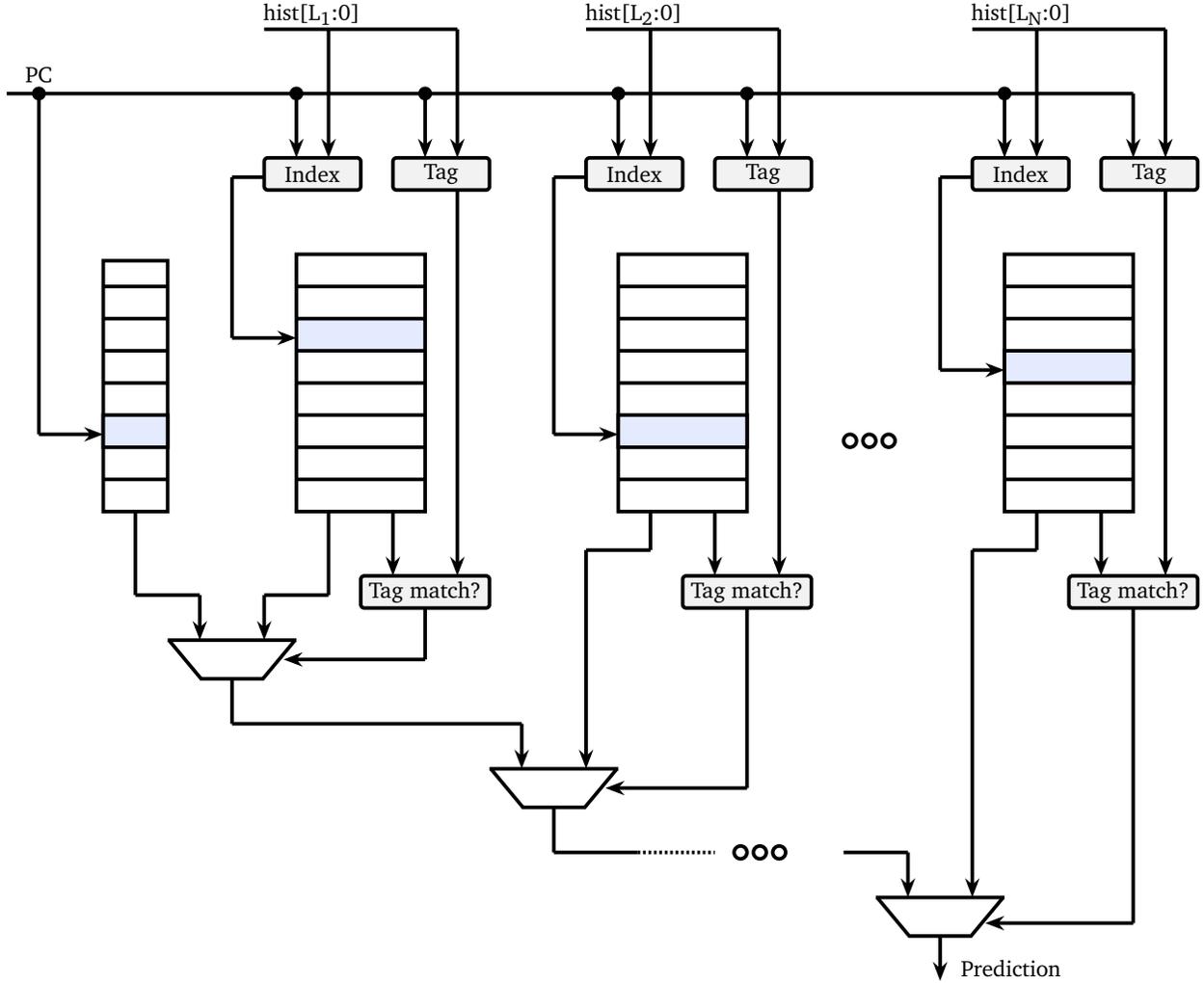

Whilst the current state of modern BP's are capable of predicting the majority of branches to near perfect accuracy, there exists classes of branches that are inherently hard to predict (HTP) by TAGE-like predictors and hence the problem of branch prediction is still considered unsolved \cite{lin2019branch, ozturk2010analysis}. The key weakness arises from attempting to identify correlations between branches in a noisy global history, or the branches themselves are uncorrelated and must rely solely on saturated counter schemes to make predictions\cite{zangeneh2020branchnet}.
Fundamentally, noise caused by HTP branches introduce a number of redundant patterns which pollute the global branch history causing TAGE-like predictors to struggle, and with non-deterministic ordering of historical patterns, more exotic BP's such as perceptron based BP's which rely on the position of a branch in the global branch history also struggle \cite{mittal2019survey, zangeneh2020branchnet, perceptron}. Problematically, HTP branches are common in most applications, but have the more prevalent effect in performance critical fields such as games, streaming services, and HFT.

The weaknesses of modern BP's stem from their need to be simple, computationally cheap and adaptive to execution phase behaviour, preventing them from capturing complex patterns seen in HTP branches \cite{zangeneh2020branchnet}. In addition, in the worst case the storage requirements for capturing complex branch patterns are exponential, and make traditional BP strategies infeasible. Whilst fundamental breakthoughs in novel BP's that build off TAGE and percepton-based BP's have become rare, Zangeneh et al proposed a convolutional neural network (CNN) based BP which can be trained offline and be capable of predicting complex and noisy branches with improved accuracy \cite{zangeneh2020branchnet}. Their work explored two CNN models, one of which was a pure software solutions and the second being a smaller hardware optimised version capable of being implemented as a practical BP, both of which showed vast reductions in MPKI on benchmarks run on TAGE-SC-L. Although prediction accuracy greatly improved for correlated branches with noisy histories, BranchNet showed poor performance on data-dependant branches and programs where mispredictions are spread across a number of static branches. The reason for the former, is that data dependant branches are dependant on input data and program phase behaviour which means there is little to no history of the branch that is correlated to the data stored in memory, which is problematic for capturing training data to make predictions \cite{zangeneh2020branchnet}. In terms of mispredictions for static branches, this is a problem regarding the sparse storage requirements for a practical CNN BP, though it may be possible to use larger models using proposed predictor virtualisation techniques \cite{burcea2008predictor}.

Though NN based BP's have the potential to be the next standard for hardware-based BP's, there is still much work to be done before commercialisation and a number of problems to be accounted for. Whilst the work by Zangeneh et al \cite{zangeneh2020branchnet} demonstrated MPKI reductions on several HTP branches on a number of benchmarks, the first challenge of using such CNN's would be to achieve some sort of generalisation across all known HTP branches, which in turn would require a substantially large model or a number of models which introduces large administrative overhead by the OS. In addition, modern OS would need to be adapted to load these CNN models on-chip prior to execution time and handle context switching, and associated context switching penalties accordingly which in turn adds complexity and overhead.

At its current state, research pertaining to novel hardware based BP's have slowed substantially, and it is likely that modern BP's have reached an asymptotic state of prediction accuracy versus complexity and implementation overhead. Whilst there may be BP's in the future that can have low misprediction rates on noisy or inherently HTP branches, there will always exist some class of branches, especially in real-time systems, that will always be completely non-deterministic and impossible to predict. Whilst this likely can never be solved by hardware, it may be possible that programmatic approaches exist that are able to tame these impossible-to-predict branches.

\subsection{C++ and Compiler Hints}

High-frequency trading (HFT) demands ultra-low latency and exceptional performance, necessitating the selection of a programming language that offers efficient execution, deterministic behavior, and fine-grained control over hardware resources. C++ has emerged as the primary language for developing critical path (path of order action) components in automated trading systems, primarily due to its design philosophy. One fundamental axiom associated with C++ is the "zero overhead principle" \cite{stroustrup1986overview, stroustrup1994design}, which ensures that developers only pay for what they use, resulting in a predictable and transparent performance model. For example, unlike high-level languages that employ garbage collectors to manage memory, C++ utilizes manual memory management. This approach avoids the unpredictable invocation times and latency associated with garbage collectors, which is critical in low-latency applications that require deterministic performance. C++ provides developers with granular control over memory using abstractions such as pointers, references, heap allocation operators, and standard library methods. However, the trade-off for low-level memory access is increased complexity and the potential for memory leaks and errors. Whilst uncommon, other languages that target the java virtual machine (JVM), for example Java, have become more popular in the HFT space using optimised GC algorithms and compilation techniques to minimise latency cost \cite{donadio2022developing}. Notably, this has become popular with leading quantitative trading firm and liquidity provider, Jane Street.

C++ features also support the shifting of execution time operations to compile time, resulting in the deferral of computational costs and the reduction of runtime latency when appropriately utilized. This capability proves to be a powerful tool for applications with stringent requirements for low latency. Templates, as a Turing-complete language feature, play a vital role in enabling this paradigm by providing type-safe parameterized blueprints, which allow the generation of specialized code by the compiler for each type-specific instantiation. Consequently, compile-time polymorphism can be achieved, albeit at the expense of flexibility and maintainability \cite{veldhuizen2003c++, vandevoorde2002c++}. Moreover, the template system can be further leveraged for performing recursive instantiations and type deductions, thereby facilitating the manipulation of types and values and enabling the execution of complex computations at compile time. It is important to note that this approach, known as "template metaprogramming," is extensively employed in low-latency settings such as High-Frequency Trading \cite{carl2017cpp, abrahams2004c++}.

In the realm of optimizing branch prediction at the language level, specific extensions provided by compilers have long existed, allowing programmers to provide hints for branch prediction. These hints enable targeted optimizations on anticipated execution paths. In GCC and Clang, this capability is manifested in the form of the \texttt{\_\_builtin\_expect} attributes, which allow programmers to specify conditions and associated probabilities (fixed as either 0 or 1) for condition evaluation at runtime \cite{clang2023, gcc2023}. It is worth noting that these so-called branch prediction hints do not directly affect the hardware-based branch prediction of processors and have not been effective since the release of Intel's Pentium M and Core 2 processors \cite{fog2012microarchitecture}. Instead, compiler built-ins optimize branch-taking by rearranging assembly code related to branches to exploit processor static prediction schemes and instruction cache effects for improved performance.

Modern processors enhance execution performance by prefetching instructions sequentially from slower to faster memory storage, such as high-speed caches in close proximity to the CPU. This prefetching is done to avoid high memory access latencies during the fetch stages \cite{cache}. When executing a code segment containing conditional statements, blocks of sequential assembly instructions associated with different branches are prefetched into the instruction cache, irrespective of how frequently the branches are executed. However, prefetched code that remains infrequently accessed throughout the program's lifespan can contaminate the instruction cache and result in cache-line fragmentation of hot code segments that are frequently executed. This can introduce jitter and latency costs \cite{tuning2018, mem2007}. To optimize branch prediction, the compiler reorganizes the underlying assembly code such that the conditional jump occurs on the least likely path. This is because modern processors initially assume that forward branches are never taken and, therefore, avoid misprediction by fetching the likely branch to the branch target  \cite{fog2012microarchitecture}. It is important to note that such static-prediction schemes are employed when the BPU encounters a branch that has not been previously visited. The dynamic prediction schemes outlined in the previous sections will begin to dictate which blocks are speculatively fetched once a branch history has been established.

\begin{figure}
    \centering

    \begin{lstlisting}[language=c++]
 if (condition)                  10af:  je     10bd
     function_1();               10b1:  call   <function_1>
 else                            (...)
     function_2();               10bd:  call   <function_2>
    \end{lstlisting}
    \vspace*{1mm}
    \caption{Comparison of C++ code without branch prediction hints (compiled with GCC on x86-64). In this case the forward branch (else) is assumed not taken and the backward branch (if) is assumed taken.}
\end{figure}

\begin{figure}
    \centering
    \begin{lstlisting}[language=c++]
 if (condition) [[unlikely]]     10af:  je     10bd
     function_1();               10b1:  call   <function_2>
 else                            (...)
     function_2();               10bd:  call   <function_1>
    \end{lstlisting}
    \vspace*{1mm}
    \caption{Comparison of C++ code with branch prediction hints (compiled with GCC on x86-64). Since the backward branch is now deemed as "unlikely", the compiler will reorganise the ASM such that the backward branch is now the forward branch and vice versa.}
\end{figure}

C++20 introduced the \texttt{[[likely]]} and \texttt{[[unlikely]]} attributes, which serve as wrappers around the original \texttt{\_\_builtin\_expect} compiler attributes and function in the same manner. Apart from these attributes, there have been no other language features attempting to optimize branch prediction \cite{likelycpp}. There is limited formal research investigating the performance impact using different benchmarks. However, some online blogs report performance gains (e.g., \cite{tuning2018} reports a 15\% increase) primarily on branches specifically designed to be highly predictable for demonstration purposes rather than from a research perspective. The fundamental problem with these attributes is that they rely on the programmer's accurate prediction of likely execution paths. While profiling and synthetic benchmark data may offer insights into hot/cold branches, programmers tend to significantly underestimate the cost of misprediction when misusing these attributes (as programmers are notoriously poor at predicting branches) \cite{infoq2023}. Furthermore, the capabilities of these attributes are confined to compile time, rendering them inflexible during runtime. Suppose a branch is indeed more likely to be taken at compile time, allowing the programmer to benefit from it if used correctly. In that case, if the likelihood of the branch changes during execution, the programmer would have no control over it and would suffer from increased latency through assembly reordering schematics. Real-time systems, which constantly need to react to live events and data, inherently exhibit such variability. Consequently, static language features like these are inadequate as effective branch misprediction mitigators. This inadequacy forms the focus of the research conducted in this work.

\subsection{High Frequency Trading}

The evolution of computer-based trading dates back several decades, starting with the introduction of fully electronic trading by NASDAQ \cite{aldridge2013high}. With the decrease in regulation and advancements in electronic exchanges and telecommunications infrastructure, high-frequency trading (HFT) has gained significant popularity, accounting for over 50\% of trading volume in equity markets \cite{MENKVELD2013712}. Defining HFT itself poses challenges. Haldane \cite{haldane} emphasizes the use of sophisticated algorithms as its main characteristic, while MacKenzie \cite{mackenzie2014sociology} and Arnoldi \cite{arnoldi2016computer} highlight the importance of speed in data processing and execution rather than the underlying strategies employed. This work primarily focuses on optimizations in proprietary automated trading systems, and hence considers HFT as a form of algorithmic trading who's strategies rely low order execution latencies to ensure profitability.

High-Frequency Trading (HFT) firms actively engage in generating trading signals, validating models, and executing trades in order to exploit inefficiencies in market micro-structure within short time frames, with the ultimate goal of achieving profitability \cite{aldridge2013high}. These firms employ diverse strategies to generate profits in financial markets. One prevalent strategy is market-making, where HFT participants continuously provide liquidity by simultaneously placing buy and sell orders, aiming to profit from the bid-ask spread. Another strategy, known as statistical arbitrage, involves capitalizing on transient deviations from fair value by identifying mispricing opportunities. Additionally, event-driven trading strategies focus on leveraging informational asymmetries that arise from significant market events \cite{HASBROUCK2013646}. While a significant body of literature exists on the economic effects of HFT (e.g., \cite{zhang2010high, brogaard2017high, brogaard2010high}), information pertaining to the engineering aspects of low-latency automated trading systems is often concealed. Nevertheless, some online conferences vaguely present important features of HFT systems, such as the networking stack, kernel bypass, custom hardware, and strategies for enhancing the speed and efficiency of production code \cite{nimrod2019cpp, carl2017cpp}.

\begin{figure}[t]
\centering
\resizebox{.7\textwidth}{!}{
\begin{circuitikz}
\tikzstyle{every node}=[font=\huge]
\draw [ fill={rgb,255:red,242; green,242; blue,242} , line width=2pt , rounded corners, ] (0,25) rectangle  node {\huge Network} (11.25,22.5);
\draw [ fill={rgb,255:red,242; green,242; blue,242} , line width=2pt , rounded corners, ] (0,21.25) rectangle (11.25,5);
\draw [ fill={rgb,255:red,229; green,235; blue,255} , line width=2pt , rounded corners, ] (1.25,20) rectangle  node {\huge UDP Network Stack} (10,17.5);
\draw [ fill={rgb,255:red,229; green,235; blue,255} , line width=2pt , rounded corners, ] (1.25,16.25) rectangle (10,13.75);
\draw [ fill={rgb,255:red,224; green,240; blue,227} , line width=2pt , rounded corners, ] (1.25,12.5) rectangle (5,10);
\draw [ fill={rgb,255:red,224; green,240; blue,227} , line width=2pt , rounded corners, ] (6.25,12.5) rectangle (10,10);
\draw [ fill={rgb,255:red,253; green,224; blue,224} , line width=2pt , rounded corners, ] (1.25,8.75) rectangle (5,6.25);
\draw [ fill={rgb,255:red,253; green,224; blue,224} , line width=2pt , rounded corners, ] (6.25,8.75) rectangle (10,6.25);
\node [font=\huge] at (5.7,15.5) {Network Switch \&};
\node [font=\huge] at (5.7,14.5) {Timestamper};
\node [font=\huge] at (3.1,11.75) {FAST};
\node [font=\huge] at (3.15,10.85) {Encoder};
\node [font=\huge] at (8.1,11.75) {FAST};
\node [font=\huge] at (8.15,10.85) {Decoder};
\node [font=\huge] at (3.1,8) {Custom};
\node [font=\huge] at (3.1,7) {App};
\node [font=\huge] at (8.1,8) {Order};
\node [font=\huge] at (8.1,7) {Book};
\draw [ line width=2pt, -Stealth] (8.15,17.5) -- (8.15,16.25);
\draw [ line width=2pt, -Stealth] (8.15,13.75) -- (8.15,12.5);
\draw [ line width=2pt, -Stealth] (8.15,10) -- (8.15,8.75);
\draw [ line width=2pt, -Stealth] (6.25,7.5) -- (5,7.5);
\draw [ line width=2pt, -Stealth] (3.1,8.75) -- (3.1,10);
\draw [ line width=2pt, -Stealth] (3.1,12.5) -- (3.1,13.75);
\draw [ line width=2pt, -Stealth] (3.1,16.25) -- (3.1,17.5);
\draw [  color={rgb,255:red,255; green,0; blue,0}, line width=2pt, -Stealth] (8.15,22.5) -- (8.15,20);
\draw [  color={rgb,255:red,255; green,0; blue,0}, line width=2pt, -Stealth] (3.1,20) -- (3.1,22.5);
\draw [ fill={rgb,255:red,253; green,224; blue,224} , line width=2pt ] (13.75,10) rectangle (15,8.75);
\draw [ fill={rgb,255:red,224; green,240; blue,227} , line width=2pt ] (13.75,11.75) rectangle (15,10.5);
\draw [ fill={rgb,255:red,229; green,235; blue,255} , line width=2pt ] (13.75,13.5) rectangle (15,12.25);
\node [font=\huge] at (18.35,12.75) {Network Layer};
\node [font=\huge] at (19,11.15) {Financial Protocol};
\node [font=\huge] at (18.9,9.35) {Application Layer};
\draw [ line width=2pt, -Stealth] (13.75,14.55) -- (15,14.55);
\draw [  color={rgb,255:red,255; green,0; blue,0}, line width=2pt, -Stealth] (13.75,16.3) -- (15,16.3);
\node [font=\huge] at (18,16.3) {10G Ethernet};
\node [font=\huge] at (17.75,14.55) {AXI-Stream};
\end{circuitikz}
}
\vspace*{5mm}
\caption{Simplified anatomy of a HFT system. Exchanges broadcast ticker data along a 10 GiB Ethernet cable to the HFT system, where the networking stack receives and processes packets in user space. Packets are typically compressed in a domain specific format for bandwidth reasons, which are then parsed into meaningful market orders by the financial protocol which are then ordered in the order book. The custom application then issues the buy and sell orders back over the network, this is the area in which the optimisations presented in this work pertain to\cite{hft-design}.  Figure adapted from \cite{hft-design}.}
\label{fig:my_label}
\end{figure}
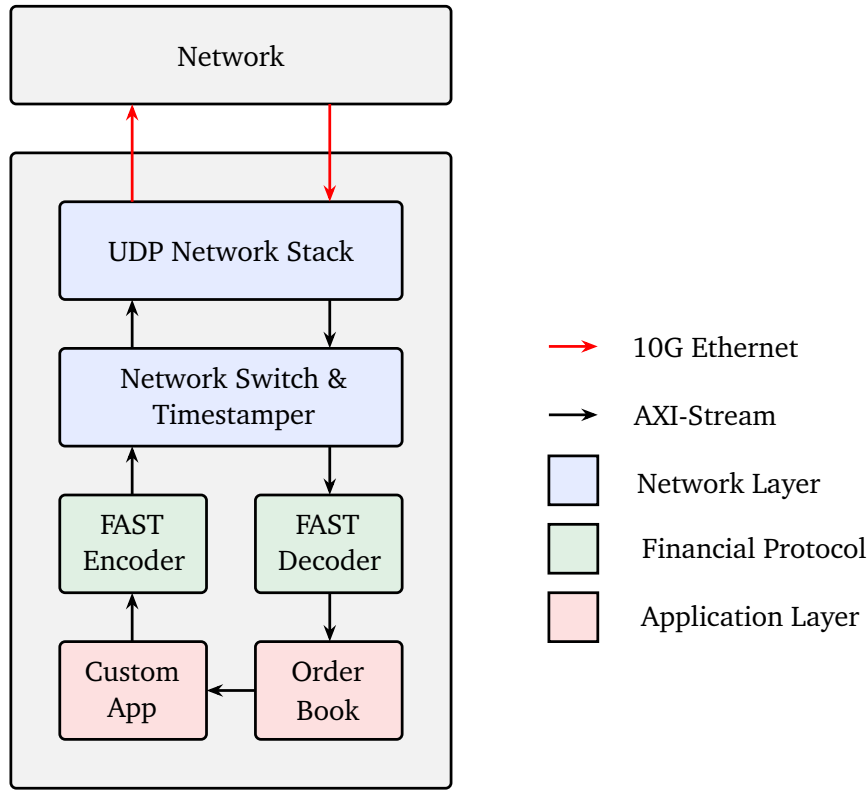

While this work primarily addresses software-based optimizations for latency-critical applications, it is crucial to underscore the significance of the hardware stack in HFT firms for trade execution. Simple trading strategies often rely heavily on the network stack, implemented in custom firmware and Field Programmable Gate Arrays (FPGA) \cite{leber2011high}, to achieve execution latencies in the nanosecond range. As proprietary software approaches maximum optimization, it is likely that the engineering focus will increasingly shift towards the hardware stack, which is still in its early stages and has considerable room for growth.

\section{Software Contribution}
\subsection{Outline}

This section is dedicated to the idea and development of semi-static conditions, outlining theory, design philosophy and optimisations that have been applied iteratively over the development process. The majority of this section is focused on developing a prototype that emulates the behaviour of a simple \texttt{if/else} statements, with some discussions given on generalisations to switch statements and non-static member functions given in later sections. Whilst discussions on portability are reserved for the evaluation stage, there was an important decision to be made about the choice of development environment for the prototype, more specifically the choice of operating system, compiler and C++ version. Since the primary applications of this language construct will find use in low-latency environments such as HFT systems, the choice of development environment was curated to reflect an industry standard. In light of this, development was done under a Linux OS (Ubuntu distribution) with GCC 13.1 and C++20. At certain stages of the development process, attention will be brought on specific Linux system calls that involve manipulating page permissions for running executables, however equivalent API's exist in other OS's (mentioned in later sections) and its fundamental mechanics is common for OS's that use paging for memory management.

\subsection{Semi-static Conditions}
Semi-static conditions can be defined as a language construct that emulate the behaviour of conditional statements, but separates condition evaluation logic and branch taking (the subsequent machine code executed on the pretext of the condition). At compile time, the underlying assembly of semi-static branch taking would resemble that of a function call with no indirection, allowing for a deterministic flow of execution with full support of compiler optimisations and hardware intrinsics. In the context of conditional branching, semi-static branch taking behaves as a static conditional statement: the condition is evaluated at compile time and does not change for the duration of program execution, which in a classical sense, be thought of as compile time template polymorphism (the branch). The \emph{semi} part is associated with the polymorphic nature of the language construct: the ability to programatically change the direction of the branch at runtime, whilst maintaining the deterministic compile-time behaviour already mentioned. With semi-static conditions, the lines between the compilation and execution phases of a program become blurred, and the nature of the executable shifts from static to somewhat polymorphic or self-modifying. This behaviour manifests itself within the branch-taking mechanism as a single/sequence of assembly instructions that redirect control flow to the respective \texttt{if} or \texttt{else} branches, controlled by branch-switching logic that performs the modification.

In the context of branch optimisation, the philosophy behind the construct is simple. The seperation of branch-switching and branch-taking logic produces an important decoupling of relatively expensive and cheap operations, allowing for more strategic and granular control over conditional branching. In this case, the branch-switching logic can be considered as an expensive operation because it involves altering assembly instructions in memory, whilst branch-taking logic can be considered cheap as it is simply a direct function call. Isolating branch-switching logic in less performance critical code paths allow conditions in performance-critical sections to be evaluated preemptively without interference, bypassing the need for branch prediction (more specifically for conditional branches) and eliminating mispredictions in the latency-critical path. When the hot-path is executed, no conditional checks are needed and the branch is executed as if it where always perfectly predicted. In instances where code paths are infrequently executed, but contain branches that are often mispredicted, semi-static conditions show promise for optimisations.

In order to realise this language construct, some key challenges are addressed in the development process, which can be broadly split into branch-changing and branch-taking logic. Branch-changing logic needs to be able to find the address of the assembly code instructions to edit in memory, and perform the editing in way that calling the branch-taking method redirects program control flow to user-specified regions based on a runtime condition. Branch-taking logic needs to ensure that control flow is redirected with minimal overhead, so it becomes comparable with the execution latency of a perfectly predicted branch, or a direct function call. At a language level, this is not only dependant upon the underlying assembly instructions that are edited, but also being able to reap the full benefits of compiler optimisations without compromising the safety of the program. On the hardware level, it is paramount that branch-taking code benefits from the same caching, instruction pre-fetching and branch target resolution effects that regular function calls or unconditional jumps do to ensure deterministic and low-execution latencies. Whilst it may be possible to achieve near-identical execution schematics to direct function calls on a language level, the cost of cache incoherence and instruction pipeline stalls can trump branch-misprediction by orders of magnitude, making the construct infeasible in low-latency settings. Therefore, it is crucial that semi-static conditions are compatible in this way with modern hardware and processors. Lastly, and more broadly, the language construct needs to be designed with ease of use in mind. This includes simple and elegant syntax, flexibility and a design that allows it to be easily portable across different architectures, compilers and operating systems.

\subsection{Prototype Development}

The first step in the development process is to establish the desired syntax of the core branch-switching and branch-taking functionality of the language construct. It is likely that this will have a significant influence on the design of semi-static conditions, so establishing how the end product is desired to look in the preliminary stages gives a clear direction in development goals. After careful consideration of simplicity and elegance, the desired usage can be seen below.\\

\begin{lstlisting}[language=c++]
void function_1() { ... }
void function_2() { ... }
(...)
BranchChanger branch(function_1, function_2);
branch.set_direction(condition);
branch.branch();

\end{lstlisting}

\noindent \\Semi-static conditions will manifest itself as the \texttt{BranchChanger} class which is instantiated by taking the addresses of two functions as arguments. These functions represent the \texttt{if} and \texttt{else} branches respectively, and their equivalent usage with conditional statements can be seen below.\\

\begin{lstlisting}[language=c++]
void function_1() { ... }
void function_2() { ... }
(...)
if (condition) 
    function_1();
else 
    function_2();
\end{lstlisting}

\noindent \\The \texttt{set\_direction} method will be responsible for controlling which of the branches is executed based on a user specified runtime condition, whilst the \texttt{branch} method will be responsible for executing the branch with minimal overhead. The signature of the \texttt{branch} method will always be identical to the functions passed as arguments when the class is instantiated, serving as a single entry and exit point for both branches. 

Now that a clear high-level design has been established, the next course of action is to implement the branch-taking functionality. Before delving into assembly instruction modification to facilitate the execution of the \texttt{if} and \texttt{else} branches, some thought needs to be given in how the \texttt{branch} method can act as a single entry and exit point for both branches. This method will not do any meaningful work, its sole purpose is to behave as a trampoline to other areas of the code segment while still being able to propagate return values and register data as if one of the branches where called directly. Its first clear responsibility is to set up the call stack in the exact way that the branches would as if they where called in isolation; since control flow will be redirected before the \texttt{branch} function has opportunity to manipulate the stack, in theory the target branches will be able to use any caller-saved data as if it where called directly. To test this theory, we can observe the disassembly for two functions with identical signatures under \textbf{-O0} optimisations so any calling behaviour is not omitted.\\

\begin{lstlisting}[language=c++]
int add(int a, int b) { ... }
int branch(int a, int b) { ... }
...
mov    esi, 2
mov    edi, 1
call   add(int, int)
...
mov    esi, 2
mov    edi, 1
call   branch(int, int)
\end{lstlisting}

\noindent \\As expected, both instances follow a standardised calling convention resulting in identical caller behaviour: arguments are pushed from right to left (in this case since there are less than 3 arguments, they are instead moved into registers as per x86 calling conventions) before the subroutine is executed. While this may seem trivial, the standardisation of calling conventions is an extremely important feature of modern compilers that can be leveraged to generalise the \texttt{branch} method in a safe and portable manner. 

Now that it's clear that functions with identical signatures observe identical caller setup, and hence the assembly generated on the callee side will be tailored to reflect this, the next course of action is to make the \texttt{branch} entry point mimic the identical signature of the branches passed into the constructor. Using class template deduction, return types and arguments of the branches passed into the constructor can be deduced at compile time and leveraged to generate the correct assembly code for the \texttt{branch} method.\\

\begin{lstlisting}[language=c++]
template <typename Ret, typename... Args>
class BranchChanger
{
    using func = Ret(*)(Args...);
    ...
    BranchChanger(func if_branch, func else_branch);
    ...
    Ret branch(Args... args);
}
\end{lstlisting}

\noindent \\The templating splits the signature into arguments and return type, where the arguments are represented by a variadic parameter pack, which can be deduced through the pointer types passed into the constructor (represented with the type alias \texttt{func} for readability). These types are then used to declare the signature of the \texttt{branch} member function, ensuring that it is identical to the \texttt{if} and \texttt{else} branches. For implementation, virtually anything can be placed inside \texttt{branch} and as long as the return type matches the signature, it will compile. Here there are two main cases to disginuish between; void and non-void return types. If the return type is non-void, returning a brace initialised object of type \texttt{Ret} will suffice for compilation, whereas void return types must be absent of non-void return values. Using \texttt{std::is\_void\_v<Ret>}, we can perform compile time type checking to circumvent this edge case and always ensure compilation for both void and non-void return types.\\

\begin{lstlisting}[language=c++]
Ret branch(Args... args)
{   
    if constexpr (!std::is_void_v<Ret>)
    {
        return Ret{};
    }
}
\end{lstlisting}

\noindent \\Now that the entry point is functional, we can double check the underlying assembly to ensure it has identical caller behavior to the branches. For this example, the branches used to instantiate the construct will be addition and subtraction functions with two integer arguments and an integer return type.\\

\begin{lstlisting}[language={[x86masm]Assembler}]
lea    rax, [rbp-16]
mov    edx, 2
mov    esi, 1
mov    rdi, rax
call   BranchChanger<int, int, int>::branch(int, int)
\end{lstlisting}

\noindent \\From the demangled function call it appears that the correct template is generated, however additional instructions have been added on the caller side. In addition to the integer arguments, an effective address is computed based on an offset from the frame pointer, which is moved into the \texttt{rdi} register after all previous arguments have been set up. From first glance it may be unclear why this is occurring, however when delving deeper into C++ calling conventions, what is happening is that an implicit \texttt{this} pointer belonging to the specific \texttt{BranchChanger} instance is being pushed onto the stack \cite{agner-call}. If the \texttt{branch} member function utilised data members specific to the parent instance then this would be necessary, however this not the case and all it does is disrupt register offsets making trampolining to regular functions infeasible. Whilst it may be possible to rectify this programmatically using inline assembly, a safer and more portable solution would be to declare the \texttt{branch} method as static since static member functions are not associated with any particular instance of a class. 

\begin{lstlisting}[language={[x86masm]Assembler}]
mov    esi, 2
mov    edi, 1
call    BranchChanger<int, int, int>::branch(int, int)
\end{lstlisting}

\noindent \\Static declarations allow the \texttt{branch} entry point to work seamlessly, however this limits the number of \texttt{BranchChanger} constructs that can be instantiated per function signature. Template specialisation allows for the differentiation of static \texttt{branch} entry points if the branches passed into the constructor have different signatures, which allows for multiple instances of semi-static conditions in a single program. However if more than one \texttt{BranchChanger} instance exists for a specific signature, this means they will share a common entry point which is problematic: two instances will be performing assembly modification on a single \texttt{branch} method which will have undefined behaviour. A way to circumvent is would be to alter the return types of the branches with other built-in or custom types to ensure that different templates are generated. Given these trade-offs, the priority for development is to work as much as possible with the compiler to avoid writing inline assembly for safety and portability reasons, and hence the final decision was to keep \texttt{branch} declarations as static.

Before moving onto implementing assembly modification, an important caveat to consider is compiler optimisations. From the compilers perspective, the \texttt{branch} method is a small function that does not produce any meaningful work making it susceptible to inlining or dead code elimination \cite{knoop1994partial}. Obviously, if this happens then the construct will be unusable, but limiting its usage to programs that are intended to be run on \textbf{-O0} defeats the purpose of it being used in high performance applications. The simplest solution would be to disable optimisations specifically on the \texttt{branch} method, which can be achieved on GCC using \texttt{pragma} directives or more elegantly using attributes.\\

\begin{lstlisting}[language=c++]
__attribute__((optimize("O0")))
static Ret branch(Args... args)
{   
    if constexpr (!std::is_void_v<Ret>)
    {
        return Ret{};
    }
}
\end{lstlisting}

\noindent \\\textbf{Assembly Editing} Now that the entry point is fully functional, development can start for the core assembly modification functionality. The target for this editing will the prologue instructions of the \texttt{branch} method, specific to each instance produced by template specialisation of the parent \texttt{BranchChanger} class. 

The first course of action is identifying the addresses of the machine code instructions to edit. In terms of the virtual address space, the machine code instructions of interest pertaining to the executable reside in the text segment, which itself is mapped by pages with read-only and execute permissions. Before we can perform any modifications, the page where the function prologue resides must be located and its permissions must be changed to read/write, otherwise the processor will raise a segmentation fault if any memory stores are attempted. Modern operating systems employ address space layout randomisation (ASLR) by randomising the base address of the virtual address space to prevent attackers from exploiting known memory addresses in executable, so locating executable pages must be deferred to runtime \cite{aga2019smokestack}. Upon construct instantiation, we can generate a pointer to the template specialised \texttt{branch} method which is essentially the logical address of the first instruction pertaining to the function. To obtain the address of the page boundary in which the function resides, we can compute the page offset modding the logical address by the page size of the system (which can be obtained using the C standard function \texttt{getpagesize}), and then subtracting this offset from the original address to align it with the lowest multiple of the page size. To change the page permissions, we can use the \texttt{mprotect} system call to alter the flags of the VMA (virtual memory area, a kernel data structure which describes a continuous section in the processes memory) corresponding to the page address previously computed \cite{linux-vma}.\\

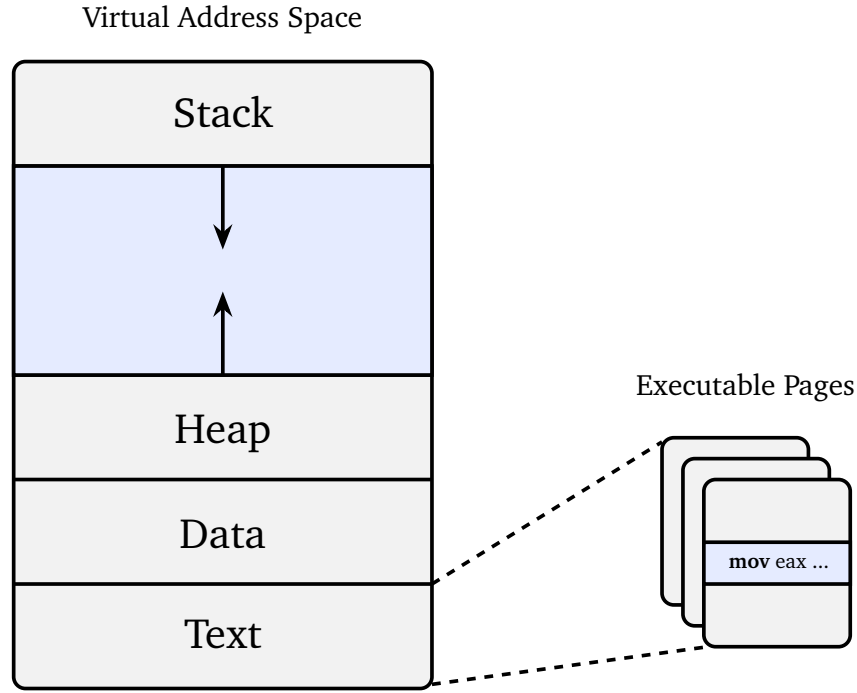
\begin{figure}[t]
\centering
\resizebox{.7\textwidth}{!}{
    \begin{circuitikz}
    \tikzstyle{every node}=[font=\Large]
    \draw [ fill={rgb,255:red,242; green,242; blue,242} , line width=1.25pt, rounded corners ] (10,15) rectangle (15,7.5);
    \draw [, line width=1.25pt](10,13.75) to[short] (15,13.75);
    \draw [, line width=1.25pt](10,11.25) to[short] (15,11.25);
    \draw [, line width=1.25pt](10,10) to[short] (15,10);
    \draw [, line width=1.25pt](10,8.75) to[short] (15,8.75);
    \draw [ fill={rgb,255:red,229; green,235; blue,255} , line width=1.25pt ] (10,13.75) rectangle (15,11.25);
    \node [font=\Large] at (12.5,14.4) {Stack};
    \node [font=\Large] at (12.5,10.55) {Heap};
    \node [font=\Large] at (12.5,9.35) {Data};
    \node [font=\Large] at (12.5,8.15) {Text};
    \draw [ fill={rgb,255:red,242; green,242; blue,242} , line width=1.25pt, rounded corners ] (17.75,10.5) rectangle (19.5,8.5);
    \draw [ fill={rgb,255:red,242; green,242; blue,242} , line width=1.25pt, rounded corners ] (18,10.25) rectangle (19.75,8.25);
    \draw [ fill={rgb,255:red,242; green,242; blue,242} , line width=1.25pt, rounded corners ] (18.25,10) rectangle (20,8);
    \draw [ line width=1.25pt, -Stealth] (12.5,11.25) -- (12.5,12.25);
    \draw [ line width=1.25pt, -Stealth] (12.5,13.75) -- (12.5,12.75);
    \draw [ fill={rgb,255:red,229; green,235; blue,255} , line width=1.25pt ] (18.25,9.25) rectangle (20,8.75);
    \draw [line width=1.25pt, dashed] (15,8.75) -- (17.75,10.45);
    \draw [line width=1.25pt, dashed] (15,7.55) -- (18.3,8);

    \node [font=\normalsize] at (18.75,11.1) {Executable Pages};
    \node [font=\normalsize] at (12.5,15.5) {Virtual Address Space};
    \node [font=\scriptsize] at (19.15,9) {\textbf{mov} eax ...};
    \end{circuitikz}
}
\vspace*{5mm}
\caption{Simplified representation of virtual address space segments, with lower segments residing at lower memory addresses. Blue segment highlighted in executable page represents an offset where a specific instruction can be found.}
\label{fig:my_label}
\end{figure}

\begin{lstlisting}[language=c++]
uint64_t page_size = getpagesize();
address -= (uint64_t)address % page_size;
mprotect(
    address, page_size, PROT_READ | PROT_WRITE | PROT_EXEC
);

\end{lstlisting}
\vspace{1pt}

Now that we have located the instructions in memory to edit (through the pointer to the \texttt{branch} function), and made this editing permissible through altering the page permissions where the function resides, we can start adding instructions to redirect control flow to the branches. With latency in mind, the scope of control flow instructions that can be used become limited to direct jumps or calls, which conventionally cannot be polymorphic without employing assembly editing. On x86 architectures, jumps and calls redirect control flow by supplying a relative 32-bit offset from the current program counter, which is reduced to a simple signed displacement arithmetic operation.\\

\begin{lstlisting}[basicstyle=\bfseries]
00000000000011a9    <foo>:
\end{lstlisting}
\vspace*{-4mm}
\begin{lstlisting}[language={[x86masm]Assembler}]
00000000000011a9:   f3 0f 1e fa          endbr64
00000000000011ad:   55                   push   rbp
00000000000011ae:   48 89 e5             mov    rbp,rsp
...   
00000000000011e3:   55                   push   rbp
00000000000011e4:   48 89 e5             mov    rbp,rsp
\end{lstlisting}
\vspace*{-4mm}
\begin{lstlisting}[basicstyle=\bfseries]
00000000000011e7:   e8 bd ff ff ff       call   11a9 <foo>   
\end{lstlisting}
\vspace*{-4mm}
\begin{lstlisting}[language={[x86masm]Assembler}]
00000000000011ec:   b8 00 00 00 00       mov    eax,0x0
\end{lstlisting}

\noindent \\Above is an example of the machine code generated for a call instruction (achieved with \textbf{objdump -d -M intel}) along with the program counter (left) and equivalent assembly (right). Focusing on the call instruction with opcode \textbf{e8}, we can see the following 4-byte displacement encoded as a signed hexadecimal value in little endian format (architecture specific). The signed 2's compliment equivalent of this displacement is -67 bytes, which is 5 bytes smaller than the displacement from the PC of the call instruction (\textbf{11e7}) to the function entry point (\textbf{11a9}). The reason for this is because the offset of relative jumps are computed from the last byte of the instruction to the first byte of the target address, which can be formally described with the equation:

\begin{center}
    Jump Offset = Target Address - Entry Point - Size of Instruction
\end{center}

Fundamentally, relative jumps are indeed branches and introduce control hazards in pipelined processors, but have subtly different prediction schemes and penalties on the micro-architectural level in comparison to conditional jumps and indirect jumps (via register values). Prediction schemes for direct jumps occur relatively early in the pipeline (processor front-end), whereas branches that have data dependencies (such as indirect jumps/calls) or are conditional on FLAGS can get deep in the pipeline (back-end) execute stages before mispredictions are discovered, and hence incur a higher penalty when previous instructions need to be flushed. For relative branches, the first line of prediction occurs in the branch target buffer (BTB) which acts as a specialised cache who's role is to predict weather the PC resolves to a branching instruction, and if so, what block to fetch next \cite{hennessy2017computer}. This is especially important for speculative prefetching: the instruction prefetcher needs to know in advance which blocks to fetch next, so if the PC is a branch, it can can steer the prefetcher to the predicted branch target and begin bringing the associated instructions into lower level caches. If predicted correctly then virtually no penalties are incurred as for conditional and data dependant branches, however the distinction occurs at the pipeline stage where mispredictions are detected and resolved. When an instruction reaches the decode stage, more information is attained surrounding the nature of the instruction. The branch address calculator (BAC) will ensure that the branches have the correct target by computing the absolute address of the PC and comparing it with the supplied target. If a direct jump is mispredicted at this stage, this means that the supplied branch target does not match the predicted, and proceeding instructions that have been incorrectly fetched will be flushed and the prefetcher will be re-steered to the correct branch target \cite{BAC-intel}. In regards to conditional branches, the same applies with the BAC however mispredictions are ultimately detected later in the execute stages which result in more severe pipeline stalls. On processors with branch order buffers (BOB), the recovery process can start before the processor pipeline has been flushed, but nevertheless the relative cost for mispredicted unconditional branches is much lower than conditional branches \cite{BAC-intel}.

\begin{figure}[t]
\centering
\resizebox{.8\textwidth}{!}{%
\begin{circuitikz}
\tikzstyle{every node}=[font=\LARGE]
\draw [ fill={rgb,255:red,229; green,235; blue,255} , line width=2pt, rounded corners ] (0,15) rectangle (11.25,10);
\draw [ fill={rgb,255:red,242; green,242; blue,242} , line width=2pt ] (1.25,13.75) rectangle (5,11.25);
\draw [ fill={rgb,255:red,242; green,242; blue,242} , line width=2pt ] (6.25,13.75) rectangle (10,11.25);
\draw [ fill={rgb,255:red,242; green,242; blue,242} , line width=2pt ] (12.5,13.75) rectangle (16.25,11.25);
\draw [ fill={rgb,255:red,242; green,242; blue,242} , line width=2pt ] (17.5,13.75) rectangle (21.25,11.25);
\draw [, line width=2pt, -Stealth] (5,12.5) -- (6.25,12.5);
\draw [, line width=2pt, -Stealth] (10,12.5) -- (12.5,12.5);
\draw [, line width=2pt, -Stealth] (16.25,12.5) -- (17.5,12.5);
\node [font=\LARGE] at (3.2,12.5) {Fetch};
\node [font=\LARGE] at (8.1,12.5) {Decode};
\node [font=\LARGE] at (14.45,12.5) {Execute};
\node [font=\LARGE] at (19.4,12.5) {Retire};
\draw [ fill={rgb,255:red,242; green,242; blue,242} , line width=2pt ] (1.25,7.5) rectangle (10,5);
\node [font=\LARGE] at (5.65,6.25) {BPU / Branch Prediction};
\draw [ fill={rgb,255:red,242; green,242; blue,242} , line width=2pt ] (1.25,8.75) rectangle (5,7.5);
\draw [ fill={rgb,255:red,242; green,242; blue,242} , line width=2pt ] (6.25,8.75) rectangle (10,7.5);
\node [font=\LARGE] at (3.15,8.15) {BTB};
\node [font=\LARGE] at (8.15,8.15) {BAC};
\draw [, line width=2pt, Stealth-Stealth] (3,8.75) -- (3,11.25);
\draw [, line width=2pt, Stealth-Stealth] (8.25,8.75) -- (8.25,11.25);
\draw [ fill={rgb,255:red,242; green,242; blue,242} , line width=2pt ] (1.25,18.75) rectangle (10,16.25);
\node [font=\LARGE] at (5.7,17.5) {Instruction / Data Caches};
\draw [, line width=2pt, dashed] (3,16.25) -- (3,13.75);
\draw [, line width=2pt](14.25,11.25) to[short] (14.25,6.25);
\draw [, line width=2pt](19.25,11.25) to[short] (19.25,6.25);
\draw [ line width=2pt, -Stealth] (19.25,6.25) -- (10,6.25);
\node [font=\LARGE] at (5.5,15.6) {Front-end};
\end{circuitikz}
}%
\vspace*{5mm}
\caption{Simplified representation of the interaction of caches and branch prediction schemes on the instruction pipeline. Unconditional branch mispredictions are resolved at the decode stage by the BAC, whereas conditional branches are resolved at the execute stage. Information of retired predicted and mispredicted branches are fed into the BPU to update the predictor.}
\label{fig:my_label}
\end{figure}
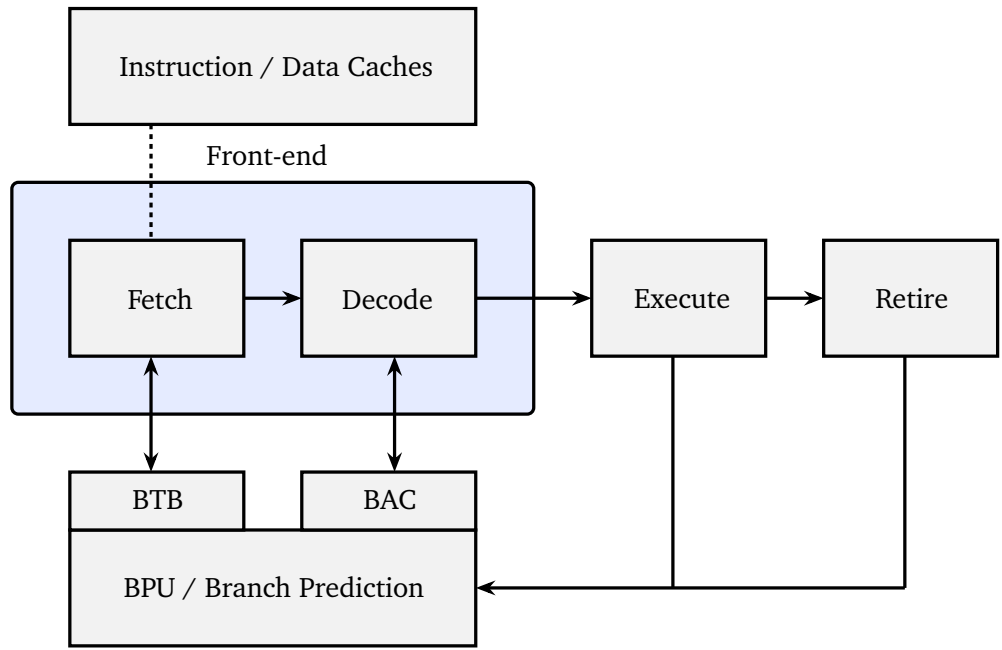

It is clear that relative jump/calls provide the cheapest means of control flow alteration, providing that the size of the jump does not exceed 2$^{32}$ bytes. The question that remains is which instruction would be most suitable from a latency and implementation perspective. Call's and jumps are very similar mechanically, with calls being a two-part atomic operation which jumps to an offset while pushing the return address onto the stack. From a latency perspective, while minimal, the additional use of the call stack pollutes data caches unnecessarily and the additional return instruction introduces more branching which is prone to mispredictions. However the main challenge comes from a development perspective; after the call is complete (within the \texttt{branch} method), the return address will be the proceeding instruction which may involve manipulating data on the stack/registers which have already been dealt with. Ensuring this does not occur for all possible signatures is tedious, and may limit future optimisations on the language construct. Using a jump will be much simpler; we can simply go straight to the branch without needing to ever complete execution of the entry point, and when the branch has finished executing, control flow will be redirected to the original calling site of the \texttt{branch} method (recall that a \texttt{ret} instruction is essentially a \texttt{jmp [reg]}). 

To implement the jump, the first thing we do is alter the opcode of the first instruction pertaining to the \texttt{branch} method to \textbf{e9} (\texttt{jmp} opcode on x86) through its pointer, and then increment it. The following 4 bytes will be reserved for the relative offsets from the current program counter. To compute the offsets, we simply use pointer arithmetic to compute the displacement in memory from the \texttt{branch} method to the respective if/else branches, then subtracting the length of the instruction from the offset as specified in the formula mentioned earlier. Next, the the integer offsets are converted into a 4-byte representation and stored in two dimensional member array (total of 8-bytes), accounting for architectural byte ordering.\\

\begin{lstlisting}[language=c++]
unsigned char offset_in_bytes[DWORD] = {
    static_cast<unsigned char>(offset & 0xff),
    static_cast<unsigned char>((offset >> 8) & 0xff),
    static_cast<unsigned char>((offset >> 16) & 0xff),
    static_cast<unsigned char>((offset >> 24) & 0xff)
};

#if __BYTE_ORDER__ == __ORDER_BIG_ENDIAN__
change_byte_ordering(offset_in_bytes);
#endif

std::memcpy(dest_array, offset_in_bytes, DWORD);
\end{lstlisting}

\noindent \\\textbf{Changing Branch Directions} At this point the development of the construct is nearly complete, leaving only the direction setting method for development. With the offsets computed and stored in a class member array, setting the branch direction would simply involve a \texttt{memcpy} of these bytes to the \texttt{branch} pointer as the destination address. Direction setting must be initially done upon instantiation, since altering the first byte of the \texttt{branch} entry point will result in a jump to an undefined location, likely causing a segmentation fault. The class was adapted to have an optional parameter in the constructor to specify the initial direction of the branch, with the default condition being true. This can be though of similar scheme to compiler branch prediction hints; the programmer can specify the likely direction which the branch would first be taken, but will still have the control to change this at any given time. The actual \texttt{set\_direction} method will use boolean indexing (the boolean being the runtime condition passed to the method) to access the bytes pertaining to the correct branch, which are copied into the 4 byte slot next to the \texttt{jmp} opcode. The boolean indexing approach is simple, and allows for active cache warming if desired.\\

\begin{lstlisting}[language=c++]
void setDirection(bool condition)
{
    std::memcpy(dest, src[condition], DWORD);
}
\end{lstlisting}

\noindent \\\textbf{Concluding Remarks} Upon testing, the semi-static conditions prototype appears to work seamlessly for varying branch signatures on all optimisation levels. Whilst it is not possible to observe the assembly editing in real time without specialised disassemblers (for example, \textbf{objdump} only shows the contents of the object file which is not edited, but rather the pages that are mapped by it), using \textbf{perf record} it is possible to observe it indirectly through the percentage cycles spent in the \texttt{branch} method. \\

\begin{lstlisting}
Percent |   
        |    BranchChanger<int, int, int>::branch(int, int)
100.00  |    push %rbp
        |    mov  %rsp,%rbp
        |    mov  %edi,-0x4(%rbp)
        |    mov  %esi,-0x8(%rbp)
        |    mov  $0x0,%eax
        |    pop  %rbp
        |    ret
\end{lstlisting}

\noindent \\Above shows the disassembly of the \texttt{branch} method with the percentage cycles spent on each instruction on the left hand side, obtained using \textbf{perf record}. The branches used in this example are simple addition and subtraction functions. The data shows that the first instruction within the \texttt{branch} constitutes all the cycles spent in the function entirely; this is the instruction that is edited to a jump and hence it is expected that this is the only instruction that executes for the duration of the program. The branches themselves have a small percentage of cycles spent relative to all other methods in the test program, which supports that the branches are in fact executed and control flow is redirected accurately, this is also confirmed by simply printing the return values to the standard output. All the above, along with correct program behaviour, suggest that the language construct works as intended. Now the core prototype is complete, thought can be given into optimisations for branch taking and branch setting, as well as additional features that expand from this core concept (switch statements, also class member functions which observe different calling behaviour than conventional functions).

\subsection{Optimisations}

So far, the prototype demonstrates a proof-of-concept, but not a final product. The overarching goal of this language construct is to provide deterministic (low standard deviation) and low latency branch taking in scenarios where misprediction rates are high. Although mispredictions on the processor level are expensive, if the branch-taking component does not perform at a similar level to perfectly predicted branches, the construct will not find use in performance sensitive environments. Prior to running benchmarks, it is crucial that we level the playing field as much as possible and make semi-static conditions competitive. 

\noindent \\\textbf{Branch Taking Optimisations} Naturally the most obvious place to start is the branch-taking method itself. In the development process, we ensured that the first instruction executed within \texttt{branch} is the jump that detours execution to one of the branches, so the processor does not waste any time executing instructions it does not need to. This is fine, however the glaring bottleneck arises from having to prevent all compiler optimisations on the method, which was implemented as a one-hot fix from preventing the function from being eliminated. Ideally, the entry point should benefit from all optimisations that regular functions do, but have the minimum amount of optimisations disabled that prevent the construct from working as intended. The first obvious approach is do only disable in-lining for the entry point; this is destructive as the compiler will place the body of the function pre-editing within the calling site which essentially does nothing. Even if it managed to inline the edited assembly, it will be completely infeasible to target the in-lined instruction within the code segment. Replacing the compiler attribute on \texttt{branch} with \texttt{\_\_attribute\_\_((noinline))} generated the following assembly under \textbf{-O3} optimisations.\\

\begin{lstlisting}[basicstyle=\bfseries]
000000000000118c:    lea  rdx,[rip+0x26d]
0000000000001193:    lea  rax,[rip+0x136]   
\end{lstlisting}
\vspace*{-4mm}
\begin{lstlisting}[language={[x86masm]Assembler}]    
000000000000119a:    sub  rax,rdx
000000000000119d:    sub  rax,0x1
00000000000011a1:    mov  DWORD PTR [rip+0x25a],eax
\end{lstlisting}

\noindent \\Surprisingly, even with the \texttt{noinline} directive the compiler still reduced the \texttt{branch} call to the \texttt{lea} instructions highlighted in bold. Upon further research into the effects of GCC compiler attributes, what appears to be happening is that the inlining prevention does in fact take place, but the compiler deems the function to have no side effects and as a result optimises out the call completely. A simple way to control this optimisation is by inlining assembly within the function body; inline assembly adds uncertainty to the compiler optimiser as it cannot determine if it has side effects on register or memory values. Adding a simple \texttt{asm("")} which does not produce any meaningful work is sufficient to prevent optimising out the function call in addition to using the \texttt{noinline} attribute.

Further testing revealed an interesting yet problematic optimisation, calls to the original \texttt{branch} method where replaced with calls to a different \texttt{branch} method which the compiler duplicated and altered a number of instructions within the body. Below is an example of both instances of the method, with demangled function names simplified for readability.\\

\begin{lstlisting}[language={[x86masm]Assembler}]
0000000000001280: <_BranchChanger_branch.constprop.0.isra.0>
0000000000001280:    ret
0000000000001281:    cs nop WORD PTR [rax+rax*1+0x0]
0000000000001288:    nop
000000000000128b:    nop    DWORD PTR [rax+rax*1+0x0]
...
0000000000001290: <_BranchChanger_branch>
0000000000001290:    endbr64
0000000000001294:    xor    eax,eax
0000000000001296:    ret
\end{lstlisting}

\noindent \\The first example of the \texttt{branch} method represents the duplicated form which is called, whereas the second example represents the method which is needed to be called to allow semi-static conditions to work. Inspecting the demangled name of the duplicated function reveals the optimisation that has been applied: interprocedual constant propagation (ICP). This optimisation is multifaceted; when the compiler recognises that a function call has some arguments passed as constants, it creates a spot-optimised clone of the function which can involve removing redundant computations and memory accesses. This can be seen in the \texttt{constprop} version; the primary instruction becomes a \texttt{ret} because the compiler can see that the \texttt{branch} method does not produce any meaningful work, the remaining instructions are included as padding to align the function on a 16-byte boundary. This padding is important especially for procedural calls since most modern processors fetch instructions on aligned 16-32 byte boundaries; fetching code after after an unconditional jump costs a few clock cycles however this delay is worsened if the branch target does not lie on a 16-32 byte boundary \cite{agner-call}. Following this, an interesting observation can be made in regards to the original \texttt{branch} method. In many instances, the function itself does not follow alignment and as a result the compiler seems to always place it at the bottom of the text segment to prevent misalignment of all other procedures in the executable. Another interesting observation is that the \texttt{constprop} version is often placed close to hot code paths in the text segment (often very close to \texttt{main}), which can reduce instruction cache fragmentation by placing contiguous subroutines relatively close to one another.

The issue of preventing constant propagation and function cloning can be easily solved by including the \texttt{optimize("no-ipa-cp-clone")} attribute in the function header. However prior analysis into the effects of ICP and procedural reordering opens some interesting avenues into possible improvements. Taking a page out of the compilers book, the first observation of ICP was the reordering of instructions such that the most important instructions reside at the function entry point with no wasted work in between. In the improved \texttt{branch} method, the preliminary instruction is typically a 4-byte \textbf{endbr64} on Intel CPU's that employ control flow enforcement technology (present on Linux with GCC and Clang), which ensures that indirect jumps/calls can only be made to functions which start in this instruction \cite{intel_cet}. In the case of semi-static conditions, it is highly unlikely that the \texttt{branch} method will find use in indirect calls due to the associated costs with indirection in general, especially in low latency settings. Given this, overwriting this preliminary instruction with a 5-byte jump was the direction taken in development, however a key thing to note is that this editing overwrites the opcode the proceeding instruction given its greater length. While it is unclear what ramifications this has on variable instruction length pre-fetching, perhaps hard coding a 5-byte jump in the entry point (and editing will not alter the length of the instruction) will be more "friendly" towards hardware semantics. In addition this may have positive implications on the BTB; from compile time the PC associated with the preliminary instruction of the \texttt{branch} method will always be a jump, meaning that in theory the BTB should always predict that a control flow instruction is present within \texttt{branch} which can save some cycles associated with mispredictions from preliminary calls. Even if later benchmarks reveal there is no observable performance gain from doing this, it does defer some work on construct instantiation.

To ensure that a unconditional jump always resides at the \texttt{branch} entry point, the \textbf{endbr64} instruction will need to be omitted by the compiler which can be done with the \texttt{nocf\_check} attribute. Since there is already an assembly instruction present within \texttt{branch} to prevent the compiler from optimising out the call, this can be simply changed to \texttt{asm("jmp 0x0")} which hard-codes a jump to an arbitrary 4-byte offset, this will be edited exclusively. Upon inspecting the disassembly, the \texttt{branch} method starts to resemble its \texttt{constprop} counterpart even more:\\

\begin{lstlisting}[language={[x86masm]Assembler}]
0000000000001280: <_BranchChanger_branch>
0000000000001280:    jmp    0 <__abi_tag-0x38c>
0000000000001285:    xor    eax,eax
0000000000001287:    ret
\end{lstlisting}

\noindent \\Interestingly, this assembly ordering seems to be maintained regardless of the function signature; the compiler seems to understand to not optimise the function call so there is full benefit of caller setup and teardown, but it also understands that the function does no useful work and optimises accordingly. The only differences that remain now between the ICP counterpart is 16-byte alignment and procedural reordering. A simple way to ensure this is including the \texttt{hot} attribute which instructs the compiler to optimise the function more aggressively and places it in a subsection of the text segment where hot code lies. This is typically done automatically with the \textbf{--vprofile-use} flag to which the compiler uses profile feedback from previous executions to determine which functions can benefit from reordering. A caveat with using this approach is it relinquishes the programmers ability to decide which functions should have priority in the hot text segment, which may decrease performance depending on the application this is integrated in. However \texttt{hot} attributes are far more common across compilers than byte-alignment directives, so from a portability standpoint it would be easier to generate the desired assembly using this method. Given these alterations, the final disassembly can be seen below:\\

\begin{lstlisting}[language={[x86masm]Assembler}]
0000000000001170: <_BranchChanger_branch>
0000000000001170:    jmp    0 <__abi_tag-0x38c>
0000000000001175:    xor    eax,eax
0000000000001177:    ret
0000000000001178:    nop    DWORD PTR [rax+rax*1+0x0]
000000000000117f:    nop
\end{lstlisting}

\noindent \\The altered version of the \texttt{branch} method, including the \texttt{hot} attribute is also shown below. Note that the \texttt{noinline} attribute is omitted since \texttt{no-ipa-cp-clone} includes this implicitly, and the function is always optimised on \textbf{-O3} to ensure the preliminary instruction is always a jump.\\ 

\begin{lstlisting}[language=c++]
__attribute__
((hot, nocf_check, optimize("no-ipa-cp-clone", "O3")))
static Ret branch(Args... args)
{   
    asm ("jmp 0x00000000");
    if constexpr (!std::is_void_v<Ret>)
    {
        return Ret{};
    }
}
\end{lstlisting}

\noindent \\The improved \texttt{branch} method was benchmarked against the prototype version with various suites, broadly split into instruction-level benchmarks with in-lined perf events and Intel cycle counters, as well as micro-benchmarks with google benchmark which involved measurements of more computationally expensive situations. More detailed methodology is explained in later sections, the purpose of these preliminary benchmarks is to ensure the proposed changes do not incur adverse effects on the language construct. Broadly on the instruction level, the improved version improved performance by several cycles across different branches, with the most signifcant contribution coming from procedual reordering associated with the \texttt{hot} attribute. For higher level measurements, some benchmarks showed performance gain by 5-10\% whereas others had identical execution times. On the instruction level it is difficult to speculate the source of this performance gain; at runtime both instances have identical execution pathways in terms of instructions so a reasonable explanation would be to improved locality between the entry point and branch targets. In larger systems with increased cache contention, the effects of alignment and locality have more of a prevalent effect on caching and prefetching which is reflected in some of the larger micro-benchmarks. Given these optimisations showed no adverse effects and showed marginal performance gain in some scenarios, they where incorporated into the final artefact.

\noindent \textbf{Branch Changing Optimisations} The current state of the direction changing method is already in quite an optimised state. There is an implicit branch for accessing the correct byte offset using a boolean index, however this is unavoidable. Fundamentally the performance of branch changing is not as important as branch taking; the whole reason for this separation is to isolate this more expensive operation from performance critical code to facilitate condition evaluation preemptively.

\subsection{Generalisations}

The focus of development has been primarily on semi-static conditions that emulate the behaviour of two-way conditional statements for conventional functions and static class member functions, and has been success-full thus fair in exploiting calling conventions to facilitate safe branch-taking. Nevertheless, the current prototype has the capability to be further generalised to work for a larger scope of branches without needing to alter the core branch-changing logic. Whilst these extensions may not find as much use for the specialised case (branch optimisation in HFT environments), they will improve the flexibility of the language construct for more general use cases outside the field of low-latency development.\\

\noindent \textbf{Class Member Functions} The current state of semi-static conditions rely on standardised calling conventions to facilitate stack setup/teardown and function argument passing by the compiler, without needing to write inline assembly code. Before extending this to class member functions, one must examine the disassembly pertaining to these invocations to understand the necessary changes that need to be implemented.\\

\begin{lstlisting}[basicstyle=\bfseries]
0000000000002436:	lea    rax,[rbp-0x60]
\end{lstlisting}
\vspace*{-4mm}
\begin{lstlisting}[language={[x86masm]Assembler}]
000000000000243a:	mov    rdi,rax
000000000000243d:	call   263a <_ZN9SomeClass3fooEv>
\end{lstlisting}
\vspace*{-4mm}
\begin{lstlisting}[basicstyle=\bfseries]
0000000000002442:	lea    rax,[rbp-0x60]
\end{lstlisting}
\vspace*{-4mm}
\begin{lstlisting}[language={[x86masm]Assembler}]
0000000000002446:	mov    rdi,rax
0000000000002449:	call   268a <_ZN9SomeClass3barEv>
\end{lstlisting}

\noindent \\The underlying assembly shows similar behaviour encountered during the development of the \texttt{branch} entry point; the effective addresses being computed (highlighted in bold) represents an implicit \texttt{this} pointer to the parent class which is the first parameter moved onto the stack. In this example both member functions are being invoked from the same instance, hence the identical offsets represented in the \texttt{lea} instruction. Propagating this behaviour to the entry point is simply the case of altering the class template to deduce the member function pointer type, and then updating the signature of the \texttt{branch} method to include the class instance within the signature, prior to the parameter pack that represents the functions arguments.\\

\begin{lstlisting}[language=c++]
template <typename Class, typename Ret, typename... Args>
class BranchChanger
{
    using func = Ret(Class::*)(Args...);
    ...  
    __attribute__
    ((hot, nocf_check, optimize("no-ipa-cp-clone", "O3")))
    static Ret branch(const Class& instance, Args... args);
}
\end{lstlisting}

\noindent \\These alterations are sufficient for making semi-static conditions work for non-static member functions without needing to change any optimisations on \texttt{branch} or the core assembly editing logic. This will become a reoccurring theme in further generalisations. This extension is contrived to work only for member functions that belong to the same class, which is expected considering the approach used to deduce the class type through templating. Nevertheless multiple instances are able to share the same entry point and have their member functions invoked through \texttt{branch}, with support for derived class methods as long as they are not overloaded (if derived class overloads a method from the base class and the base class pointers are passed to the constructor, only the base methods will be executed). 

\noindent \\\textbf{Switch Statements} The design of the language construct make it seem that generalisation to n-ary conditional statements would be simple; simply change the template parameters and the offset storage array to reflect the number of branches. However the syntactic requirements complicate template deduction. If it was possible to alias parameter packs directly, this would be a simple task of deducing the function pointer signature from a variadic pack of pointers, extracting the return types and arguments externally and aliasing them from within the class. Unfortunately C++20 does not support this directly, and trying to work around this by using containers such as \texttt{std::tuple} to hold the arguments is non-trivial, since there is also the task of extracting these types and forwarding them to either a pointer or the \texttt{branch} template declaration.

The solution to this is to break the \texttt{BranchChanger} class into a base class and derived class. The base class will be partially specialised to extract the regular or class member function signature (as seen so far), with the sole purpose of generating the \texttt{branch} method through template deduction and hense locating the bytes to edit.\\

\begin{lstlisting}[language=c++]
template <typename T>
class branch_changer_aux {};

template <typename Ret, typename... Args>
class branch_changer_aux { ... };

template <typename Class, typename Ret, typename... Args>
class branch_changer_aux { ... };
\end{lstlisting}

\noindent The derived class will be the actual \texttt{BranchChanger} which the programmer will interact with. The class itself has variadic template parameters representing a number of function pointers with identical signatures, these will be the branches which can range from 2 to $\infty$. Using \texttt{std::common\_type}, we can deduce the actual function signature type from the parameter pack which is used to instantiate the correct base class through CRTP.\\

\begin{lstlisting}[language=c++]
template <typename... Funcs>
BranchChanger : public branch_changer_aux
<typename std::common_type<Funcs...>::type> { ... }
\end{lstlisting}

\noindent \\The only adaptation needed will be the constructor, which will need to expand the parameter pack and iterate over all pointers passed to the constructor, computing relative offsets and storing them element-wise. C++20 supports unpacking these types into into a \texttt{std::vector} directly using brace-initialised fold expressions:\\

\begin{lstlisting}[language=c++]
BranchChanger(const Funcs... funcs)
{
    using ptr_t = typename std::common_type<Funcs...>::type
    std::vector<ptr_t> pack = { funcs... };
    for (int i = 0; i < pack.size(); i++) { ... }
    (...)
}
\end{lstlisting}

\noindent \\After a bit of hideous template meta-programming, the language construct becomes fully generalised to work for any number of branches, ans both regular and member functions. Template deduction is completely abstracted from the programmer without the need of manually writing out types, providing and elegant and affluent interface for easy use and integration. This concludes the development of semi-static conditions, the remainder of development time was focused on productionising the construct into a library, with a focus on portability across different operating systems and compilers. This stage is rather dull and not worth discussing, most of the complexity arose from creating pre-processor macros to ensure different system calls and optimisations are enabled based on the users OS and compiler flags specified.

\section{Benchmarks and Applications}
\subsection{Outline}
This section is dedicated to benchmarking the core operations that comprise semi-static conditions, exploring the effects self-modifying assembly instructions on performance, and investigating applications in both HFT and more general use cases. The experiments outlined will leverage a number of benchmarking suites ranging from Google Benchmark to custom performance counters that offer the fine granularity required to measure instruction level effects. The proceeding section is dedicated to outlining the experimental methods employed in obtaining these results for transparency and reproduciblity purposes. All measurements have been collected on an Intel(R) Core(TM) i7-10700 CPU (2.90GHz) with 256-kilobyte L1 instruction and data caches, 2-megabyte L2 caches and 16-megabyte L3 caches. Measurements collected will be architecture specific but have nevertheless been tested on similar architectures with Intel processors and have had consistent performance patterns with varying numbers. Tests have primarily been focused on semi-static conditions with regular or static member functions as branches given the large search space that exists with these kind of experiments.

\subsection{Experimental Method}
Conventional microbenchmarking frameworks such as Google Benchmark are useful for high level performance measurements where test cases are sufficiently long enough to measure observable differences in latency. However they often fall short for higher resolution measurements involving instruction level micro-benchmarks. When expected differences in performance manifest at the cycle level, the overhead associated with running these frameworks in conjunction with timer resolutions and standard errors mean that any observable differences in latency are \emph{hidden} by background noise, and often results become more influenced by the measurement taking rather than the actual measurements. This section is dedicated to outlining the experimental methods employed in capturing these sensitive measurements, based heavily on work done by Agner Fog, Matt Godbolt, and Intel as part of their microbenchmarking guides for i7 processors \cite{agner-optimise, godbolt-micro, intel-bench}.

\noindent \textbf{Clock Cycle Measurements} Benchmarks for code comprised of small numbers of assembly instructions where conducted using architecture specific timestamp counters, in this case RDTSC was used. Measurements are taken by reading the processors timestamp counter at two intervals with the code to benchmark in between, it is important to note that RDTSC counts reference cycles rather than core clock cycles due to CPU throttling effects. On super-scalar processors instructions are executed out-of-order and in-parallel to optimise penalties associated with different instruction latencies. This is problematic for such measurements; there is no guarantee that the RDTSC instruction is called in the precise temporal order that is specified programtically, and measurements may include other assembly instructions that are not intended to be measured. To resolve this, serialising instructions can be used in conjunction with RDTSC to force the CPU to complete all preceding instructions before continuing execution. Examples of serialising instructions are CPUID and LFENCE, in all tests LFENCE is used as it has a lower overhead and does not clobber the output registers of RDTSC. An example setup can be seen below, this snippet has been adapted from the official Intel microbenchmarking guide for i7 processors \cite{intel-bench}.\\

\begin{lstlisting}[language=c++]
_mm_lfence();
uint64_t start = __rdtsc();
_mm_lfence();

code_to_measure();

_mm_lfence();
uint64_t end = __rdtsc();
_mm_lfence();
uint64_t cycles = end - start;
\end{lstlisting}

\noindent \\Compiler optimisations often reorder assembly which complicates measurement taking. While there is no set way to ensure this, a trial and error approach was taken by padding instructions before the measurement taking and cross checking the disassembly to see if the necessary code is placed between the RDTSC calls.

Measuring taking itself does incur some overhead. To account for this, prior to benchmarking a background measurement is taken by running the above code with no instructions in between the RDTSC calls for many iterations (often in the order of 10$^7$), from which a mean latency is computed and subtracted from all proceeding benchmarks (excluding outliers). The benchmarks themselves are also run for many iterations since measurements tend to fluctuate around a mean value after sufficient warm-up time, due to variance in CPU frequency and individual instruction latencies. Therefore, data collected for benchmarks are processed as distributions rather than fixed computed values, which is beneficial for reasoning about latency standard deviations (important for HFT) and observing hardware level effects that contribute to this (e.g. branch mispredictions). 

\begin{figure}[!tbp]
  \centering
  \begin{minipage}[b]{0.49\textwidth}
    \includegraphics[width=\textwidth]{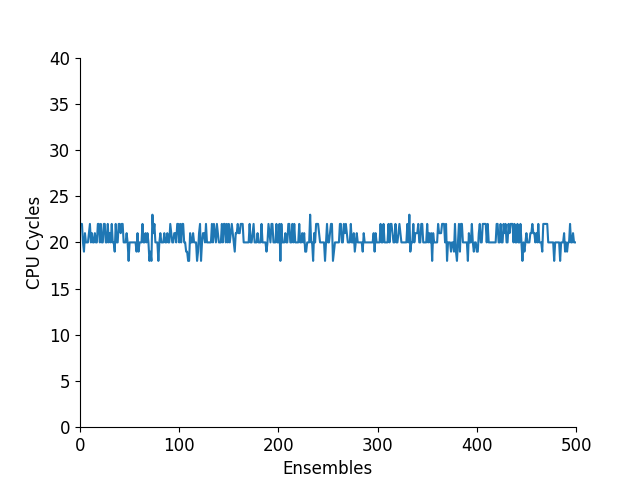}
    \caption*{\textbf{(a)} Measurement fluctuations}
  \end{minipage}
  \hfill
  \begin{minipage}[b]{0.49\textwidth}
    \includegraphics[width=\textwidth]{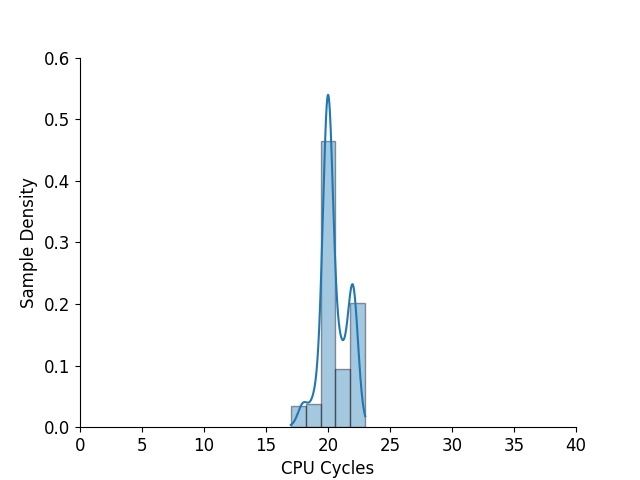}
    \caption*{\textbf{(b)} Sample distribution}
  \end{minipage}
  \caption{CPU cycle measurements of RDTSC overhead.}
\end{figure}

\noindent \\\textbf{Profiling} CPU performance counters are incredibly useful in identifying sources for particular hot-spots during program execution, and in the context of this research, identifying granular hardware effects that contribute to observed latencies. Perf was primarily used for performance profiling. A downside to this is that perf traditionally profiles the entire executable, rather than small subsets of it, meaning that any small observable changes hardware counters that are expected become enveloped in the overall noise of the system. Luckily, linux offers a API to access a subset of perf performance counters inside the executable using the \texttt{perf\_event\_open} system call, allowing for small pieces of code to be profiled in isolation (similar to RDTSC). Events are set up using the following code:\\

\begin{lstlisting}[language=c++]
struct perf_event_attr attr;
attr.type = PERF_TYPE_HARDWARE;
attr.config = PERF_COUNT_HW_INSTRUCTIONS;
attr.disabled = 0;
attr.exclude_kernel = 1;
attr.exclude_idle = 1;
attr.exclude_hv = 1;
attr.exclude_guest = 1;
\end{lstlisting}

\noindent \\The \texttt{type} and \texttt{config} attributes are used to select the performance counters which are generally the only parameters that are changed between tests. The remaining attributes offer finer control over sampling and are configured to exclude external noise from measurement taking. The actual profiling code can be shown below which has been adapted from the linux documentation of \texttt{perf\_event}:\\

\begin{lstlisting}[language=c++]
fd = perf_event_open(&attr, getpid(), -1, -1, 0);
ioctl(fd, PERF_EVENT_IOC_RESET, 0);
ioctl(fd, PERF_EVENT_IOC_ENABLE, 0);

code_to_profile();

ioctl(fd, PERF_EVENT_IOC_DISABLE, 0);
rc = read(fd, &count, sizeof(count));
\end{lstlisting}

\noindent \\Whilst this approach offers the best solution to granular profiling, the drawback is that the \texttt{perf\_event} API only offers a small subset of performance counters that perf offers. In experiments that utilise profiling, this inline approach is used when applicable, whilst the command line approach is used when performance counters that cannot be obtained using \texttt{perf\_event} are needed. In this case, code is often kept to a minimum to reduce measurement noise and often run alongside a baseline to extrapolate differences in performance counters.

\noindent \\\textbf{Microbenchmarking Frameworks} Where applicable, Google benchmark was used to gather latency data for less fine grained events. Google benchmark automatically configures the number of iterations the benchmark is run to get a stable estimate.

\subsection{Benchmarks}
This section is dedicated to benchmarking the branch-changing (\texttt{set\_direction}) and branch-taking (\texttt{branch}) methods for semi-static conditions, exploring the effects of self-modifying code and deducing optimal usage. Often, measurement distributions do not follow standard distributions due to skewness, so non-parametric tests are conducted on small subsets of the samples where applicable.

\noindent \\\textbf{Branch-changing Benchmarks} This set of tests is concerned with exploring the instances where altering assembly instructions in memory cause performance degradation and how they can be avoided.\\

\noindent The first test benchmarks the performance of \texttt{set\_direction} versus an equivalent 4-byte \texttt{memcpy} to non-executable memory. For fairness, a class was created with identical data members to semi-static conditions which where initialised with random bytes to represent some runtime deduced data. The \texttt{set\_direction} method in the baseline class is identical to the one in \texttt{BranchChanger}.\\

\begin{lstlisting}[language=c++]
class Baseline 
{
private:
    unsigned char* bytecode;
    unsigned char bytes[2][DWORD];
    
public:
    Baseline() 
    {
        bytecode = new unsigned char[DWORD];
        _random_bytes(bytes[0]);
        _random_bytes(bytes[1]);
    }

    void set_direction(bool condition) 
    {
        std::memcpy(bytecode, bytes[condition], DWORD);
    }
};
\end{lstlisting}

\begin{figure}[t]
    \centering
    \begin{minipage}{0.49\textwidth}
    \includegraphics[width=\textwidth]{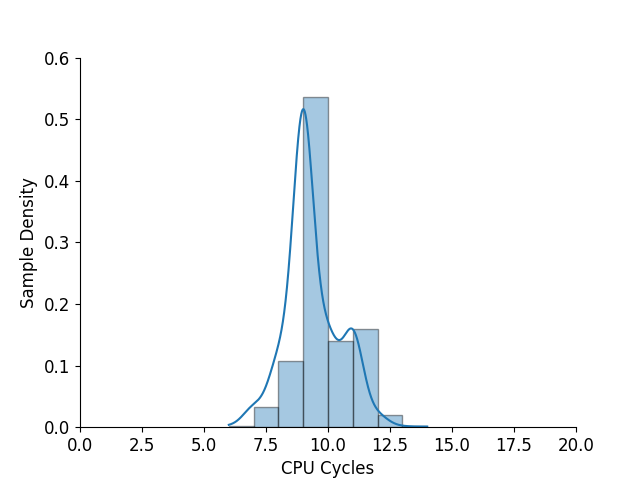}
    \caption*{\textbf{(a)} Baseline (M=9, SD=1)}
    \end{minipage}\hfill
    \begin{minipage}{0.49\textwidth}
    \includegraphics[width=\textwidth]{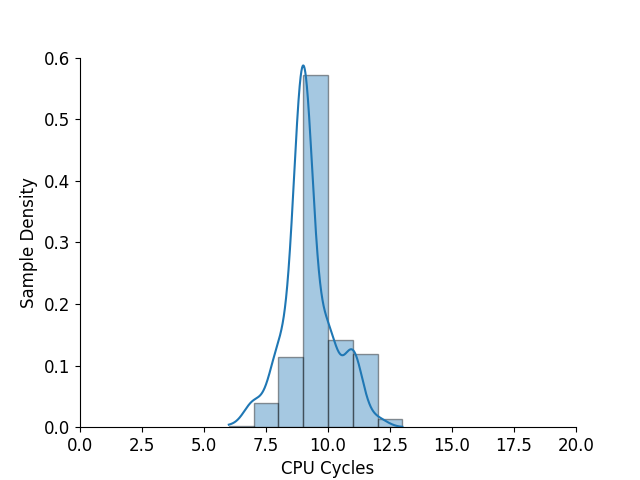}
    \caption*{\textbf{(b)} Branch (M=9, SD=1)}
    \end{minipage}\par
    \vskip\floatsep
    \begin{minipage}{0.49\textwidth}
    \includegraphics[width=\textwidth]{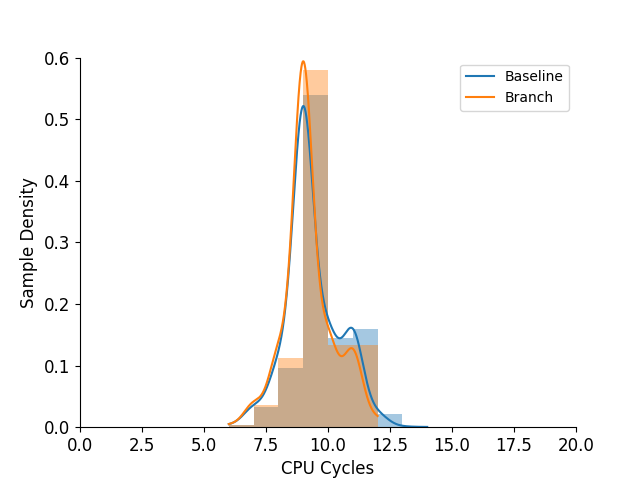}
    \caption*{\textbf{(c)} Comparison (P$>$0.5)}
    \end{minipage}
\caption{Benchmark results in CPU cycles for branch-changing overhead versus an equivalent 4-byte \texttt{memcpy} to non-executable memory.}
\end{figure}

\noindent \\Interestingly, writing to executable memory on its own does not incur any additional penalties with respect to the baseline, which is reflected by the near identical distributions in execution latencies. It is understood that modern processors tolerate self-modifying code (SMC) but are in no way friendly towards it, often initiating full pipeline and trace cache clears which can cause penalties in the hundreds of cycles \cite{intel-optimize}. The actual semantics of how processors detect SMC is unclear, however the general consensus in architecture forums and patents point towards a "snooping" mechanism which is initiated by store instructions into executable memory addresses. These snoops compare physical addresses of in-flight store instructions with entries in instruction cache-lines to see if the store location corresponds to instructions in executable memory. If there is an address match, the SMC clear is initiated and new instructions are fetched from memory to lower level instruction caches \cite{patent-smc}. Understanding this the results make sense; since there is no branch-taking occurring (where SMC occurs) there are no traces of the associated instructions in instruction cache lines, pre-fetch queues, or i-TLB and hence the physical address check fails to initiate SMC clears.

Following these observations, the next test involved benchmarking the semi-static conditions \texttt{set\_direction} method followed by branch-taking, with the baseline having \texttt{branch} replaced with a direct call to one of the functions passed to the constructor. The goal is to try and trigger SMC machine clears following previous discussions, and measure their associated penalties.\\

\begin{figure}[t]
  \centering
  \begin{minipage}[b]{0.49\textwidth}
    \includegraphics[width=\textwidth]{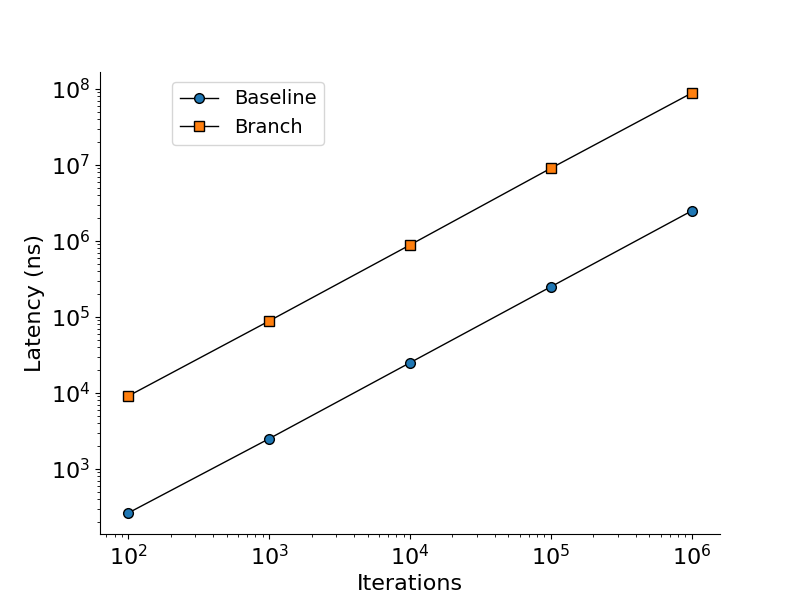}
    \caption*{\textbf{(a)} Execution latency}
  \end{minipage}
  \hfill
  \begin{minipage}[b]{0.49\textwidth}
    \includegraphics[width=\textwidth]{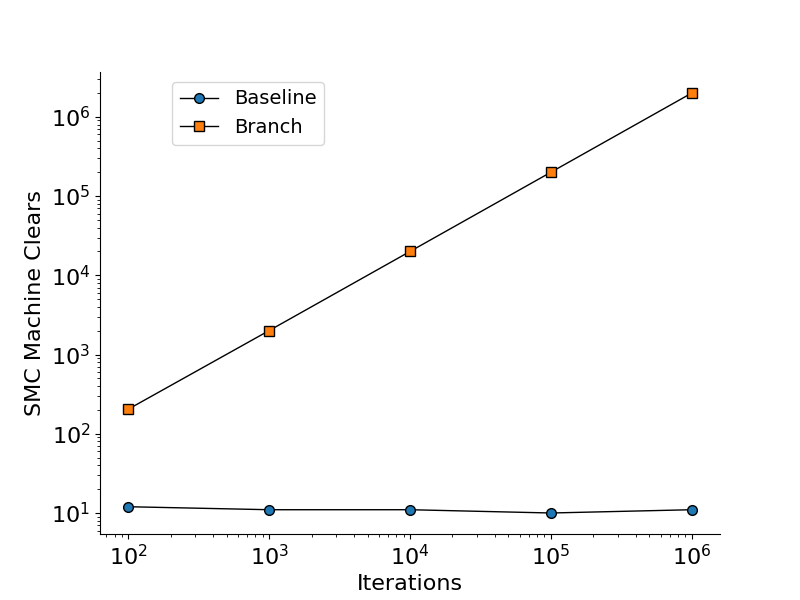}
    \caption*{\textbf{(b)} SMC machine clears count}
  \end{minipage}
  \caption{Latency and SMC machine clear count measurements for branch-changing followed by immediate branch-taking using semi-static conditions.}
\end{figure}

\begin{lstlisting}[language=c++]
__attribute__((optimize("O0")))
void if_branch() { return; }  

__attribute__((optimize("O0")))
void else_branch() { return; }  
(...)
for (int i = 0; i < iterations; i++)
{
    branch.set_direction(condition);    
    branch.branch()   
}
\end{lstlisting}

\noindent As expected, the presence of branch-taking close to the branch-chaining method triggers large numbers of SMC machine clears. The penalty of this is quite severe; extrapolating the execution latencies reveal that SMC machine clears multiply running times by 30-40x in this benchmark, with approximately 2 clears occurring per iteration on average. The additional overhead of SMC machine clears seems to be approximately 100 cycles which is consistent with approximations made by Agner and Intel, and are in agreement with various benchmarks made on architectural forums. Nevertheless the machine clear trigger seems to have deterministic behaviour; the number of SMC clears scale linearly with iterations, which is reflected in the overall execution latency.  It can be said for certain that executing modified assembly instructions relatively soon after editing initiate these clears, which outlaw the use of semi-static conditions inside tight loops even if branches are poorly predicted.

\begin{figure}[t]
  \centering
  \begin{minipage}[b]{0.49\textwidth}
    \includegraphics[width=\textwidth]{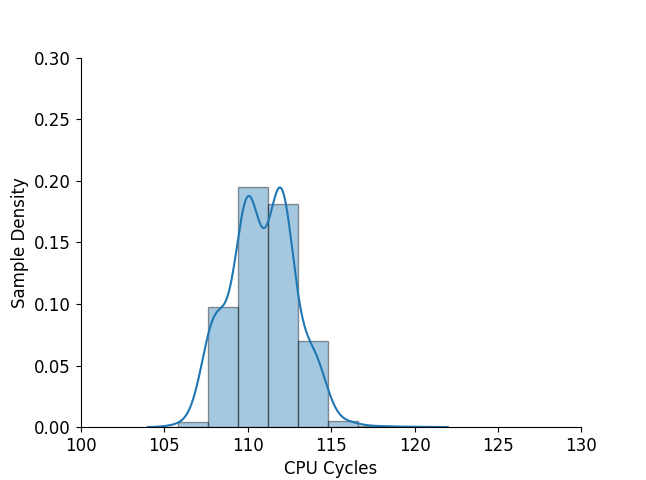}
    \caption*{\textbf{(a)} SMC penalty (M=111, SD=2)}
  \end{minipage}
  \hfill
  \begin{minipage}[b]{0.49\textwidth}
    \includegraphics[width=\textwidth]{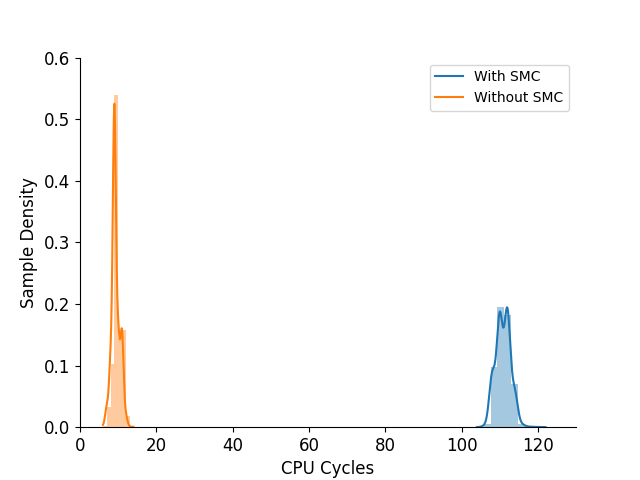}
    \caption*{\textbf{(b)} Comparison (P=0)}
  \end{minipage}
  \caption{CPU cycle measurements for branch-changing with SMC machine clear penalty.}
\end{figure}

Whilst the cost of such machine clears and their trigger are easy to reason about and are deterministic, the actual segments of assembly where these penalties manifest are not. This is problematic; the total cost of the branch-changing method are propagated to areas of code outside of itself in the form of SMC penalties, which introduces uncertainty in execution latencies for code that may be performance critical (such as \texttt{branch}!). An interesting observation is that there seems to a "lag" period from when the \texttt{set\_direction} method is executed to where the SMC cost starts to manifest, assuming that instructions are executed in the temporal order in which they appear in the disassembly, which is consistent with behaviour observed by \emph{Ragab et al}. The process of initiating the pipeline snoop to triggering the SMC clear takes several cycles since it involves instruction stream walks and i-TLB checks, resulting in a transient execution window of stale instructions caused by the de-synchronisation of the store buffer and instruction queue \cite{ragab2021rage}. Ideally, there should be some large enough buffer within the branch-changing method to contain this cost exclusively within \texttt{set\_direction} for more deterministic execution latencies. From a prevention standpoint, strong serialising instructions such as \texttt{CPUID} and \texttt{SERIALISE} where inserted into \texttt{set\_direction} to see if they where capable of preventing the SMC trigger, however this was not the case. In previous discussions we established that SMC triggers are caused by comparisons of in-flight or executed store instructions with physical addresses in instruction caches, so the ineffectiveness of serialising instructions is clear. Whilst \texttt{CPUID} and \texttt{SERIALISE} force the processor to complete all previous instructions and even drain the store buffer to prevent reordering, they do not have influence on instruction caches from which SMC checks are conducted, further supporting the presence of such "snooping" mechanism. 

It is possible to manually flush instruction cache lines pertaining to \texttt{branch} using the \texttt{\_mm\_clflush} intrinsic which shows some promise in minimising SMC clears when assembly modification is temporally closer to branch-taking, reducing such clears to approximately one per edit. Localility in this sence is very important; the closer assembly editing is to branch-taking, the more prominent the effect of SMC clears since the stale instructions have now polluted the pipeline, caches and instruction queues and thus require more machine clears to rectify. In the event that branch-changing occurs right before branch-taking, cache flushes do not seem to have a net positive effect which supports the notion that locality (in terms of instructions queued from the point to modification to the point where the modified code is executed) of SMC is the determining factor for the severity of associated penalties. Interestingly, Intel's optimization manuals do in fact recommend that SMC should not share the same 1-2Kib sub-page for speculative prefetching and execution reasons, further supporting this notion \cite{intel-optimize}. To test this hypothesis further, an artificial buffer was created which comprised of a cache flush followed by some computational work, acting as a temporal barrier between assembly modification and branch taking.\\

\begin{lstlisting}[language=c++]
__attribute__((optimize("O0")))
void _smc_buffer()
{
    _mm_clflush(address_of_branch);
    uint64_t buffer[DWORD * 4];
    for (int i = 0; i < DWORD * 4; i++)
        buffer[i]++;
}
\end{lstlisting}

\noindent \\Executing the following code directly after assembly modification is indeed sufficient in halving SMC clears, the same behaviour observed when adding cache flushes relatively close to branch-taking. The cache flush prevents additional machine clears due to physical address matches between store instructions and instruction cache data, whilst the computational work provides a sufficient buffer within the instruction pipeline to prevent similar matches with in-flight instructions. Usage of such a buffer will be optional to the programmer, but would also require some sort of active cache warming with \texttt{branch} to ensure that memory access penalties do not propagate to the hot-path. A simple optimisation to minimise clears would be to conditionally check if condition passed to \texttt{set\_direction} is the same to the current direction already set; in the current state, assembly modification is performed even when it isn't needed and always initiates machine clears. The associated overhead with this buffer incorporated within \texttt{set\_direction} prevents it from being used tight loops where branch taking also occurs; even if these penalties did not exist, the 9 cycle latency of copying bytes would likely see no performance benefit even if branches are often mispredicted. Whilst it is hard to quantify the amount of computational work required between assembly editing and branch-taking to minimise these penalties, if branch-changing occurs sufficiently far from the hot-path, SMC clears are likely to be minimal.

\noindent \\\textbf{Branch-taking Benchmarks} This set of tests is concerned with comparing the efficiency of the branch-taking method with conventional direct function calls, and exploring its implications on the BTB.

\noindent \\The first test investigated the overhead of \texttt{branch} versus a direct function call. The function call latency served as the baseline and was the same function that was executed through the \texttt{branch} entry point, so the branch direction was never changed.\\

\begin{lstlisting}[language=c++]
__attribute__((optimize("O0")))
void if_branch() { return; } 

__attribute__((optimize("O0")))
void else_branch() { return; }  
\end{lstlisting}

\noindent \\Measurements where taken in the following manner, repeated for 10$^7$ iterations.\\

\begin{lstlisting}[language=c++]
void measurement()
{
    start_measurement();
    branch.branch();
    stop_measurement();
}
\end{lstlisting}

\begin{figure}[t]
    \centering
    \begin{minipage}{0.49\textwidth}
    \includegraphics[width=\textwidth]{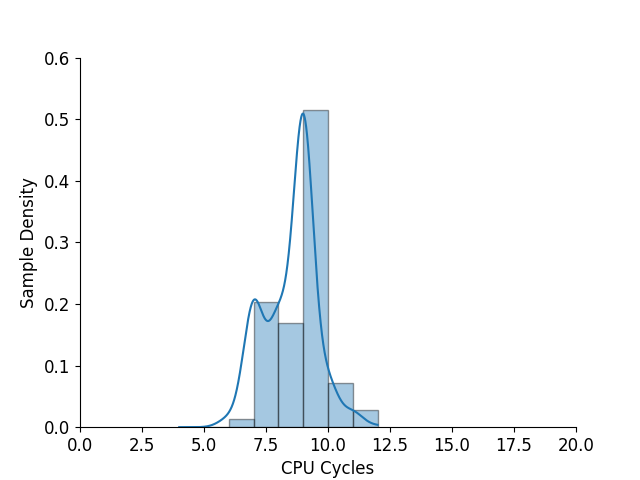}
    \caption*{\textbf{(a)} Baseline (M=9, SD=1)}
    \end{minipage}\hfill
    \begin{minipage}{0.49\textwidth}
    \includegraphics[width=\textwidth]{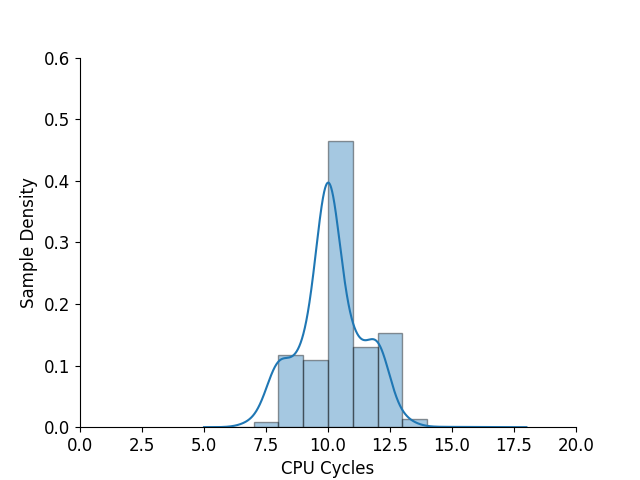}
    \caption*{\textbf{(b)} Branch (M=10, SD=1)}
    \end{minipage}\par
    \vskip\floatsep
    \begin{minipage}{0.49\textwidth}
    \includegraphics[width=\textwidth]{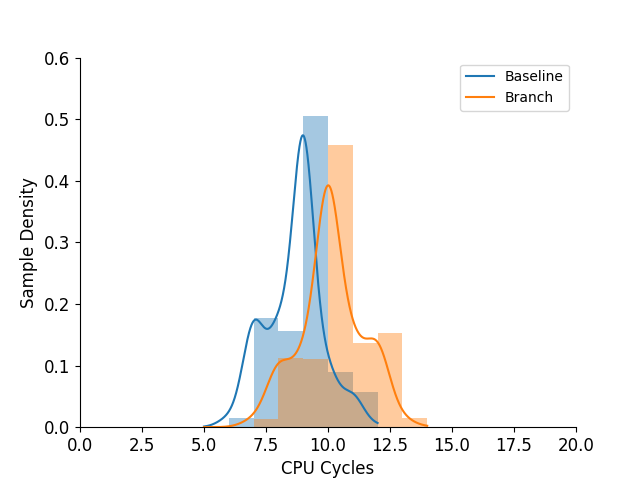}
    \caption*{\textbf{(c)} Comparison (P$<$0.000001)}
    \end{minipage}
\caption{Benchmark results in CPU cycles for branch-taking overhead versus a conventional direct function call.}
\end{figure}

\noindent \\Under the same conditions branch-taking has virtually identical overhead to a regular function call, with equal standard deviations in execution latency. The small difference in overhead may be attributed to the additional \texttt{jmp} instruction that resides within the prologue of \texttt{branch}, which is the only difference between the execution pathways of the baseline and semi-static conditions. An experiment to measure the latency of a single relative \texttt{jmp} on this specific architecture would validate the hypothesis, however this is tedious and unnecessary. Using Agner's instruction benchmarks for Intel and AMD CPU's (note that the CPU that the benchmarks where run on is part of the 10\textsuperscript{th} generation Comet Lake family, this was unavailable in Agner's instruction tables so reference latencies from Ice Lake where used), the typical latency of a relative jump is 1-2 cycles which falls within the range of the observed differences \cite{agner-instr}. Given the large number of iterations that the benchmark was run and that the branch direction was never changed, it is likely that the instructions that where measured where hot in lower-level caches with optimal usage of other hardware semantics, skewing the latency of the additional \texttt{jmp} towards the minimum value, and thus supporting the hypothesis. 
 
The implications of these results are multifaceted. The current implementation of branch-taking seems to be optimised to the theoretical limit; the additional overhead incurred seems to be caused exclusively by the additional control flow instruction that is inserted, which fulfills the overarching goal for its implementation. In low-latency environments such as HFT, low standard deviations are important for deterministic execution times and hence are a focus in development. The branch-taking method has shown that it has virtually identical standard deviations in execution latency as an isolated function call, offering programmers the assurance of determinism with respect to the current state-of-the-art.

In the development portion of this work, we discussed the difference in prediction schemes for conditional and unconditional branches. In theory, when the branch direction is changed, the BTB entry corresponding the jump instruction within the \texttt{branch} prologue becomes invalidated due to an incorrect branch target. Whilst the BTB will still accurately predict if the PC is indeed a jump, the stale branch target will likely be detected by the BAC which will initiate a pipeline flush and re-steer the prefetcher. To observe this behavior, we run the following testing suite with \textbf{perf stat -e baclears.any:u} which counts the number of BPU front-end re-steers initiated by user-space code:\\

\begin{lstlisting}[language=c++]
for (int i = 0; i < iterations; i++)
{
    branch.set_direction(condition);
    branch.branch();
    condition = !condition;
}
\end{lstlisting}

\begin{figure}[t]
  \centering
  \begin{minipage}[b]{0.49\textwidth}
    \includegraphics[width=\textwidth]{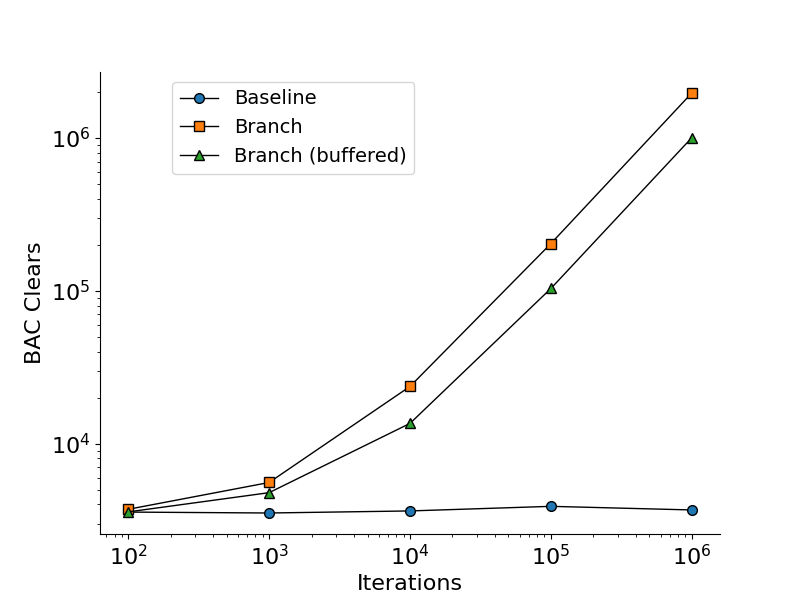}
    \caption*{\textbf{(a)} BAC clears count}
  \end{minipage}
  \hfill
  \begin{minipage}[b]{0.49\textwidth}
    \includegraphics[width=\textwidth]{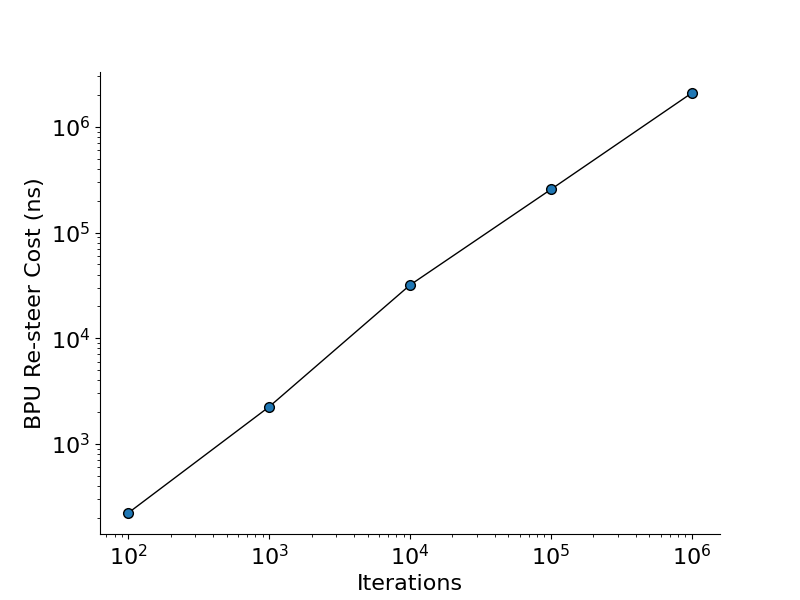}
    \caption*{\textbf{(b)} Overhead of BAC clears}
  \end{minipage}
  \caption{BAC clear counters for continuously changing branch targets (branch) versus static branch targets (baseline, \texttt{set\_direction} is always true), total overhead is calculated by subtracting the baseline latency from the benchmark. Branch (buffered) in (a) represents some computational work between \texttt{set\_direction} and \texttt{branch}.}
\end{figure}

\noindent \\Similar to SMC machine clears, BAC corrections appear to increase linearly with iterations when the branch direction is continuously changed in deterministic fashion. An interesting observation is that introducing a buffer of computation between branch-changing and branch-taking calls halves the number of corrections, averaging one clear per iteration. This suggests that Intel BTB's are updated atomically upon the retirement of branch macro-instructions; without the buffer the measurement loop is sufficiently small in terms of computation such that the modified \texttt{jmp} from the next iteration enters the pipeline before the current \texttt{jmp} gets retired. Since the BAC does not update BTB entries, but rather re-steers the prefetcher to the correct branch target, it initiates two re-steers per iteration due to stale BTB entries. The penalty of the correction also scales linearly, averaging an additional 2.2ns per iteration which equates to approximately \textbf{6 cycles} on this particular architecture, less than half the cost of a conditional branch misprediction (presumed around 13 cycles on Skylake CPU's) \cite{agner-instr}.

Though less severe, branch-taking using semi-static conditions does in-fact impose misprediction penalties, however it can be mitigated from the hot-path. The misprediction is essentially a one time cost when switching branch-directions, once the BTB has been corrected with the updated branch target, all successive calls \texttt{branch} will incur zero penalties and be executed with minimal latencies. This allows programmer to do effective 'warming' in the cold-path which is not possible with conditional statements; branch prediction is facilitated through capturing histories of branches based on their program counter and correlating them with global patterns. Adding conditional statements in the cold path in an attempt to 'warm' the BPU for the identical hot-path code will likely have little since the BPU treats both branches separately based on PC. Whilst it may capture some correlation, introducing more branches pollutes the global history can also introduce more noise. Conversely, calling \texttt{branch} in the cold path to warm the BTB works since control flow is redirected to the PC with the stale \texttt{jmp}, facilitating the correction preemptively and also brings the associated branch target (the functions being branched to) into lower level instruction caches. This forming of warming can also be used to isolate the SMC clear penalty discussed in the \texttt{set\_direction} benchmarks since the modified code is being executed temporally close to assembly editing. Using HFT as an example, this can be realised by sending 'dummy orders' through the \texttt{branch} method after the branch-direction has been set to ensure warming occurs.\\

\begin{lstlisting}[language=c++]
void cold_path()
{
    (...)
    do_some_work();
    branch.set_direction(condition);
    branch.branch(dummy_order);
    do_some_more_work();
    (...)
}
\end{lstlisting}

\noindent \\\textbf{Interim conclusions} Results obtained from the preceding test suite provide valuable insight into the implications of using semi-static conditions on the hardware level, which in turn help deduce optimal usage. The majority of the cost with using this language construct is manifested within the \texttt{branch\_changer} method in terms of machine clears caused by self modifying code, whilst branch-taking (the latency critical part) has extremely low overhead with deterministic execution latencies. In the preliminary sections of this work, it was hypothesised that optimal usage would involve separating branch-changing and branch-taking into cold and hot paths respectively, providing a means for pre-empative condition evaluation and branchless execution. The benchmarking results validate this use case, showing immense promise for branch optimisation in code paths that are infrequently executed, but contain branches that are poorly predicted. In addition, the tests brought novel investigations into modern Intel processor semantics, exploring behaviour relating to SMC and branch-prediction pertaining to unconditional branches, which in turn can help low-latency developers optimise their code paths respectively.

\subsection{Applications}
This section builds off the previous findings, exploring applications where semi-static conditions outperforms the current state of the art: conditional statements with branch prediction. Tests are primarily focused on HFT applications, which can be applied conceptually to more general use cases.\\

\noindent \textbf{Hot-path optimisation in HFT} In contrary to traditional engineering terminology, the "hot-path" in the context of HFT refers to a section of code which is executed relatively infrequently, but when it is executed it needs to be extremely fast. This is typically the time taken from receiving market data to sending an order, with the whole process taking just several microseconds to execute. Lying at the heart of the trading system, and probably the most critical piece of code in terms profitability, this area is the target for most optimisation. Since it executed relatively infrequently, the remainder of the trading system is often designed with cache warming measures that keep data used in the critical path hot in low level caches to avoid memory access penalties. However this poses a challenge for branch prediction; branches in the hot path may have limited histories with complex patterns, resulting in mispredictions. Here, semi-static conditions are applied to optimise branch-taking for this use case.

The test suite comprised of a tight measurement loop using RDTSC counters where runtime conditions where randomly generated using the Mersenne Twister Engine. For an infinite series of randomly generated booleans, it is expected that roughly half will be true and false, however since measurements are non-infinite there will likely be skews towards one boolean in benchmarks. To ensure testing is fair, both branches need to have the same execution latencies. The choice of branches to benchmark initially where 64 byte memory copies and bit flips to a volatile \texttt{struct}, the scenario representing message passing to custom firmware (such as network cards and FPGA's) in a HFT system.\\

\begin{lstlisting}[language=c++]
__attribute__((noinline))
void send_order(unsigned char* message)
{
    std::copy(message, message + 64, FPGA.payload);
    FPGA.flag = !FPGA.flag;
}
\end{lstlisting}

\noindent \\Emulating the hot-path in terms of infrequent execution is especially challenging from a benchmarking point of view. Percentage cycles spent in the measurement zone (the hot-path) where used as a proxy, and this was minimised by adding com-

\begin{figure}[h]
    \centering
    \begin{minipage}{0.49\textwidth}
    \includegraphics[width=\textwidth]{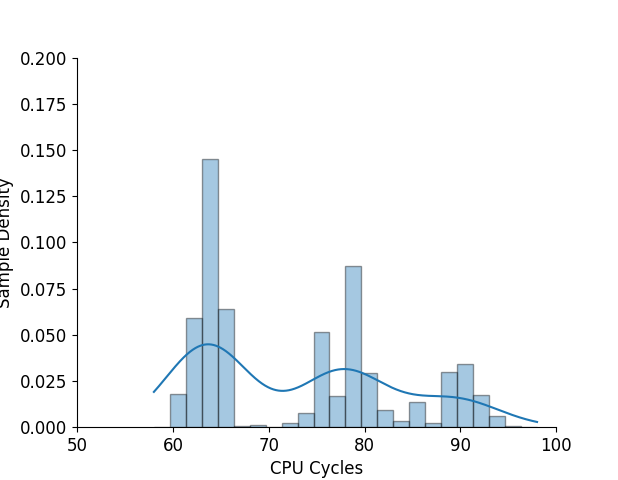}
    \caption*{\textbf{(a)} Conditional statements without cache warming (M=75, SD=10)}
    \end{minipage}\hfill
    \begin{minipage}{0.49\textwidth}
    \includegraphics[width=\textwidth]{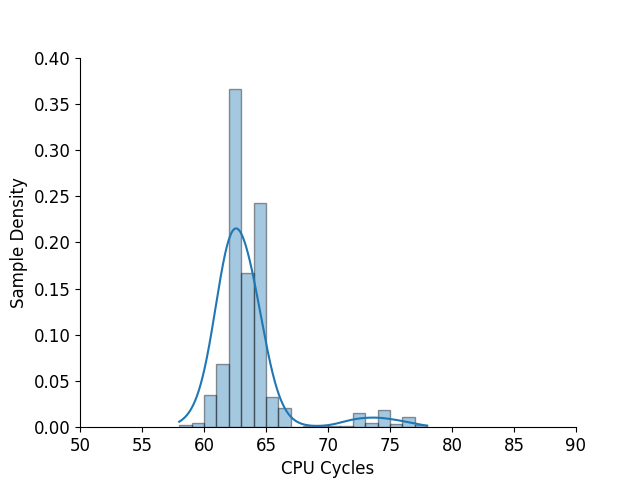}
    \caption*{\textbf{(b)} Semi-static conditions without cache warming (M=63, SD=3)}
    \end{minipage}\par
    \vskip\floatsep
    \begin{minipage}{0.49\textwidth}
    \includegraphics[width=\textwidth]{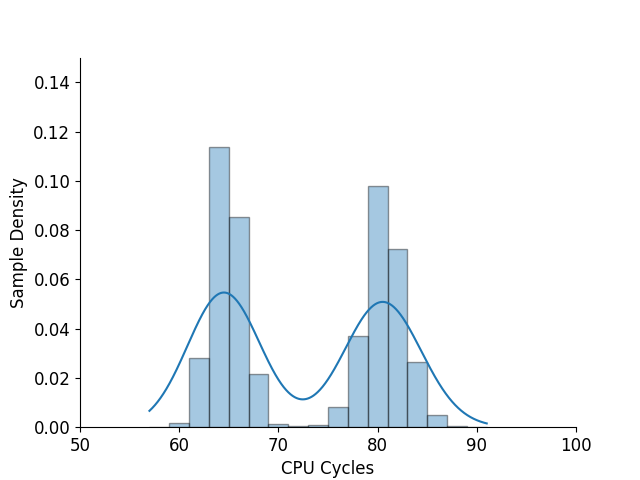}
    \caption*{\textbf{(c)} Conditional statements with cache warming (M=68, SD=8) }
    \end{minipage}\hfill
    \begin{minipage}{0.49\textwidth}
    \includegraphics[width=\textwidth]{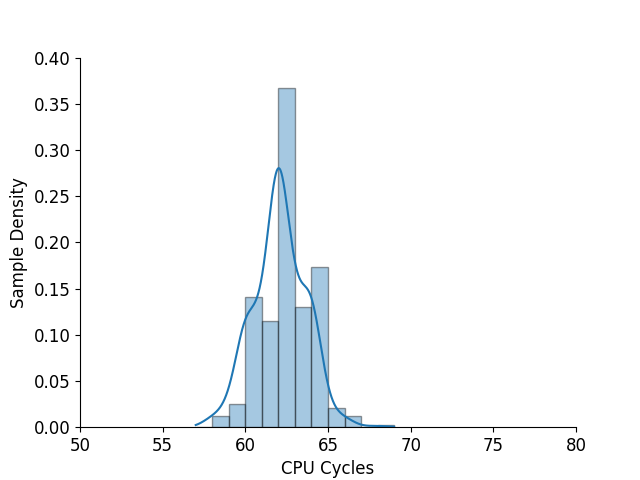}
    \caption*{\textbf{(d)} Semi-static conditions with cache warming (M=62, SD=2)}
    \end{minipage}\par
    \vskip\floatsep
    \begin{minipage}{0.49\textwidth}
    \includegraphics[width=\textwidth]{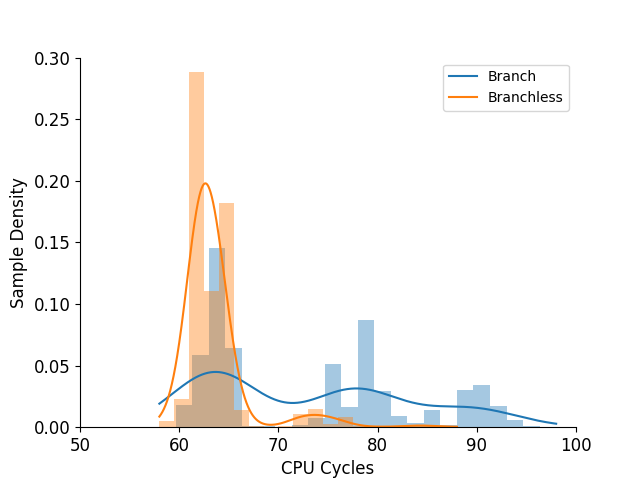}
    \caption*{\textbf{(e)} Comparison without cache warming }
    \end{minipage}\hfill
    \begin{minipage}{0.49\textwidth}
    \includegraphics[width=\textwidth]{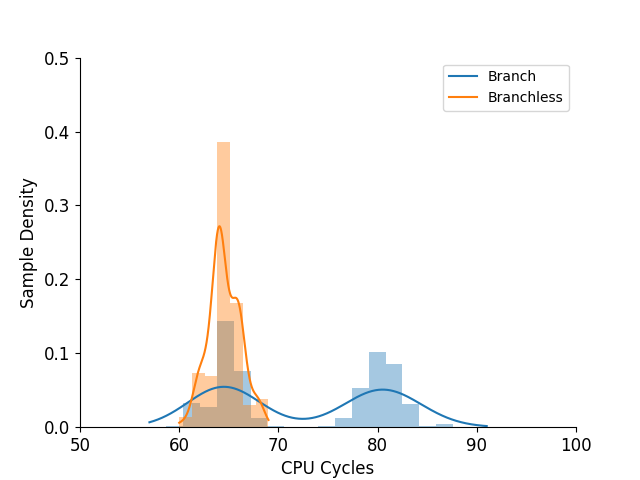}
    \caption*{\textbf{(f)} Comparison with cache warming}
    \end{minipage}\par

\caption{CPU cycle measurements of conditional branching (branch) versus semi-static conditions (branchless) with and without cache warming.}
\end{figure}

\noindent putational work in the form of randomly generating messages and running pricing calculations outside the hot-path. The randomly generated messages where passed to the branches in attempt to emulate real-time message passing, and measurements where taken with and without cache warming.

The results shown are quite remarkable. Across the board semi-static conditions produce a tight uni-modal distribution with median latencies and standard deviations much lower than conditional branching. In terms of conditional branches, the sample densities elegantly model the latencies of predicted and mispredicted branches as a bimodal distribution which contribute to the larger median and standard deviation in cycles. In tests without cache warming, both conditional branching and semi-static conditions have a small distribution of measurements in the 70-90 cycle range presumably due to cache effects caused by message passing, which seems to disappear completely when cache warming is employed. Looking closer at the bimodal nature of conditional statement execution latencies (M=65 SD=2, M=78 SD=2 without cache warming and M=64 SD=2, M=80 SD=2 with cache warming respectively), the difference in median values between both distributions is 13-16 cycles which are in good agreement with estimations made by Agner for i7 processors which seem to have a penalty of 13-18 cycles \cite{agner-optimise}. Through using semi-static conditions, programmers gain the benefit branch-taking latencies comparable (in this case even slightly better!) with perfectly predicted branches, saving 2-4ns on average and up to 6ns if the branch is always mispredicted, with more deterministic execution latencies manifesting with low standard deviations.

Subsequent tests with other branches reveal the same behaviour for both conditional statements and semi-static conditions; whilst the distributions where centred around different medians due to the varying execution of different branches, the relative offsets between the distributions remained the same. Interestingly usage of the [[likely]] and [[unlikely]] branch prediction hints had no effect in mitigating misprediction rate for conditional branching. The reason for this is clear, compiler hints simply reorganise the assembly code to aid the processors static predictor in taking the more likely execution path, but since conditions are random and neither path is more likely being taken it has no net positive effect. That being said, it does seem that semi-static conditions are better than branch prediction hints for this use case. Further tests added more computational work to the hot path to observe this behaviour when branching is surrounded by more comlex logic, an example of the measuring suite for conditional branches is seen below.\\

\begin{lstlisting}[language=c++]
void measure_hot_path()
{
    begin_measurement();
    (...)
    do_some_calculations();
    if (condition)
        send_order(message);
    else
        adjust_order(message);
    do_some_work();
    (...)
    end_measurement();
}
\end{lstlisting}

\begin{figure}[t]
    \centering
    \begin{minipage}{0.49\textwidth}
    \includegraphics[width=\textwidth]{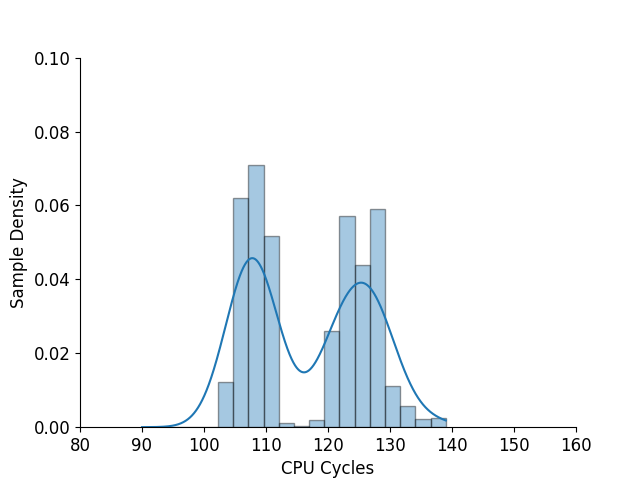}
    \caption*{\textbf{(a)} Conditional statements (M=120, SD=10)}
    \end{minipage}\hfill
    \begin{minipage}{0.49\textwidth}
    \includegraphics[width=\textwidth]{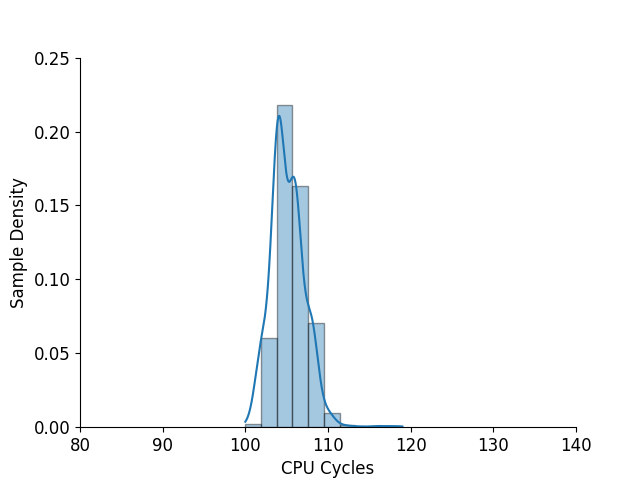}
    \caption*{\textbf{(b)} Semi-static conditions (M=104, SD=3)}
    \end{minipage}\par
    \vskip\floatsep
    \begin{minipage}{0.49\textwidth}
    \includegraphics[width=\textwidth]{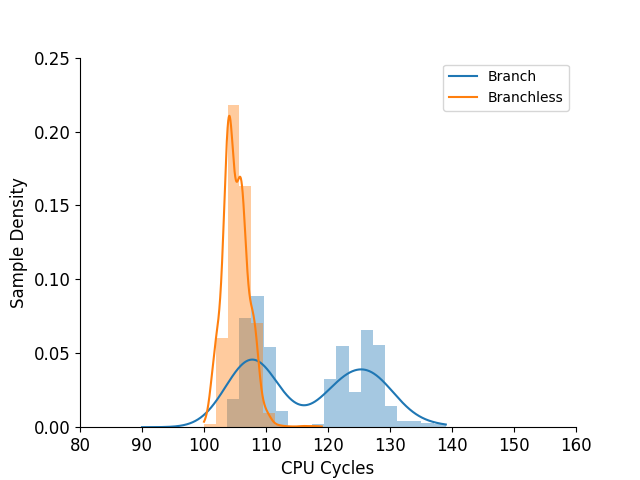}
    \caption*{\textbf{(c)} Comparison}
    \end{minipage}
\caption{CPU cycle measurements for conditional branching versus semi-static conditions with updated hot-path.}
\end{figure}

Cycle distributions seem to follow an identical pattern to the previous results regardless of the additional of extra computational work. It seems that misprediction cost for this case increased slightly for conditional branches to 18 cycles (M=108 SD=2 for predicted and M=126 SD=4 for mispredicted) which suggested that branch mispredictions can impact surrounding code resulting in higher penalties.

In the case of n-ary conditional statements in the form of if-else chains or switch statements, unpredictable conditions yield similar cycle distributions to the preceding examples. Using the current methodology where random conditions are generated in the range of $0$ to $n-1$ where $n$ is the number of branches, as $n\rightarrow \infty$ the misprediction rate tends to 1 and hence distributions become uni-modal and skewed to higher cycle numbers. This is entirely expected: if conditions are random and constantly changing, then the probability of predicting the correct branch is inversely proportional to the number of branches, which tends to zero as the number of branches tends to infinity. Typically on GCC, large if-else chains or switch statements are optimised into jump tables, which organise branches at distinct memory locations in a particular structure (binary tree for example), and utilise indirect jumps to computed offsets to facilitate branch taking. This additional indirection does incur more significant penalties when mispredicted owing to the additional data dependencies in the pipeline; extrapolating the median latencies from the predicted (M=13, SD=2) and mispredicted (M=31, SD=3) branches, the misprediction cost is estimated to be 18 cycles in this example! 

\begin{figure}[t]
    \centering
    \begin{minipage}{0.49\textwidth}
    \includegraphics[width=\textwidth]{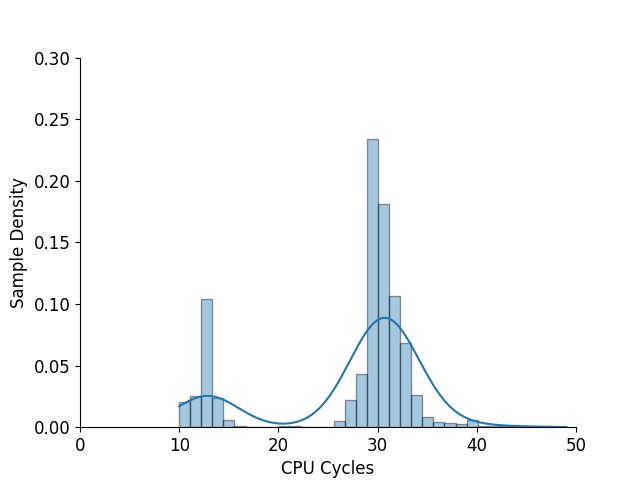}
    \caption*{\textbf{(a)} Switch statements (M=30, SD=8)}
    \end{minipage}\hfill
    \begin{minipage}{0.49\textwidth}
    \includegraphics[width=\textwidth]{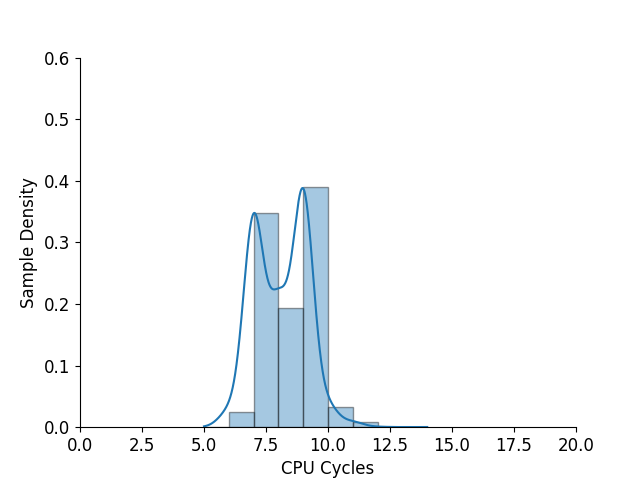}
    \caption*{\textbf{(b)} Semi-static conditions (M=8, SD=1)}
    \end{minipage}\par
    \vskip\floatsep
    \begin{minipage}{0.49\textwidth}
    \includegraphics[width=\textwidth]{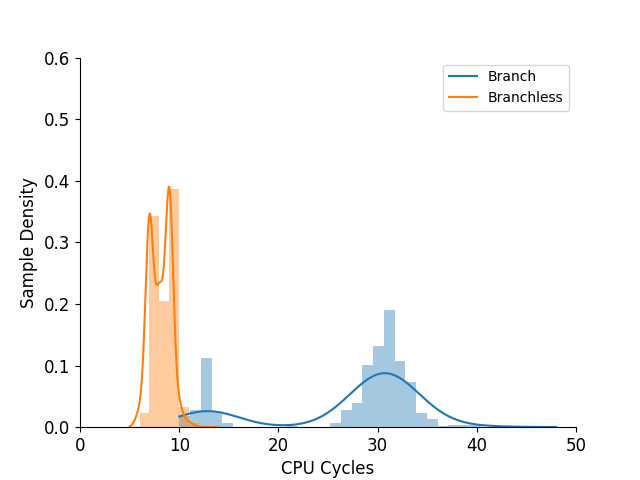}
    \caption*{\textbf{(c)} Comparison }
    \end{minipage}
\caption{CPU cycle measurements for a 5 case switch statement versus semi-static conditions with unpredictable conditions in the hot-path. Empty functions used as branches for ease of testing.}
\end{figure}

It is clear that for this particular use case, semi-static conditions offer superior performance to conditional branching when misprediction rates are high and branch-changing can be isolated in cold code paths. In the context of HFT, speed is paramount, and shaving off several nanoseconds per branch in the critical path offers an edge in order execution latency against competitors and can result in more profitable trading.\\

\begin{figure}[t]
    \centering
    \begin{minipage}{0.49\textwidth}
    \includegraphics[width=\textwidth]{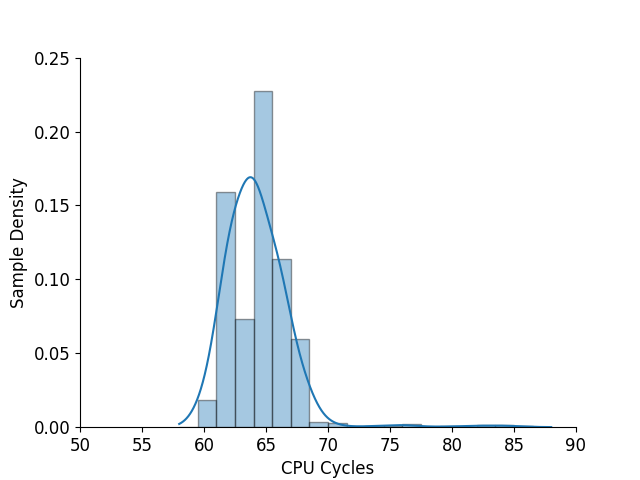}
    \caption*{\textbf{(a)} Conditional statements (M=64, SD=3)}
    \end{minipage}\hfill
    \begin{minipage}{0.49\textwidth}
    \includegraphics[width=\textwidth]{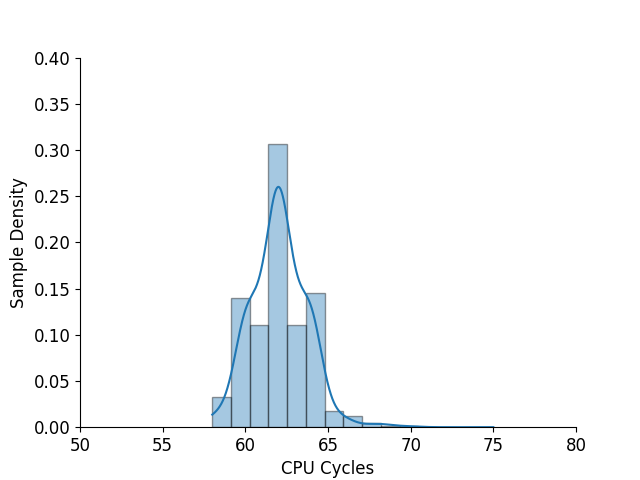}
    \caption*{\textbf{(b)} Semi-static conditions (M=62, SD=2)}
    \end{minipage}\par
    \vskip\floatsep
    \begin{minipage}{0.49\textwidth}
    \includegraphics[width=\textwidth]{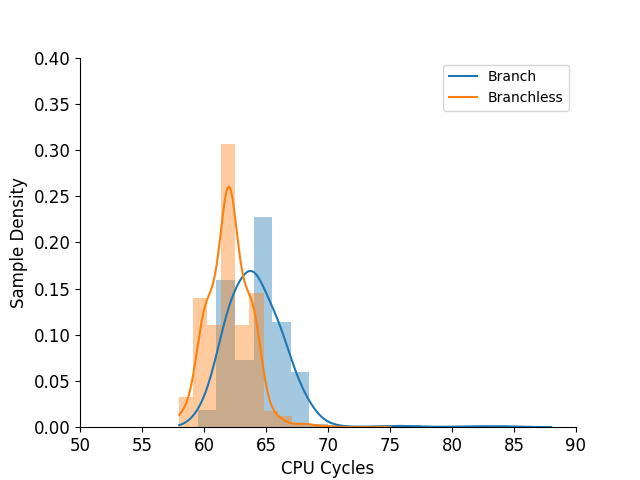}
    \caption*{\textbf{(c)} Comparison (P$<$0.000001)}
    \end{minipage}
\caption{CPU cycle measurements for conditional branching versus semi-static conditions with predictable branching, conditions change every 1000 iterations.}
\end{figure}

\noindent \textbf{General use cases} Whilst hot-path optimisation of mispredicted branches is not necessarily tied to HFT, but can be expanded to general performance critical sections of code in various applications (gaming, aerospace, infrastructure ect), an interesting investigation would be for general use cases where branches are not necessarily unpredictable (in terms of conditions). Earlier investigations revealed that branch-taking has comparable latency with isolated function calls, and in comparison to conditional statements, have marginally less assembly to execute to facilitate branch taking. The following investigations outline instances where these subtle differences manifest in increased performance, using the same methodology of separating branch-changing from hot-path measurements. To simulate predictable branches, the same test suite used for specialised applications can be adapted to changing the boolean conditions based on a regular interval, rather than randomly generating them.

\begin{figure}[t]
    \centering
    \begin{minipage}{0.8\textwidth}
    \includegraphics[width=\textwidth]{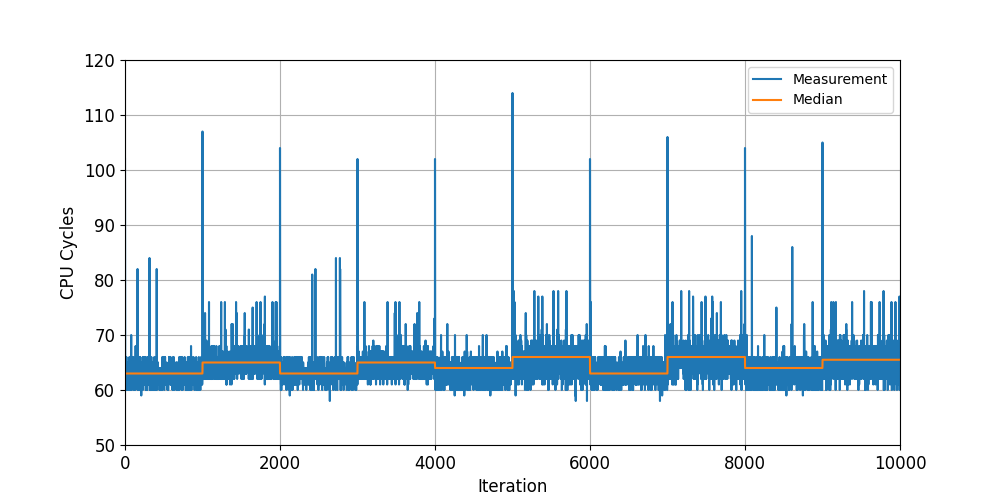}
    \caption*{\textbf{(a)} Conditional statements}
    \end{minipage}\par
    \vskip\floatsep
    \begin{minipage}{0.8\textwidth}
    \includegraphics[width=\textwidth]{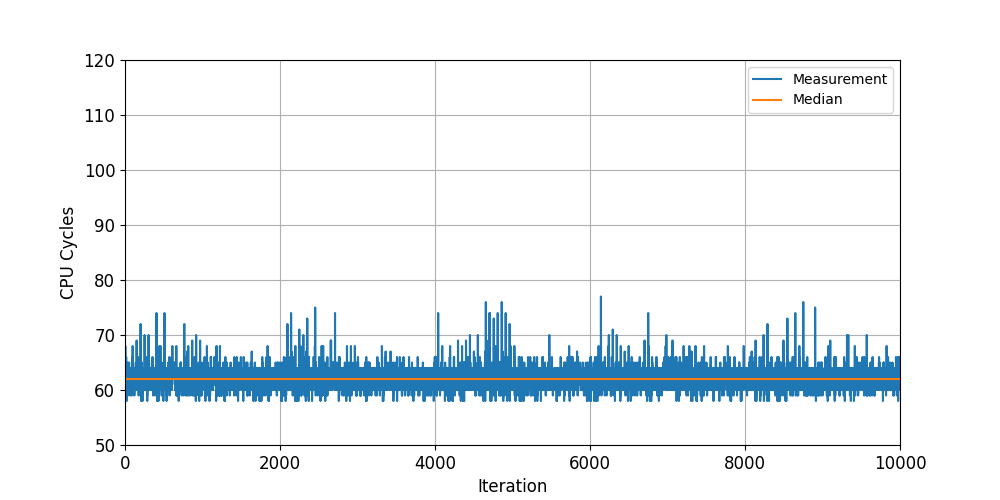}
    \caption*{\textbf{(b)} Semi-static conditions}
    \end{minipage}
\caption{CPU cycle measurements per iteration for conditional branching and semi-static conditions.}
\end{figure}

Even when branches are predictable, semi-static conditions are slightly more performant in terms of branch-taking. In contrast with earlier tests, the latency of conditional statements for, a uni-modal distribution since the misprediction rate is close to zero, nevertheless the extension of the trend-line to the 80-85ns region shows there are mispredictions, but are rare. What is fascinating is that the core distribution of measurements for conditional branching is shifted to higher latencies in comparison to semi-static conditions, when execution latencies for predicted branches should be the same. The key questions that arise are what is the cause of this shift, and how significant is the contribution of mispredictions to the overall median latency at different switching intervals. To answer the former, a subset of measurements for both semi-static and conditional branches can be extracted and plotted to visualise the execution latencies per iteration.

The results reveal some remarkable behaviour. Firstly, the misprediction penalty can be seen distinctly by sharp increases in latency at regular 1000 iteration intervals, this is indeed directly caused by changing the branch direction and seems to be corrected relatively quickly, which is indicative of an n-bit prediction scheme. Every time the branch direction changes, the median shifts by 2 or 3 cycles forming the observed the observed saw-tooth pattern. Upon inspection of the underlying assembly, this behaviour is an artefact of the compiler reordering the conditional statement's assembly such that the forward branch (if) experiences slightly lower execution latencies than the backward branch (else) due to additional jumps around the code segment.\\

\begin{lstlisting}
000000000000305c:     mov    rdi,QWORD PTR [rip+0x42e5]     
0000000000003063:     shl    rdx,0x20
0000000000003067:     mov    rsi,rax
000000000000306a:     or     rsi,rdx
000000000000306d:     cmp    BYTE PTR [rip+0x42dc],0x0
\end{lstlisting}
\vspace*{-4mm}
\begin{lstlisting}[basicstyle=\bfseries]
0000000000003074:     je     3090 <_measure+0x40>
0000000000003076:     call   2c00 <send_order>
\end{lstlisting}
\vspace*{-4mm}
\begin{lstlisting}
(...)
\end{lstlisting}
\vspace*{-4mm}
\begin{lstlisting}[basicstyle=\bfseries]
0000000000003090:     call   2c40 <_adjust_pricing>
0000000000003095:     jmp    307b <_measure+0x2b>
\end{lstlisting}

\noindent \\These kind of assembly ordering semantics have even more prevalent effects for switch statements (5-6 cycles faster!) which can be seen by comparing the distributions of semi-static conditions and predicted branches for switch statements. This is mainly due to the additional assembly required perform the necessary computations when preparing to traverse jump tables:\\

\begin{lstlisting}
00000000000041fd:      shl    rdx,0x20
0000000000004201:      mov    rbx,rax
0000000000004204:      or     rbx,rdx
0000000000004207:      cmp    QWORD PTR [rip+0x5161],0x4
\end{lstlisting}
\vspace*{-4mm}
\begin{lstlisting}[basicstyle=\bfseries]
000000000000420f:      ja     4235 <_measure+0x45>
\end{lstlisting}
\vspace*{-4mm}
\begin{lstlisting}
0000000000004211:      mov    rax,QWORD PTR [rip+0x5158]
0000000000004218:      lea    rdx,[rip+0x1e05]
000000000000421f:      mov    rdi,QWORD PTR [rip+0x5142]
0000000000004226:      movsxd rax,DWORD PTR [rdx+rax*4]
000000000000422a:      add    rax,rdx
\end{lstlisting}
\vspace*{-4mm}
\begin{lstlisting}[basicstyle=\bfseries]
000000000000422d:      jmp    *rax
0000000000004230:      call   3ed0 <_send_order_1>
\end{lstlisting}
\vspace*{-4mm}
\begin{lstlisting}
(...)
0000000000004249:      nop    DWORD PTR [rax+0x0]
\end{lstlisting}
\vspace*{-4mm}
\begin{lstlisting}[basicstyle=\bfseries]
0000000000004250:      call   3eb0 <_send_order_n>
0000000000004255:      jmp    4235 <_measurev+0x45>
\end{lstlisting}
\vspace*{-4mm}
\begin{lstlisting}
0000000000004257:      nop    WORD PTR [rax+rax*1+0x0]
(...)
\end{lstlisting}

\noindent \\When comparing this to the underlying assembly generated for semi-static conditions, it becomes very apparent why branch-taking for the latter exhibits lower and more deterministic execution latencies.\\

\begin{lstlisting}
000000000000309d:     mov    rdi,QWORD PTR [rip+0x42a4]    
00000000000030a4:     mov    rbx,rax
00000000000030a7:     shl    rdx,0x20
00000000000030ab:     or     rbx,rdx
\end{lstlisting}
\vspace*{-4mm}
\begin{lstlisting}[basicstyle=\bfseries]
00000000000030ae:     call   2b00 <_branch_>
\end{lstlisting}

\noindent \\In the context of general usage for 'static branches' (conditions that change infrequently), the results have several implications. The power of semi-static conditions has always been the ability to change the direction of a branch programatically. At compile time, the programmer has little influence over such re-orderings, and although an informed prediction can be made about the more likely direction of the branch (forward or backward), if this where to change at runtime, the programmer will be paying at least 2-3 cycles per iteration which will start to add up. This allows for the \texttt{set\_direction} method to be used more freely; rather than isolating it in the cold path, if the branch direction is changed relatively infrequently then the cost of code modification will amortise itself over many iterations of cheap branch taking.

\begin{figure}[t]
    \centering
    \begin{minipage}{0.49\textwidth}
    \includegraphics[width=\textwidth]{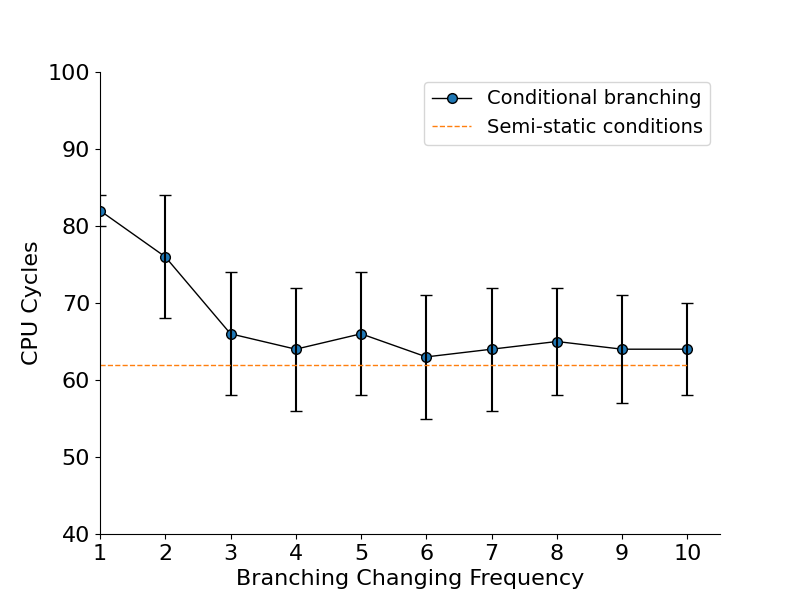}
    \caption*{\textbf{(a)} Latency}
    \end{minipage}\hfill
    \begin{minipage}{0.49\textwidth}
    \includegraphics[width=\textwidth]{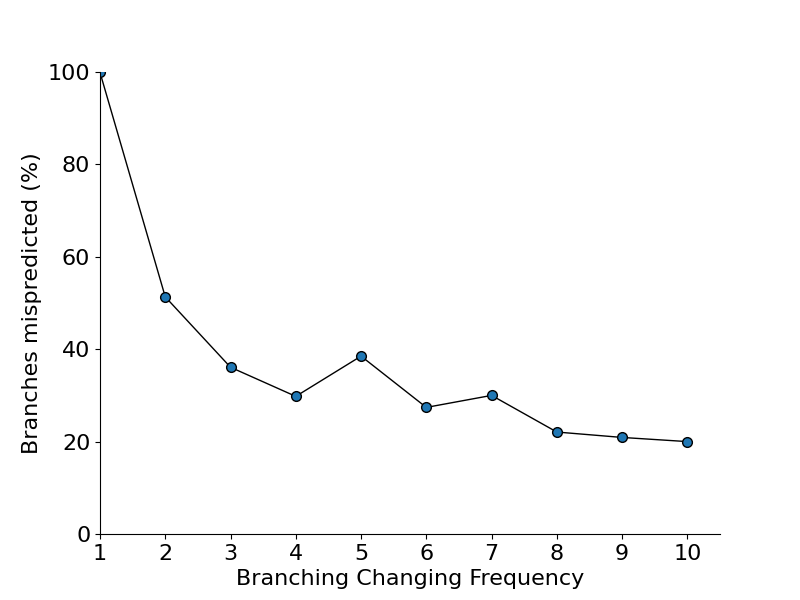}
    \caption*{\textbf{(b)} Misprediction rates}
    \end{minipage}
\caption{CPU cycle measurements and respective misprediction rates for conditional branching at varying branch changing frequencies (number of iterations per condition change).}
\end{figure}

In regards to contributions from branch mispredictions, the median latency and standard deviations for conditional branching where measured for varying branch changing frequencies (in terms of number of iterations passed before the condition is changed), along with the associated misprediction rates. These results are quite unexpected; even for the branch changing every every iteration one would expect the BPU to spot this relatively simple pattern of taken not-taken, however this is not the case. Even when the conditions change at regular intervals in a predictable manner, the organisation of where these instructions lie within the executable effect the BP's ability to make predictions based on history correlations. In these benchmarks, the actual 'path' that is being measured resides in a function to help maintain assembly ordering for fair and precise measurements, which is called within a tight loop which performs a significant amount of computation per iteration. This can add noise, however the more likely reason is that the BP is unable to correlate the iteration count with the condition being evaluated (at regular intervals). The actual predictive mechanism can be deduced through the results; when branches are changed every iteration, misprediction rates are close to 100\% which results in significantly higher latencies but tighter standard distributions since all branches are mispredicted. As branch changing frequency decreases, the misprediction rate follows as well as the median latency, but standard deviations increase increase since measurement distributions contain a mix of predicted and mispredicted branches. From the data, it appears that there are 1.5-2 mispredictions per condition change, which is synonymous with a 2-bit saturated counter prediction scheme (hinting towards TAGE-like predictors used in modern Intel processors)! When misprediction rates are high, the associated penalties are the major contributor to increased latencies and standard deviations. When branches are better predicted and conditions change relatively infrequently, differences in latency are attributed to the underlying assembly generated by the compiler, resulting in more subtle performance differences.

So far general application benchmarks have been conducted on the cycle level, an interesting investigation is to see how they manifest in larger systems. The next test involves a multi-threaded benchmark where branch-directions are changed at regular time intervals, the branches perform a relatively simple computation and store the result in an array to prevent optimisation. Whilst the example is simplistic, it represents a system that relies polling events on a worker thread which effect conditions that are evaluated in continuous loops, an example of such application is feature-flag selection for larger code-bases.\\

\begin{lstlisting}[language=c++]
static void benchmark(benchmark::State& s)
{
    int results[2];
    std::thread worker(poll_events);
    for (auto _ : s)
    {
        for (int i = 0; i < iterations; i++)
        {
            if (condition)
                results[flag] += action_1();
            else
                results[flag] += action_2();
        }
    }
    worker.join();
}
\end{lstlisting}

\begin{figure}[t]
    \centering
    \begin{minipage}{0.49\textwidth}
    \includegraphics[width=\textwidth]{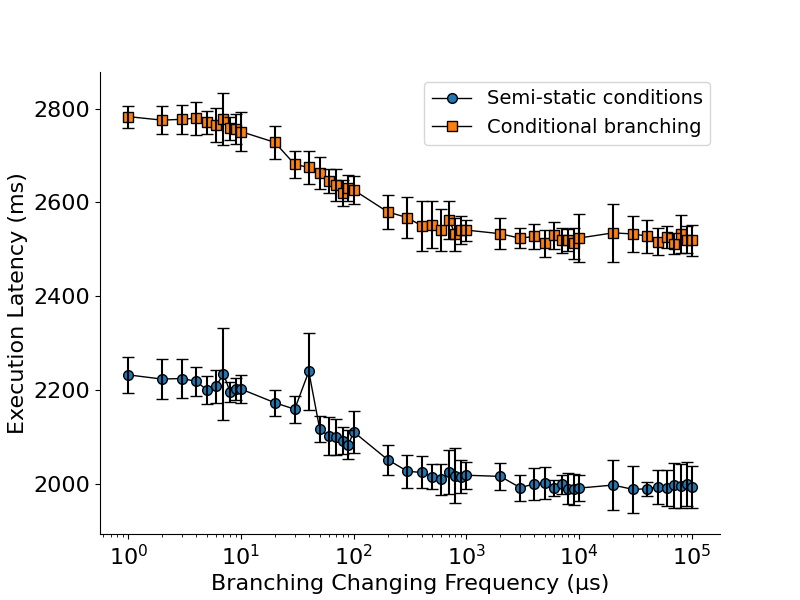}
    \caption*{\textbf{(a)} Without \texttt{std::mutex}}
    \end{minipage}\hfill
    \begin{minipage}{0.49\textwidth}
    \includegraphics[width=\textwidth]{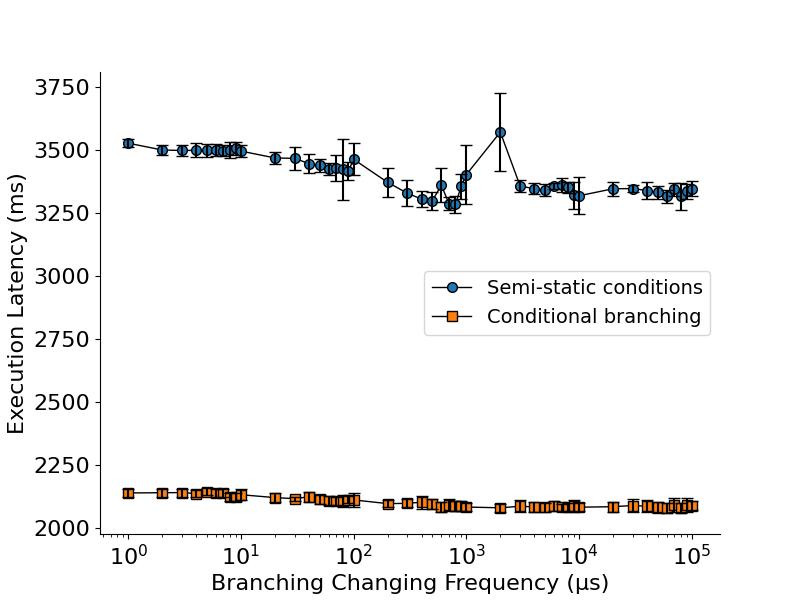}
    \caption*{\textbf{(b)} With \texttt{std::mutex}}
    \end{minipage}
\caption{Benchmark results for semi-static conditions versus conditional branching in multi-threaded application where branches change at regular time intervals. Mutex's have been applied to semi-static conditions exclusively.}
\end{figure}

\noindent \\Even though the observed changes are relatively small, occasionally they can manifest themselves as large performance gains, however this will be very dependant on the system in which semi-static conditions is integrated in. When using the language construct in a multi-threaded environments, there is a chance that the wrong branch is executed since assembly editing is not thread safe. Using synchronisation will prevent this, however it results in large performance degradation. Regardless, the proposed language construct shows immense promise in performance optimisation in general settings; any places where misprediction rates cause performance bottlenecks or branch-taking is slow due to surrounding code, semi-static conditions offer a convenient alternative to more efficient branch-taking. 

\subsection{Summary of Conclusions}
The foregoing investigations have not only provided a transparent performance analysis of the various methods that comprise semi-static conditions, but have been successful identifying numerous use cases where mispredicted branches can be optimised heavily with the use of the language construct.

By employing a variety of sophisticated testing suites originating from architectural forums and prior literature, we effectively delved into the underlying factors driving the performance of branch-changing and branch-taking methods. Concerning branch-changing, the minimal isolated cost of assembly editing was evident; however, an important discovery was the substantial impact of the code's locality that undergoes modification on the hardware penalties induced by SMC detection. The mechanics behind this phenomenon are well comprehended, along with the associated penalties and strategies for alleviation. This comprehension led to the realization that, for branch-changing, it is optimal to confine such modifications to less critical, cold code paths. Leveraging preemptive conditional evaluation, the control over branch directions can be guided away from performance-sensitive code sections. As for branch-taking, which invariably exists within the critical path, performance was optimized to the theoretical limit. The only disparities observed where in additional instruction latencies when compared to conventional calls. While there is a slight initial penalty for branch-taking after assembly modification, it was discerned that this setback could be mitigated through BTB warming, a possibility not readily achievable in the same manner with conditional branch prediction.

Using these newfound insights to compare against the current state-of-the-art practices, it becomes evident that under circumstances involving occasional branch execution combined with frequent mispredictions, the approach of semi-static conditions emerges as the preferred choice. The strategy of employing semi-static conditions demonstrates advantages in terms of both execution speed and consistency across all such scenarios, yielding the possibility of notable performance improvements in real-world application due to the influence of neighboring logic. In a broader context, the proposed methodology once again displays potential by offering slightly improved performance even when branches are largely predictable. However, more substantial discrepancies in performance become apparent in more extensive conditional constructs, which could translate to significant gains in larger-scale systems.

\section{Evaluation}

This section will evaluate the software contribution based on the prototype that we have developed, along with the experimental methods employed to accurately benchmark the efficiency of semi-static checks. We then proceed to evaluate the overall approach taken to developing semi-static checks in comparison to alternative viable solutions. We then analyse the safety of the language construct through a combination of static and runtime analysis approaches, these methods are also employed to focus more on reliability in terms of behaviour and synchronisation. We then touch on portability to different architectures and operating systems, and focus on the experimental methods employed for efficiency benchmarks and outlines more appropriate tests for industry settings.

\subsection{Overall Approach}

The goal of semi-static checks is to provide programmers control over conditional branching, reaping the performance benefit of direct method invocation whilst being able to change which method is called at runtime based on a condition. This kind of behaviour cannot be attained through conventional means available in the C++ standard; particular \emph{"branches"} can be generated at compile-time based on conditions through template instantiation, but the ability to do this with runtime generated conditions is fundamentally impossible since templates are a compile-time phenomenon. Mechanisms for switching between function calls exist without explicit conditional branching exist in the form virtual inheritance facilitated by v-table look-ups and function pointer de-referencing. However in reality this is slower than conditional branching in most scenarios even when branches are often mispredicted, and in the case for virtual functions, a good optimiser will often speculatively de-virtualise calls which reduce to conditionals. At the hardware level, these types of indirect calls require not only predictions for conditions but also data in the form of memory addresses (for example, a v-table lookup takes roughly 3 memory accesses before the \texttt{call} address is resolved), and as a result suffer from higher misprediction penalties and are more vulnerable to adverse cache-related effects in oppose to regular conditional statements which have a smaller instruction cache footprint.

This lack of flexibility meant that only assembly editing can facilitate the desired behaviour: fast deterministic branch-taking controlled by the programmer through a slower auxiliary interface. The reality of this is multifaceted. SMC is non-standard compliant and is widely considered a poor programming practice owing to complex maintenance, debugging and portability, despite it being used abundantly in debuggers and the Linux kernel. Given these concerns, the goal of development was to minimise the assembly editing component whilst maximising the simplicity of design for maintenance and portability reasons. However this is not the only viable approach.\\

\begin{figure}[t]
\centering
\resizebox{\textwidth}{!}{%
\begin{circuitikz}
\tikzstyle{every node}=[font=\LARGE]

\draw [ fill={rgb,255:red,242; green,242; blue,242} , line width=2pt, rounded corners ] (-115,32.5) rectangle (-110,30);
\draw [ fill={rgb,255:red,242; green,242; blue,242} , line width=2pt, rounded corners ] (-108.75,32.5) rectangle (-103.75,30);
\draw [ fill={rgb,255:red,242; green,242; blue,242} , line width=2pt, rounded corners ] (-101.25,32.5) rectangle (-96.25,30);
\node [font=\LARGE] at (-112.5,31.2) {Copy bytecode};
\node [font=\LARGE] at (-106.25,31.6) {Process and};
\node [font=\LARGE] at (-106.25,30.8) {adjust offsets};
\draw [ fill={rgb,255:red,242; green,242; blue,242} , line width=2pt, rounded corners ] (-107.75,28.25) rectangle (-104,27.25);
\draw [ fill={rgb,255:red,242; green,242; blue,242} , line width=2pt, rounded corners ] (-108,28.5) rectangle (-104.25,27.5);
\draw [ fill={rgb,255:red,242; green,242; blue,242} , line width=2pt, rounded corners ] (-108.25,28.75) rectangle  node {\LARGE Branches} (-104.5,27.75);
\node [font=\LARGE] at (-98.75,31.95) {Allocate};
\node [font=\LARGE] at (-98.75,31.25) {executable};
\node [font=\LARGE] at (-98.75,30.4) {memory};
\draw [ fill={rgb,255:red,229; green,235; blue,255} , line width=2pt, rounded corners ] (-100.95,34.5) rectangle  node {\LARGE set\_direction} (-96.5,33.5);
\draw [ fill={rgb,255:red,242; green,242; blue,242} , line width=2pt, rounded corners ] (-101.25,28.5) rectangle (-96.25,26);
\node [font=\LARGE] at (-98.75,27.5) {Write bytecode};
\node [font=\LARGE] at (-98.75,26.75) {to memory};
\draw [ fill={rgb,255:red,242; green,242; blue,242} , line width=2pt, rounded corners ] (-95,32.5) rectangle (-90,30);
\draw [ fill={rgb,255:red,229; green,235; blue,255} , line width=2pt, rounded corners ] (-94.5,34.5) rectangle  node {\LARGE branch} (-90.75,33.5);
\node [font=\LARGE] at (-92.6,31.7) {Execute};
\node [font=\LARGE] at (-92.6,30.85) {bytecode};
\draw [ fill={rgb,255:red,242; green,242; blue,242} , line width=2pt, rounded corners ] (-95,28.5) rectangle  node {\LARGE Free memory} (-90,26);
\draw [line width=2pt, -Stealth, dashed] (-103.75,31.25) -- (-101.25,31.25);
\draw [, line width=2pt](-112.5,30) to[short] (-112.5,28.25);
\draw [ line width=2pt, -Stealth] (-112.5,28.25) -- (-108.25,28.25);
\draw [ line width=2pt, -Stealth] (-110,31.25) -- (-108.75,31.25);
\draw [ line width=2pt, -Stealth] (-98.75,30) -- (-98.75,28.5);
\draw [ line width=2pt, -Stealth] (-98.75,33.5) -- (-98.75,32.5);
\draw [ line width=2pt, -Stealth] (-92.5,33.5) -- (-92.5,32.5);
\draw [line width=2pt, -Stealth, dashed] (-96.25,31.25) -- (-95,31.25);

\draw [ -Stealth] (-92.5,30) -- (-92.5,28.5);
\end{circuitikz}
}
\vspace*{5mm}
\caption{State diagram of semi-static conditions internals for both setup and execution using JIT-style runtime assembly generation.}
\end{figure}
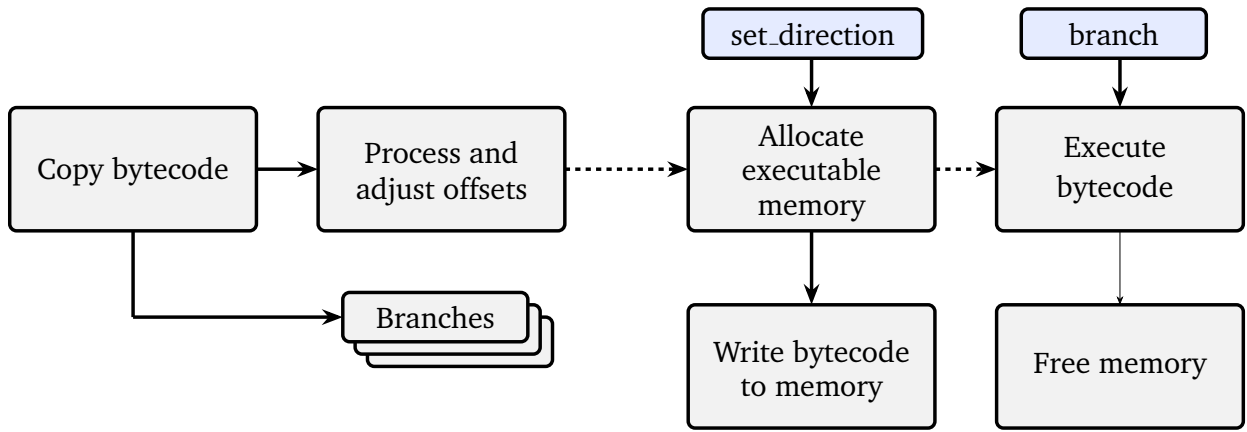

\noindent \textbf{Run-time code generation (JIT)} An alternative approach to SMC facilitated branch-changing is a just-in-time (JIT) approach, which has become popular in modern compilers and interpreters. Whilst JIT is generally more widely accepted form of runtime assembly manipulation, there are some inherent issues that make it inferior to our approach. The first is complexity. We show the process of preparing the byte-code associated with the branches to actually executing them, with the majority of the complexity residing in the first two stages. Copying the byte-code is error prone: whilst it is simple to find the preliminary instructions for the function, being able to accurately determine when the function is \emph{"finished"} in memory requires architecture specific complex logic. For example, using a \texttt{ret} opcode as a proxy would not work as there can be multiple exit points, and differentiating an opcode from a byte offset would require accounting for instruction length which introduces a substantial amount of administrative overhead. Then comes the challenge of adjusting position dependant instructions within this byte-code to work, such as PC relative instructions, which further increases development complexity. The best way to do this would be to have an in house JIT compiler as part of the language construct, which is impractical.

Even if hypothetically the former challenges where addressed, the eminent problem that would prevent JIT from being used in low-latency settings is the means of which it can be executed in modern C++.

\begin{lstlisting}[language=c++]
void* new_page = mmap(nullptr, 4096, 
    PROT_READ | PROT_WRITE | PROT_EXEC, 
    MAP_PRIVATE | MAP_ANONYMOUS, 
);
std::memcpy(new_page, bytecode, sizeof(bytecode));
(...)
int (*add)(int, int) = (int (*)(int, int))new_page;
int result = add(5, 3);
\end{lstlisting}

\noindent \\Above is a sample of two necessary components required required to facilitate JIT-style assembly execution in C++: memory allocation and execution. From user space, the only way to execute this newly generated byte-code is through casting the address of the newly created executable page to a function pointer and defencing it, which is much more expensive at runtime than conditional branching. Whilst runtime code generation could offset SMC machine clears since instructions are not being modified and executed, rather just generated, the overhead of pointer de-referencing in latency critical paths defeats the whole purpose of the language construct.\\

\noindent \textbf{Assembly editing} In comparison to JIT, the method used to develop semi-static conditions has many advantages. The first is clearly simplicity as runtime assembly editing using an intermediary trampoline function (\texttt{branch} method) is facilitated through a 4-byte \texttt{memcpy} at runtime. The simplicity of the concept directly translates to simplicity in development which is reflected by the overall size of software artefact. In terms of maintainability, which is important for a library which employs architecture specific optimisations, a simple implementation will easier adapt to future changes. From a low-latency application perspective, less code translates to less assembly instructions which have smaller instruction cache footprints, a favourable property for systems that prioritise keeping as much latency-critical code in lower level caches. The second advantage is branch-taking speed; using the assembly editing approach the branch-taking overhead is simply the overhead of a relative jump, which has superior prediction schemes from pointer de-referencing in JIT. A downside of the current approach is SMC penalties incurred during branch direction changing. Even in a hybrid approach where the trampoline can be hard-coded on separate executable pages, editing existing assembly is unavoidable. Whilst avoiding SMC penalties seems impossible using this scheme on modern architectures, it does open some interesting investigations in mitigation schemes, but nevertheless the trade-off for superior branch-taking ought to be favourable for these sorts of applications.
 
\subsection{Safety}

Assembly editing gives rise to undefined behaviour with respect to the C++ standard, and can bring forth security vulnerabilities. The first issue that can arise is that one or more of the branches lies at a signed displacement greater than $2^{32}$ bytes from the \texttt{jmp} instruction within the \texttt{branch} method. In this instance the program will redirect control flow to an area in the code segment which does not belong to any of the branches, resulting in adverse behaviour. Instances like this ought to be caught out as early as possible. To simulate this, two arbitrary function pointers with a displacement greater than $2^{32}$ are created and passed into the \texttt{BranchChanger} constructor as such:\\

\begin{lstlisting}[language=c++]
    using ptr = int(*)(int, int);
    ptr func_1;
    ptr func_2 = reinterpret_cast<ptr>(
        reinterpret_cast<intptr_t>(func_1) + 
        (static_cast<intptr_t>(1) << 34)
    );
    BranchChanger branch(func_1, func_2);
\end{lstlisting}

\noindent \\As a result the following runtime exception is raised upon instantiation:\\

\begin{itemize}
    \item[] terminate called after throwing an instance of 'branch\_changer\_error'\\

    what(): Supplied branch targets (as function pointers) exceed a 2GiB displacement
    from the entry point in the text segment, and cannot be reached with a 32-bit
	relative jump. Consider moving the entry point to different areas in the text
    segment by altering hot/cold attributes.\\

    Aborted (core dumped)
\end{itemize}

\noindent \\Another form of undefined behaviour stems from more than one instance of semi-static conditions being present at any given time during program execution. In this scenario, the conflicting instances of \texttt{BranchChanger} will share the same \texttt{branch} entry point due to template specialisation and compete for assembly editing of a single \texttt{jmp}. As a result, branches that do not belong to the immediate \texttt{BranchChanger} instance may be executed. When multiple instances do exist, the following error is raised:\\

\begin{itemize}
    \item[] terminate called after throwing an instance of 'branch\_changer\_error'\\

    what(): More than once instance for template specialised semi-static conditions detected. Program terminated as multiple instances sharing the same entry point is dangerous and results in undefined behaviour (multiple instances write to same function.\\

    Aborted (core dumped)
\end{itemize}

\noindent \\The former exception handling measures are effective at detecting the main behaviours associated with assembly editing that can result in detrimental effects later on in program execution. This, along with other higher level behaviour exceptions, have been integrated into a rigorous automated testing suite part of the library, allowing programmers to test the build prior to usage for peace of mind.

In the context of security, the primary vulnerability that arises using semi-static conditions is through changing executable page permissions to read/write/execute. Often times during program execution, multiple pages and hence multiple address ranges are vulnerable to memory modification from other processes, allowing attackers to inject their own code, access sensitive data or tamper with sensitive information. From a mitigation standpoint, eliminating this vulnerability completely is not possible, since page permissions need to be altered at least temporarily to facilitate assembly editing. What can be done is minimise the transient window of vulnerable address ranges by performing page permission modification and reversion exclusively within the \texttt{set\_direction} method, in this case an auxiliary version named \texttt{set\_direction\_safe}.\\

\begin{lstlisting}[language=c++]
    void set_direction_safe(bool condition)
    {
        if (condition != direction)
        {
            change_page_perimissions(bytecode, 
                PROT READ | PROT WRITE | PROT EXEC
            );
            std::memcpy(src, dest[condition], DWORD);
            change_page_perimissions(bytecode, 
                PROT READ | PROT EXEC
            );
            direction = condition;
        }
    }
\end{lstlisting}

\noindent \\Using this method, programmers can be ensured that the risk of such exploits are minimised. To ensure that \emph{safe-mode} is respected throughout the program lifetime, programmers can add the following flag upon compilation:\\

\begin{lstlisting}[language=c++]
    -DSAFE_MODE
\end{lstlisting}

\noindent \\However this comes with a cost. In a low-latency setting, using the secure branch-changing method introduces higher execution times and larger standard deviations owing to the two system calls used to alter page permissions, in addition to increased pressure on instruction-caches. When thinking of general cases, the more expensive cost of branch-changing has negative implications on amortisation, and will likely limit the use in scenarios where branches are well predicted. This is the inherit trade off in low-latency settings; security versus speed. In commercial software that is exposed directly to the user, such security ought to be included, however in the case of secretive proprietary software (such as trading systems) such security measures may not be necessary. In light of this, both API's are exposed to the programmer, whom can make an informed decision what to include.

\subsection{Reliability}

In terms of reliability, the main areas of focus in the context of semi-static conditions are correctness and consistency of behaviour. From a high-level perspective, an exhaustive formal proof that shows semi-static conditions has equivalent behaviour to conditional statements is unnecessary. In the context of semi-static conditions, static analysis of program behaviour does not bear much meaning since the program state itself is not static due to self-modifying assembly instructions! To evaluate correctness a test suite can be set up by running a tight loop that (1) changes the branch direction and (2) takes the branch successively and see if the wrong branch is taken based on the runtime condition.\\

\begin{lstlisting}[language=c++]
    while (run)
    {
        branch.set_direction(condition);
        branch.branch(counter);
        condition = !condition;
    }
\end{lstlisting}

\noindent \\The branches will have a similar function prologue which determines if the correct branch is executed using logical expressions:\\

\begin{lstlisting}[language=c++]
    void true_branch(int& counter)
    {
        counter += (condition ^ true);
        (...)
    }
\end{lstlisting}

\noindent \\Under these conditions, in a single threaded environment semi-static conditions will \textbf{always} exhibit correct behaviour, however this is not always the case in multi-threaded setting. Fundamentally, what is occurring are write operations to memory (producer) which are executed by the CPU (consumer) in some interleaved order giving rise to race conditions. The extent of how often the wrong branch is executed is heavily dependant upon how often context switching occurs followed by branch taking and how large the branches are in terms of assembly instructions, which is often very difficult to predict. For general use cases, wrong branches are executed relatively rarely (close to 0), whereas more chaotic systems where context switches occur every several microseconds or so result in much more inconsistent behaviour.

The absence of thread safety without synchronisation is a key drawback of the assembly editing approach. With the use of synchronisation, correct behaviour is assured however performance degrades significantly which can be seen from the examples. Although the overall focus in multi-threaded environments where limited, it would be interesting to observe if serialising instructions such as \texttt{lfence} and \texttt{mfence} can be used as a synchronisation mechanism unilaterally without compromising the performance of branch-taking in future investigations.

\subsection{Usage and Portability}

The software artifact is packaged as a static library which can be incorporated into project using the CMake build system. The decision to package semi-static checks as a static library over a dynamic library is due to performance reasons; static libraries present all code in the executable at compile time whereas dynamic libraries need to be loaded by the OS at runtime. Using the dynamic library methods require an extra layer of indirection through symbol table look-ups, and parts of the library themselves are brought into memory on-demand which is prone to cache misses and page faults, all of which contribute to jitter which is undesirable in low-latency setting. The drawbacks of using static libraries are increased compilation times and the size of the executable, which is made more prominent given the template-heavy implementation of semi-static conditions.

Cross platform builds on C++ are notoriously messy. CMake abstracts the differences between various build systems and compilers on different platforms which simplifies development and usage greatly. From a portability standpoint, the CMake pre-processor can be configured to selectively include architectural specific headers into the final build which is extremely beneficial given the low-level nature of the library. To begin the build (assuming the library is cloned into the same directory), users simply first specify a build directory as such:\\
\begin{itemize}
    \item[] \$ cmake -E make\_directory "build"
\end{itemize}
\noindent \\Next, the build system files can be generated in the newly created directory with\\
\begin{itemize}
    \item[] \$ cmake -E chdir "build" cmake ../
\end{itemize}
\noindent \\Then finally built with\\
\begin{itemize}
    \item[] \$ cmake --build "build"
\end{itemize}
\noindent Subsequent tests can be run to validate the build system using\\
\begin{itemize}
    \item[] \$ cmake -E chdir "build" ctest
\end{itemize}
\noindent \\To use the library, the generated archive must be compiled and linked against the branch library (libbranch.a) as such\\
\begin{itemize}
    \item[] \$ g++ mycode.cpp -std=c++17 -isystem semi-static-conditions/include
      -Lsemi-static-conditions/build -lbranch -o mycode
\end{itemize}
\noindent \\This necessitates the entire setup required to use the library, the only requirements being a relatively new version of CMake (version 3.2 and above) to facilitate the building! The core \texttt{BranchChanger} construct can be used through including the following header:\\

\begin{lstlisting}
    #include <branch.h>
\end{lstlisting}

\begin{table}[t]
\centering

\begin{tabular}{lcccccc}
\toprule
\multirow{2}{*}{Compiler} & \multicolumn{2}{c}{Windows} & \multicolumn{2}{c}{MAC} & \multicolumn{2}{c}{Linux} \\
\cmidrule(lr){2-3} \cmidrule(lr){4-5} \cmidrule(lr){6-7}
 & x86-64 & ARM & x86-64 & ARM & x86-64 & ARM \\
\midrule
\cellcolor{lightgray}{GCC} & \cellcolor{lightgray}{\checkmark} & \cellcolor{lightgray}{\ding{55}} & \cellcolor{lightgray}{\ding{55}}  & \cellcolor{lightgray}{\ding{55}}  & \cellcolor{lightgray}{\checkmark}  & \cellcolor{lightgray}{\checkmark}  \\
MSVC & \checkmark & \ding{55} & \ding{55} & \ding{55} & \checkmark & \ding{55} \\
\cellcolor{lightgray}{Clang} & \cellcolor{lightgray}{\checkmark} & \cellcolor{lightgray}{\ding{55}} & \cellcolor{lightgray}{\ding{55}}  & \cellcolor{lightgray}{\ding{55}}  & \cellcolor{lightgray}{\checkmark}  & \cellcolor{lightgray}{\checkmark}  \\
\bottomrule
\end{tabular}
\vspace*{2mm}
\caption{Compatibility matrix for the semi-static conditions library. Ticks are given to platforms where the library has been tested and is functional, and crosses are given to untested platforms and/or no functionality.}
\end{table}

\noindent \\Portability was a challenge from a development perspective. The implementation of semi-static checks utilises OS specific system calls, compiler specific attributes, inline assembly, and architecture specific offset computations which can all manifest in different combinations depending on the users system configuration. In terms of achieving the correct build, a dedicated single header file was utilized to define the platform using a series of pre-processor directives. These directives are employed throughout the library to enable the incorporation of architecture-specific, compiler-specific, and OS-specific code. It is possible to delegate this process entirely to CMake, however such an approach would necessitate users to specify the platform using a series of compiler flags which is rather tedious, and therefore avoided. 

For this initial version, popular compilers and operating systems where targeted for compatibility, on both x86-64 and ARM architectures. Builds where tested by invoking the library using the CMake steps shown above on different systems, results of the tests (including if the library was unable to tested) are shown in Table 5.1. Unfortunately, semi-static conditions is not portable on Apple silicon owing to the Hardened Runtime OS security feature which prevents page permissions from being changed to write/execute. Some of these security features can be disabled allowing for binary editing of runtime-allocated pages using the JIT approach, however as discussed, does not meet the latency requirements for this construct.

\subsection{Experimental Method}

A great focus of this work was to evaluate the performance of semi-static conditions against the current state of the art and rationalise the observed behaviour and inefficiencies observed from a micro-architectural perspective. Due to the sensitivity of the measurements, careful consideration was taken in developing the measuring suite that prioritises the accuracy and reproducibility of measurements. The choice of measurement instruments where sensible; reference cycle counters where sufficient in producing accurate measurements allowing for the observation of subtle behaviours on the hardware level owing to its low usage overhead. However a number of improvements can be made to the overall production setup in the context of OS tuning and networking. 

Given the primary objective of this work, which centers around exploring innovative branch optimization techniques in a HFT context, it becomes imperative that the OS chosen for conducting performance benchmarks mirrors an industrial setup to the greatest extent possible. The selection process for the system was driven by a preference for a Linux-based High-Performance Computing (HPC) environment featuring a Intel i7 processor. Additionally, a cloud-based Virtual Machine (VM) with an Intel Xeon chip was utilized. It is, however, important to note that in both cases, complete root privileges were not attainable. This limitation had significant implications, particularly in the domain of kernel tuning. Specific adjustments related to CPU scaling and scheduling could not be replicated to the same extent as they would be in a genuine HFT system. This discrepancy is noticeable in the variations and distributions observed in the recorded measurements. To address this challenge, a nuanced approach was adopted. The measurement suite was meticulously tailored to execute numerous iterations, allowing the CPU to stabilize at an optimal operating frequency. This strategic approach effectively minimized measurement variance, enabling subsequent interpretation through statistical tests. While it remains true that with meticulous kernel tuning, the necessity for extensive statistical treatment could be substantially diminished, leading to clearer and more refined data, the outcomes presented  successfully fulfilled the intended communication goals. Despite the existing constraints, the data provided here aptly conveyed the essential insights and findings.

\begin{figure}[t]
    \centering
    \includegraphics[width=\textwidth]{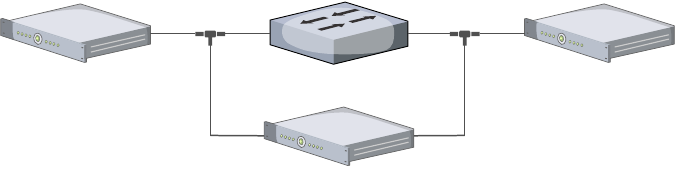}
    \vspace*{1mm}
    \caption{Ideal productionised setup of benchmarking suite for HFT applications. Left server replays market data across a high speed ethernet cable, the switch in the middle is equipped with high precision time-stamps, server on the right is the production system to be measured, and server in the middle computes response times. Adapted from \cite{carl2017cpp}.}
    \label{fig:enter-label}
\end{figure}

The second improvement is of a more comprehensive nature, and its attribution extends to the entire system. An inherent limitation associated with microbenchmarking in isolation, particularly when examining minute differences at the low-nanosecond scale, is that the tests themselves possess an artificial nature, deviating from the essence of a genuine production environment. Despite the endeavors to craft test suites that emulate a pseudo-realistic production environment in terms of computational workload, the actual representation of behavioral changes remains somewhat imprecise when contrasted with a fully functional High-Frequency Trading (HFT) system. Yet, it's important to acknowledge the inherent limitations in addressing this issue. The proprietary nature of trading firms' code renders it a closely guarded trade secret, largely due to its direct influence on profitability. The epitome of an ideal experimental arrangement, a concept initially proposed by Carl Cook as the most robust methodology for micro-benchmarking latency enhancements within a trading system \cite{carl2017cpp}. It's worth noting, however, that such a setup warrants its own comprehensive analysis and entails substantial complexity and associated costs.

\section{Conclusions}

We have shown that software-level branch optimisation can be achieved using a novel language construct, semi-static conditions, offering superior execution latencies to the current state of the art in both specialised and generalised scenarios. There is no doubt that the methodologies applied to facilitate this optimisation are unconventional. The use of assembly editing has long been shunned from a software development and micro-architectural perspective, however it seems that revisiting this old yet interesting nuance of low-level development has opened new doors for low-latency optimisation. 

The notion of semi-static checks is simple, separate branch-taking from condition evaluation, however as we have seen throughout this work this is quite a surface level interpretation. What really is happening is a trade of branch-prediction schemes on the hardware level which is facilitated through assembly editing. In semi-static checks, the conditional statement is re-engineered as polymorphic relative jump instruction which can be controlled directly by the programmer through an auxiliary interface. As an artefact of this modification, the idea of execute-stage branch mispredictions are completely eliminated, and replaced with less severe BAC correction ocurring earlier in the decode stage. Unlike conditional branching, BAC corrections caused by stale BTB entries can be mitigated from latency critical code through pre-empative branching, a new type of active \emph{'BPU warming'} that guarantees a BTB correction upon instruction retirement and circumvents any associated penalties for the current branch direction indefinitely. The result? A powerful decoupling that optimises branch-taking to the theoretical limit. Our results showcased the superior performance of semi-static conditions against conditional branching when often mispredicted, with lower execution latencies and more deterministic standard deviations, an desirable property for HFT. Even when branches are predicted well, semi-static conditions show faster branch-taking as a consequence of a cleaner control path and fewer instructions on the machine code level, allowing the expensive branch-changing method to be amortised in single and multi-threaded scenarios.

The phenomenon of semi-static checks brings forth many avenues for further investigation, particularly around the application of self-modifying binaries for program optimisation, and minimising any adverse hardware effects. We took a detailed look at the behaviour of writing instructions into executable memory and revealed that locality between assembly editing and the assembly being executed results in severe processor penalties in the form of SMC machine clears. Though it is understood that locality in terms of caching, prefetching and paging has a proportional effect on the severity of SMC penalties, this investigation was unable to properly quantify the effect. Further investigations into understanding this would allow the branch-changing portion of the construct to be further optimised and make it more flexible to use cases, in addition this would be beneficial from a fundamental perspective in understanding the effects of SMC fully. In addition, the general idea of making targeted and granular changes in assembly to optimise various language level intrinsics should be investigated more broadly. Perhaps the general idea of substituting more BPU-friendly control flow instructions and modifying them can be applied to many different types of branches, more specifically dynamic dispatching and any form of register jumps.

In sum, the absence of language level-optimisations for hardware based branch prediction has birthed a new toolkit in tackling optimisation problems in low-latency setting. With this work pioneering the use of assembly editing for branch optimisation, we hope it serves as a benchmark for future investigations that apply this long-neglected and interesting programming practice to more complex problems surrounding optimisation.
\bibliographystyle{vancouver}


\end{document}